\documentclass[%
  aip,
 amsmath,amssymb,
 reprint,%
]{revtex4-1}

\usepackage{graphicx}
\usepackage{dcolumn}
\usepackage{bm}
\usepackage{subfigure}
\usepackage[utf8]{inputenc}
\usepackage[T1]{fontenc}
\usepackage{mathptmx}
\usepackage{etoolbox}
\usepackage{color}

\usepackage{lineno}
\usepackage{amsmath}
\usepackage{bm}
\usepackage{subcaption} 
\usepackage{hyperref}
\usepackage{float}
\usepackage{algorithm}
\usepackage{algpseudocode}
\usepackage{placeins}
\usepackage{tikz}
\usetikzlibrary{shapes}

\newcommand\bnabla{\bm{\nabla}}

\makeatletter
\def\@email#1#2{%
 \endgroup
 \patchcmd{\titleblock@produce}
  {\frontmatter@RRAPformat}
  {\frontmatter@RRAPformat{\produce@RRAP{*#1\href{mailto:#2}{#2}}}\frontmatter@RRAPformat}
  {}{}
}%
\makeatother
\begin{document}

\preprint{AIP/123-QED}

\title{An open-source, adaptive solver for particle-resolved simulations with both subcycling and non-subcycling methods}
\author{Xuzhu Li}
\affiliation{Research Center for Astronomical Computing, Zhejiang Laboratory, Hangzhou 311100, China}
\affiliation{School of Mechanical Engineering, Hefei University of Technology, Hefei 230009, China}
\author{Chun Li}%
\affiliation{School of Energy and Power Engineering, Lanzhou University of Technology, Lanzhou, Gansu 730050, China}%

\author{Xiaokai Li}%
\affiliation{School of Physical Science and Technology, ShanghaiTech University, Shanghai 201210, China}%

\author{Wenzhuo Li}%
\affiliation{Advanced Propulsion Laboratory, Department of Modern Mechanics, University of Science and Technology of China, Hefei 230026, China}%

\author{Mingze Tang}
\affiliation{School of Aeronautics, Northwestern Polytechnical University, Xi'an 710072, China }

\author{Yadong Zeng}
\altaffiliation{Correspondence author: zengx372@utexas.edu}
\affiliation{Department of Computer Science, University of Texas at Austin, Texas 78712, USA}

\author{Zhengping Zhu}
\altaffiliation{Correspondence author: zhuzhp@zhejianglab.edu.cn}
\affiliation{Research Center for Astronomical Computing, Zhejiang Laboratory, Hangzhou 311100, China}

\date{\today}

\begin{abstract}
We present the IAMReX, an adaptive and parallel solver for particle-resolved simulations on the multi-level grid.~The fluid equations are solved using a finite-volume scheme on the block-structured semi-staggered grids with both subcycling and non-subcycling methods.~The particle-fluid interaction is resolved using the multidirect forcing immersed boundary method.~The associated Lagrangian markers used to resolve fluid-particle interface only exist on the finest-level grid, which greatly reduces memory usage.~The volume integrals are numerically calculated to capture the free motion of particles accurately, and the repulsive potential model is also included to account for the particle-particle collision.~We demonstrate the versatility, accuracy, and efficiency of the present multi-level framework by simulating fluid-particle interaction problems with various types of kinematic constraints.~The cluster of monodisperse particles case is presented at the end to show the capability of the current solver in handing with multiple particles.~The source code and testing cases used in this work can be accessed at~\url{https://github.com/ruohai0925/IAMR/tree/development}.~Input scripts and raw postprocessing data are also available for reproducing all results.

\end{abstract}

\maketitle














\section{Introduction} \label{S:1}
Particle-laden flows are of common occurrence in natural and industrial applications~\cite{balachandar2010turbulent,brandt2022particle}, such as sediment transport, turbidity currents and fluidized bed reactors.~The comprehension of the physics underlying particle-turbulence interactions is crucial for these applications.~For particles that are smaller than the Kolmogorov scale, point-particle simulations have provided deep insight into the interactions~\cite{squires1990particle,wang1993settling,ferrante2003physical,vance2006properties,zhao2010turbulence,lee2015modification,li2016modulation,wang2019two,zheng2021modulation}.~For particles that are larger than the Kolmogorov scale, particle-resolved simulations are widely utilized to study the interactions~\cite{pan1997numerical,bagchi2003effect,burton2005fully,shao2012fully,picano2015turbulent,wang2016flow,wang2022direct,wang2023drag}.~The immersed boundary method has become a popular approach for particle-resolved simulations due to its ability to avoid the time-consuming regeneration of Eulerian grids for moving boundaries.~It typically utilizes structured Cartesian grids that are fixed in time.~The no-slip boundary condition on the particle surface is satisfied implicitly by applying a volumetric forcing to the flow around the particle surface~\cite{mittal2005immersed,sotiropoulos2014immersed,griffith2020immersed,verzicco2023immersed}.

Different techniques have been developed to derive the volumetric forcing.~One such category is the feedback forcing technique~\cite{goldstein1993modeling,saiki1996numerical}, the volumetric forcing is calculated by a system of virtual springs and dampers attached to the particle surface $\bm f=\alpha \int_{0}^{t} (\bm u - \bm u_b)d\tau +\beta (\bm u - \bm u_b)$, where $\alpha$ and $\beta$ are two free parameters and $u_b$ is particle surface velocity.~The penalty technique~\cite{angot1999penalization,specklin2018sharp} can be regarded as a special instance of the feedback technique, where the parameters are set to $\alpha=0$ and $\beta=1/K$.~The undesirable feature of the feedback technique is that those two free parameters are determined based on the flow conditions.~Additionally, the characteristic time scales of the spring-damper system severely restrict the computational time step~\cite{lai2000immersed,lee2003stability}.~Another category is the direct forcing technique~\cite{fadlun2000combined,griffith2005order,uhlmann2005immersed}, in which the particle surface is discretized using Lagrangian markers.~Each marker experiences a Lagrangian interface force derived from the difference between the desired and actual velocities at the particle interface.~The volumetric forcing is then calculated through spreading the Lagrangian interface force.~Compared with the feedback technique, the direct forcing technique is more versatile since it eliminates stability constrains and does not require empirical constants~\cite{lai2000immersed,lee2003stability}.~However, the direct forcing technique is based on a single Lagrangian marker.~When applied to multiple markers, the direct forcing on each Lagrangian marker will be affected by its neighbors, which may not enforce the no-slip boundary condition well.~The multidirect forcing technique~\cite{luo2007full,kempe2012improved,breugem2012second} is developed as a remedy to this problem.~The no-slip boundary condition is more accurately satisfied by several applications of direct forcing via an explicit iterative procedure.~In this work, we implement the multidirect forcing technique on a semi-staggered grid, which avoids the checkboard issue of the collocated grid~\cite{martin2000cell,martin2008cell} and can resolve multiple particles~\cite{almgren1998conservative}.

Owing to the above advantages of the direct forcing immersed boundary (DFIB) method, it has been successfully utilized to study the interaction between thousands of particles and near-wall turbulence via particle-resolved sediment transport simulations.~\citet{ji2013direct,ji2014saltation} investigated the statistical features of the near-wall turbulence and saltation particles in sediment transport.~\citet{kidanemariam2014direct,kidanemariam2017formation,kidanemariam2022open} investigated the formation of sediment patterns in sediment transport.~\citet{scherer2022role} investigated the role of turbulent large-scale motions in forming sediment patterns in sediment transport.~\citet{vowinckel2016entrainment} investigated the mechanism of particle entertainment over an erodible bed.~\citet{zhu2022particle} investigated the probability distribution functions of several saltation parameters in sediment transport.~\citet{jain2020effect} investigated the sediment transport with non-spherical particles.~Although particles accumulate near the sediment bed in sediment transport, those simulations employed uniform grids across the entire computational domain to resolve not only the sediment bed but also the particle-free region further away from it, resulting in a huge amount of computation.


Great efforts were made to reduce the number of Eulerian grids and Lagrangian markers required by the DFIB method.~The DFIB method employs a Dirac delta function to interpolate the fluid velocity from an Eulerian grid onto Lagrangian markers as well as spreading the forcing in the opposite direction.~The commonly used Dirac delta function proposed by~\citet{roma1999adaptive} requires uniform Eulerian grids to conserve total force and torque, which substantially increases the number of Eulerian grids, especially for the channel flow.~Because the turbulence scale near the wall is much smaller than that in the channel center.~By employing the reproducing kernel particle method (RKPM)~\cite{liu1995reproducing} to modify the Dirac delta function, the DFIB method can be applied to non-uniform Eulerian grids, significantly reducing the Eulerian grids required by the DFIB method~\cite{pinelli2010immersed,akiki2016immersed,jang2017immersed}.~Furthermore,~\citet{akiki2016immersed} proposed a dynamic non-uniform distribution of Lagrangian markers on a sphere, which resulted in a 76.96\% reduction in the number of Lagrangian markers compared with uniform distribution in the particle-resolved simulation of 640 monodisperse spherical particles randomly distributed in the channel flow.

In addition to the non-uniform mesh approach described above, another idea to reduce the Eulerian cells requirement is to employ a multi-level grid and utilize the adaptive mesh refinement (AMR) technique for the DFIB method.~AMR is a highly effective computational technique for tackling the complexities of fluid flows~\cite{berger1984adaptive,berger1989local}.~It stands out for its ability to dynamically adjust grid resolution based on the evolving solution, optimizing computational resources precisely where they are most needed.~For the particle-resolved simulation, one can refine the grid cells near the particle interface and/or at wake of particles where the velocity gradient is large.~Unlike static mesh refinement, which maintains a fixed grid hierarchy, AMR can refine and coarsen the grid as needed.~There are three primary types of Adaptive Mesh Refinement (AMR).~The first is cell-based refinement, where each cell that meets refinement criteria is divided into four (in 2D) or eight (in 3D) smaller cells, organized in a quad- or oct-tree structure~\cite{guittet2015stable,mirzadeh2016parallel,popinet2003gerris}.~Libraries such as p4est~\cite{BursteddeWilcoxGhattas11} and libMesh~\cite{libMeshPaper} efficiently support this approach.~The second type is patch-based refinement, which generalizes the cell-based method but requires logically rectangular regions, often called grids, patches, or boxes.~This strategy, also known as quad-tree or oct-tree patch-based refinement, constructs fine patches of a minimum size in each dimension.~The FLASH library~\cite{fryxell2000flash} supports this type.~The third type, also patch-based, organizes data into levels of refinement based on mesh resolution.~Unlike tree-structured methods, this approach constructs variable-sized patches that are logically rectangular, which makes it relatively easy to use the domain decomposition method for parallelization~\cite{gunney2016advances}.~Equations on the nested patches can also be solved efficiently utilizing the multigrid (MG) solver~\cite{almgren1998conservative}.A number of open-source libraries, such as AMReX~\cite{zhang2019amrex,zhang2020amrex}, ForestClaw~\cite{burstedde2014forestclaw}, Chombo~\cite{colella2009chombo}, et al., support this approach.~We note the last two types of AMR also fall into the category of two "block-structured refinement".~In this work, we exclusively investigate the third type of AMR and build our adaptive solver on a block-structured framework~\cite{zhang2019amrex,zhang2020amrex}.

Some previous studies have explored the combination of AMR with particle-resolved simulations.~For instance, ~\citet{bhalla2013unified} integrated the distributed Lagrange multiplier (DLM) immersed boundary (IB) method with block-structured AMR, demonstrating the accuracy of their approach through test cases involving a single particle in the single-phase flow.~\citet{zeng2022subcycling} also incorporated the DLM algorithm within a collocated AMR grid framework.~However, their validation was limited to single-particle scenarios.~Compared with the generation of markers in~\cite{uhlmann2005immersed,zhu2022particle,kempe2012improved}, a known limitation of the DLM algorithm is that Lagrangian markers must be placed within all particles, leading to increased computational costs and memory usage.~Additionally, ~\citet{bhalla2014fully} simulated the dielectrophoretic motion of particles in microfluidic channels, while ~\citet{li2009mesh} combined tree-structured AMR with unstructured grids to simulate spray particles in multiphase flows.~\citet{nangia2019dlm} treated a point absorber (a type of wave energy converter) as a particle and added a spring-damping system to study its energy absorption efficiency in waves.~However, in those studies, the flow solutions were updated using a composite time-stepping approach, where the discretized equations for velocity and pressure were coupled across coarse-fine grid boundaries and solved simultaneously at multiple levels.~This coupling constrained the time step to the finest grid spacing to maintain numerical stability.~In this work, we develop an adaptive AMR framework that allows level-by-level advancement, using both subcycling and non-subcycling methods, for particle-resolved simulations.~Since an AMR framework for particle-resolved simulations involving multiple particles is still lacking, developing such an open-source framework would improve the capability of the DFIB method and help us have a deeper comprehension of interactions between multiple large particles and turbulence.


The focus of the present paper is to develop an adaptive level-by-level AMR framework for the DFIB method, which greatly reduces its Eulerian grid cells and the computational time for particle-resolved simulations.~Our open-source framework can efficiently simulate multiple particles within the flow field using either subcycling or non-subcycling methods.~The remainder of this paper is organized as follows: we start with the mathematical formulation of the fluid-particle system in Section~\ref{S:2}, including the operators used in the Lagrangian-Eulerian interaction.~Next, both the single-level and multi-level advancement algorithms are presented in Section~\ref{S:3}.~We first describe the numerical discretization of the single level in Section~\ref{S:singleleveladvancementfluid} in which different types of kinematic constraints are considered (Session~\ref{S:typesofsolidconstraints}).~For the multi-level time advancement in Session~\ref{S:multileveladvancement}, we compare the subcycling and non-subcycling methods (Session~\ref{S:52}) and highlight the benefits of the synchronization operations (Session~\ref{S:sync}).~We then briefly introduce our open-source framework IAMReX in Session~\ref{S:IAMReX}.~The particle-related validation cases that highlight the accuracy, efficiency, and robustness of our adaptive solver are then given in Section~\ref{sec:Result}.~Finally, the conclusions and future work are given in Section~\ref{sec:Conclusion}.




\section{Mathematical formulation} \label{S:2}

This section describes the governing equations for a fluid-particle system occupying a three-dimensional multi-level Cartesian grid $\Omega \subset \mathbb{R}^{3}$.~The upper-left part of Fig.~\ref{fig:multilevelsolid} shows a schematic of two particles on a three-level grid with AMR.~When a schematic is sliced, the particles can be seen distributed on the finest level from the bottom-left corner of Fig.~\ref{fig:multilevelsolid}.~The momentum and material incompressibility equations are described using a fixed Eulerian coordinate system $\mathbf{x}=\left(x_{1}, x_{2}, x_{3}\right) \in \Omega$.~The markers attached to the particle are described using a Lagrangian coordinate system, where $\mathbf{s}=\left(s_{1}, s_{2}, s_{3}\right) \in \Omega_{\rm c}$ denotes the fixed material coordinate system attached to the structure and $\Omega_{\rm c} \subset \mathbb{R}^{3}$ is the Lagrangian curvilinear coordinate domain.~The position of the particle at time $t$ is $\mathbf{X}(\mathbf{s}, t)$; it occupies a volumetric region $V_{\mathrm{b}}(t) \subset \Omega$.~The equations of motion of the coupled fluid-particle system are

\definecolor{mycolorblack}{RGB}{0,0,0}
\definecolor{mycolorpurple}{RGB}{128,0,128}
\definecolor{mycolororange}{RGB}{255,165,0}
\definecolor{mycolorgreen}{RGB}{0,128,0}

\newcommand{\markercircle}{\raisebox{0.5pt}{\tikz{\node[draw,scale=0.5,circle,fill=mycolorblack](){};}}}
\newcommand{\markercirclegreen}{\raisebox{0.5pt}{\tikz{\node[draw,scale=0.5,circle,fill=mycolorgreen](){};}}}
\newcommand{\markersquare}{\raisebox{0.5pt}{\tikz{\node[draw,scale=0.4,regular polygon, regular polygon sides=4,fill=mycolororange,rotate=0](){};}}}
\newcommand{\markersquaretwo}{\raisebox{0.5pt}{\tikz{\node[draw,scale=0.4,regular polygon, regular polygon sides=4,fill=mycolororange,rotate=0](){};}}}

\begin{figure}[ht]
	\centering
	\includegraphics[width=1.0\linewidth]{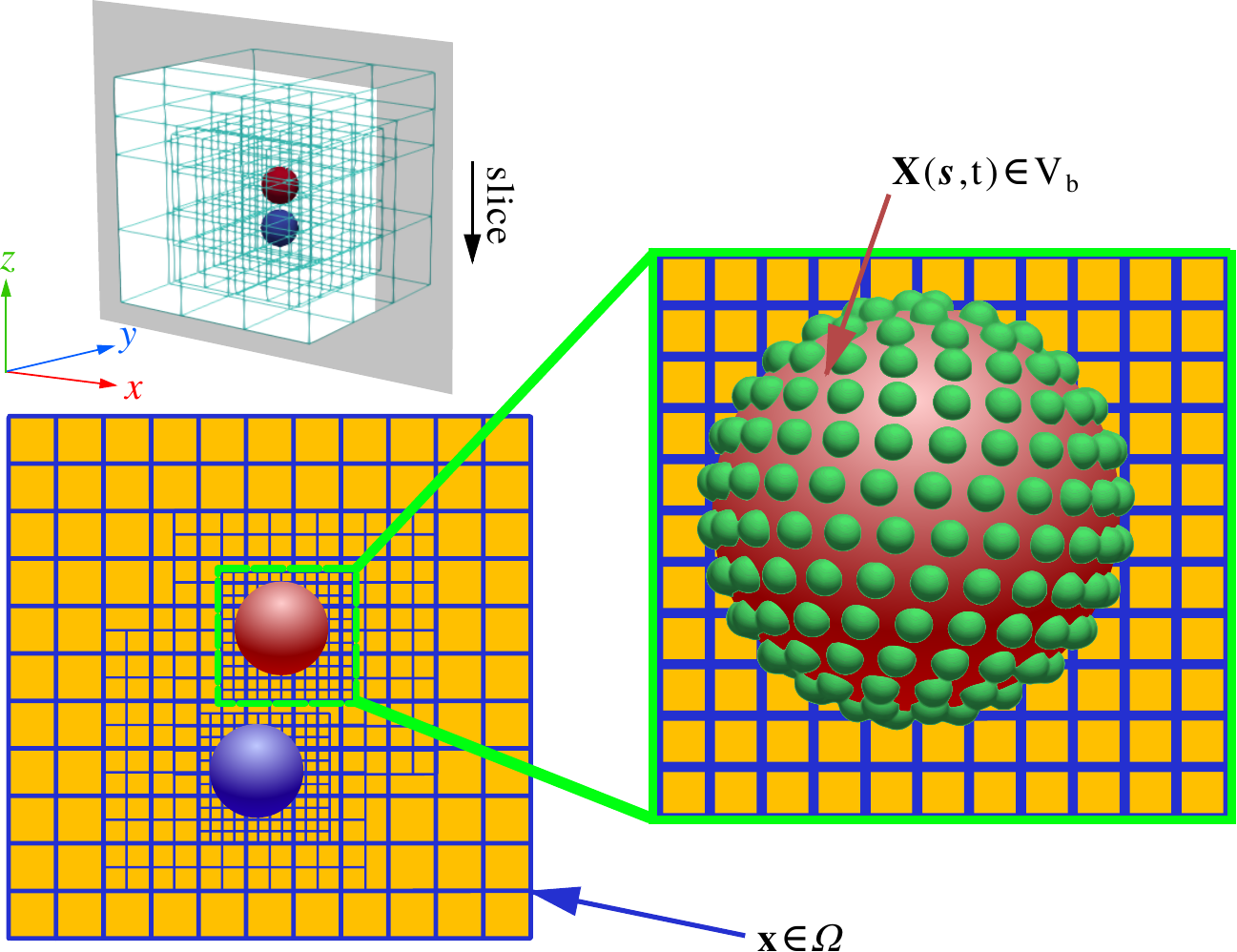}
	\caption{Upper-Left: schematic of two solid particles on a three-dimensional multi-level Cartesian grid.~Bottom Left: a slice of the upper-left schematic.~Right: schematic of Eulerian grid cells and Lagrangian markers.~The Eulerian grid cells (\protect\markersquare, orange) discretize the $\Omega$ region, and the Lagrangian markers (\protect\markercirclegreen, green) discretize the $V_{\mathrm{b}}(t)$ region.}
	\label{fig:multilevelsolid}
\end{figure}


\begin{linenomath*} \begin{equation}\label{eq:ns}
\begin{aligned}
\rho_{f} \left(\frac{\partial \mathbf{u}}{\partial t}(\mathbf{x}, t)+ \bnabla \cdot \left(\mathbf{u}(\mathbf{x}, t) \mathbf{u}(\mathbf{x}, t) \right) \right)=-\bnabla p(\mathbf{x}, t)+\\
\bnabla \cdot \left[\mu_{f} \left( \bnabla\mathbf{u}(\mathbf{x}, t) + \bnabla\mathbf{u}(\mathbf{x}, t)^{T}\right)  \right] +\rho_{f}\mathbf{g}+\mathbf{f}_\mathrm{c}(\mathbf{x}, t),
\end{aligned}
\end{equation} \end{linenomath*}

\begin{linenomath*}
\begin{equation}\label{eq:div}
\begin{aligned}
\bnabla \cdot \mathbf{u}(\mathbf{x}, t)=0,
\end{aligned}
\end{equation} \end{linenomath*}

\begin{linenomath*}
\begin{equation}\label{eq:f}
\begin{aligned}
\mathbf{f}_{\mathrm{c}}(\mathbf{x}, t)=\int_{V_{\mathrm{b}}(t)} \mathbf{F}_{\mathrm{c}}(\mathbf{s}, t) \delta(\mathbf{x}-\mathbf{X}(\mathbf{s}, t)) \mathrm{d} \mathbf{s},
\end{aligned}
\end{equation} \end{linenomath*}

\begin{linenomath*}
\begin{equation}\label{eq:x}
\begin{aligned}
\frac{\partial \mathbf{X}}{\partial t}(\mathbf{s}, t)=\mathbf{U}(\mathbf{s}, t),
\end{aligned}
\end{equation} \end{linenomath*}

\begin{linenomath*}
\begin{equation}\label{eq:u}
\begin{aligned}
\mathbf{U}(\mathbf{s}, t)=\int_{V_{\mathrm{b}}(t)} \mathbf{u}(\mathbf{x}, t) \delta(\mathbf{x}-\mathbf{X}(\mathbf{s}, t)) \mathrm{d} \mathbf{x}.
\end{aligned}
\end{equation} \end{linenomath*}

Here, $\mathbf{u}(\mathbf{x}, t)$ is the Eulerian velocity of the coupled fluid-particle system, $p(\mathbf{x}, t)$ is the pressure, $\rho_{f}$ is the Eulerian density field, and $\mu_{f}$ is the dynamic viscosity of the fluid-structure system.~The gravitational acceleration is written as 
$\mathbf{g}=\left(g_{1}, g_{2}, g_{3}\right)$.~In Eq.~(\ref{eq:ns}),
$\mathbf{f}_\mathrm{c}(\mathbf{x}, t)$ represents the Eulerian force density, which accounts for the presence of the solid in the domain.~$\delta(\mathbf{x})=\Pi_{i=1}^{3} \delta\left(x_{i}\right)$ represents the three-dimensional Dirac delta function, which is employed to exchange the information between the Eulerian quantity and Lagrangian quantity.~Specifically, Eq.~(\ref{eq:f}) converts the Lagrangian force density $\mathbf{F}_{\mathrm{c}}(\mathbf{s}, t)$ to an equivalent Eulerian force density $\mathbf{f}_{\mathrm{c}}(\mathbf{x}, t)$, in an operation that is referred to as \emph{force spreading}.~Eq.~(\ref{eq:u}) maps the Eulerian velocity $\mathbf{u}(\mathbf{x}, t)$ to the Lagrangian marker velocity $\mathbf{U}(\mathbf{s}, t)$, in an operation that is referred to as \emph{velocity interpolation}.~For notational convenience, we denote the force spreading operation in Eq.~(\ref{eq:f}) as

\begin{linenomath*}
\begin{equation}\label{eq:fs}
\mathbf{f}_{\mathrm{c}}=\boldsymbol{\mathcal{S}}[\mathbf{X}] \mathbf{F},
\end{equation} \end{linenomath*}
where $\boldsymbol{\mathcal{S}}[\mathbf{X}]$ is the force spreading operator.~Similarly, the velocity interpolation operation in Eq.~(\ref{eq:u}) is written in shorthand notation as

\begin{linenomath*}
\begin{equation}\label{eq:vi}
\mathbf{U}=\boldsymbol{\mathcal{J}}[\mathbf{X}] \mathbf{u},
\end{equation} \end{linenomath*}
where $\boldsymbol{\mathcal{J}}[\mathbf{X}]$ is the velocity interpolation operator.~We note that \emph{force spreading} and \emph{velocity interpolation} work together to satisfy no-slip boundary conditions at the fluid-solid interface.~As shown in~\cite{peskin2002immersed,nangia2019dlm}, these two coupling operators also conserve energy as long as $\boldsymbol{\mathcal{S}}$ and $\boldsymbol{\mathcal{J}}$ are adjoint.

\section{Numerical Discretization} \label{S:3}

This section gives the numerical discretization of Eq.~\ref{eq:ns}-\ref{eq:u}.~We first describe the discretization of the fluid system on the single level in section~\ref{S:singleleveladvancementfluid} and then discuss two types of kinematic constraints in section~\ref{S:typesofsolidconstraints}.~The discretization and advancement on the multi-level grid with AMR are detailed in section~\ref{S:multileveladvancement}, in which both subcycling and non-subcycling methods are used.~The open-source code IAMReX (Section ~\ref{S:IAMReX}) is introduced at the end.        

\subsection{Single-level advancement} \label{S:singleleveladvancementfluid}

\begin{figure}[ht]
    \includegraphics[width=0.65\linewidth]{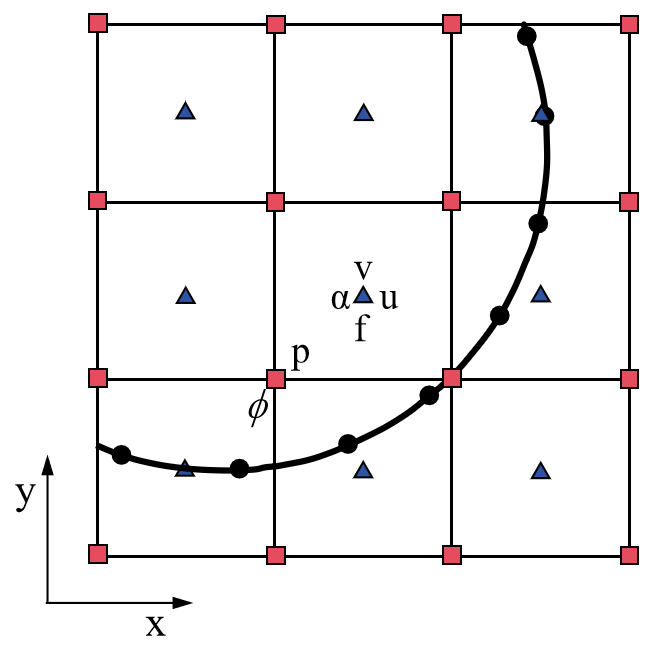}
    \caption{Sketch of the two-dimensional semi-staggered grid and variable locations.~The blue triangles, red squares, and black circles represent cell-centered variables, node-centered variables, and interface Lagrangian markers.}
    \label{fig:semi-staggered-grid}
\end{figure}

To solve the partial differential equations of Eq.~\ref{eq:ns}-\ref{eq:u}, the canonical projection~\cite{chorin1967numerical, almgren1998conservative} is applied to the semi-staggered grid.~As shown in Fig.~\ref{fig:semi-staggered-grid}, the fluid velocity ($u$ and $v$), the Eulerian force $f$, and particle volume fraction $\alpha$ are located at the cell center.~The pressure $p$ and level set function $\phi$ are at the node center.~The temporal and spatial discretizations of equations for single-level advancement are considered here.~At the time $t^{n}$, the Eulerian velocity $\mathbf{u}^{n}$ and pressure $p^{n-1/2}$ are known.~The particle position $\mathbf{X}^{n}_l$ and velocity $\mathbf{U}\left(\mathbf{X}^{n}_l\right)$ are also available.~The time advancement during the interval $[t^{n}, t^{n+1}]$ proceeds as follows.

\textbf{Step 1}: The intermediate velocity $\widetilde{\mathbf{u}}^{*,n+1}$ is solved semi-implicitly as
\begin{linenomath*}
\begin{linenomath*} \begin{equation} \label{eq:31}
\begin{aligned}
\rho_{f}\left(\frac{\widetilde{\mathbf{u}}^{*,n+1}-\mathbf{u}^{n}}{\Delta t}+\bm{\nabla} \cdot \left(\mathbf{u}\mathbf{u} \right)^{n+\frac{1}{2}}\right)=-\bm{\nabla} p^{n-\frac{1}{2}}+\\
\frac{1}{2}\left(\bm{\nabla} \cdot \mu \bm{\nabla}{\widetilde{\mathbf{u}}^{*,n+1}}+
\bm{\nabla} \cdot \mu \bm{\nabla} {\mathbf{u}}^{n}\right)+\rho_{f}\mathbf{g},
\end{aligned}
\end{equation}\end{linenomath*}
\end{linenomath*}
where the convective term $\bm{\nabla} \cdot \left(\mathbf{u}\mathbf{u} \right)^{n+\frac{1}{2}}$ is calculated using the second-order Godunov scheme~\cite{almgren1998conservative,sussman1999adaptive, sverdrup2018highly, zeng2023consistent}.~In this step, only the pure fluid system is solved and no particle-related influence is included.

\begin{figure}[ht]
    \includegraphics[width=0.65\linewidth]{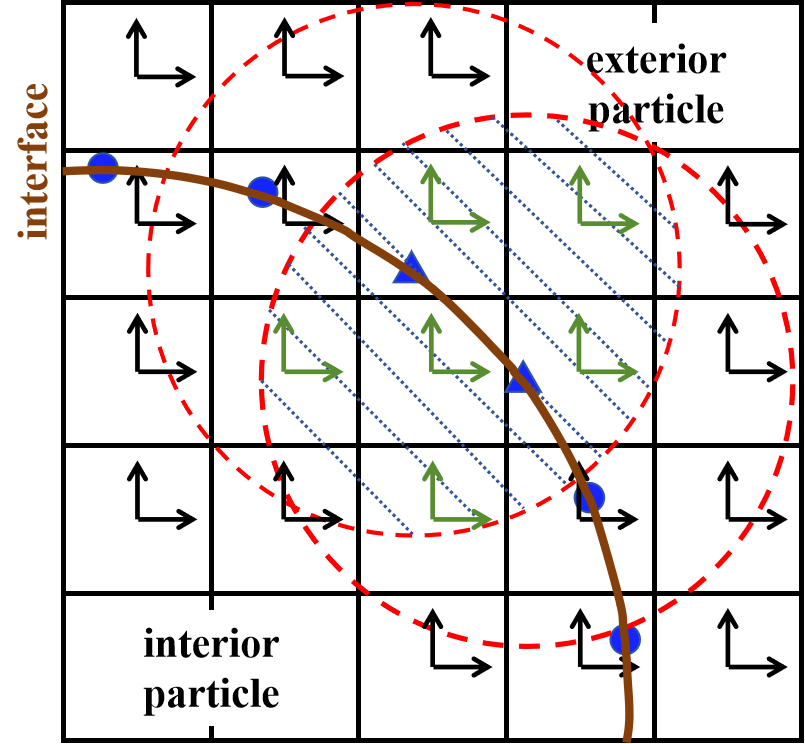}
    \caption{Illustration of the diffuse distribution of the IBM force around the particle interface.}
    \label{fig:interpolation}
\end{figure}

\textbf{Step 2}: The updated velocity $\widetilde{\mathbf{u}}^{*, n+1}$ needs to be corrected to satisfy the no-slip boundary condition at the fluid–particle interfaces $\partial V_{\mathrm{b}}(t)$.~This step is divided into four substeps~\cite{kempe2012improved,breugem2012second} in Algorithm~\ref{mlr}.~We first interpolate the intermediate Eulerian Velocity obtained from \textbf{Step 1} to the Lagrangian Velocity of markers.~The Lagrangian forces are then calculated based on the desired velocity at the interface and the intermediate velocity.~Next, the Eulerian forces are obtained from the spreading of Lagrangian forces by using either the three-point or four-point delta function~\cite{kempe2012improved,breugem2012second,zeng2022subcycling}.~As shown in Fig~\ref{fig:interpolation}, the Eulerian cells, enclosed by red dashed circles, refer to grid areas that are influenced by two blue markers.~These two Lagrangian markers also share some intersected areas, which are marked by green arrows.~Finally, the Eulerian velocity is corrected by the updated Eulerian Force.

\begin{figure}[ht]
    \includegraphics[width=0.8\linewidth]{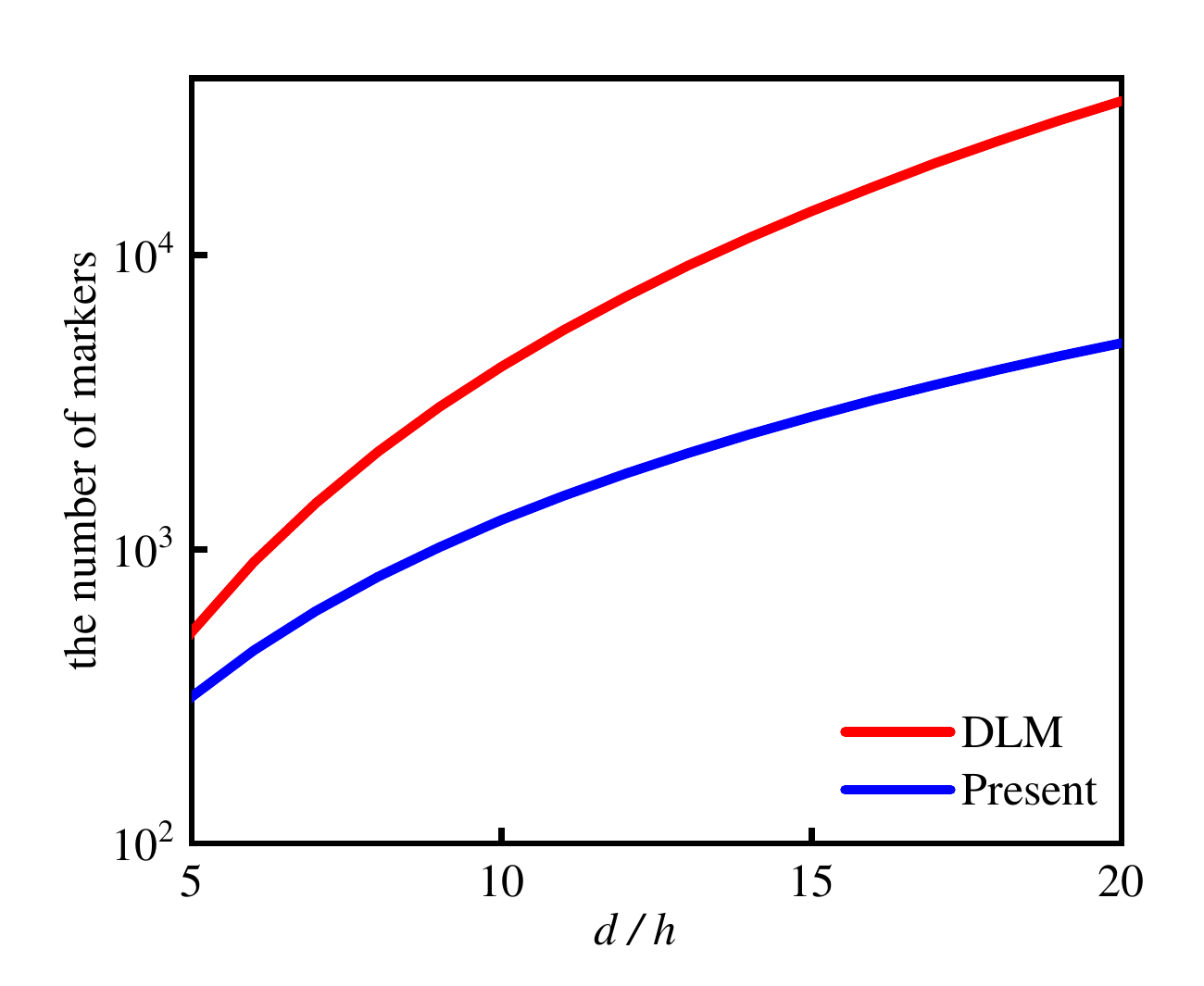}
    \caption{The number of markers change with the $d/h$ for a single particle scenario, where $d$ is the particle diameter and $h$ is the grid spacing.}
    \label{fig:DLM_DFIBM}
\end{figure}

In Algorithm~\ref{mlr} mentioned above, four points need to be noted.~First, the Lagrangian markers only exist on the finest level during the Eulerian-Lagrangian interaction process.~This brings the benefits of memory saving since particle-related information does not need to be stored on coarser levels.~Second, the Lagrangian markers only distribute on the surface of particles.~This is different from the DLM method in~\cite{bhalla2013unified,nangia2019dlm}, in which the markers also appear inside the particle and there is one marker per Eulerian grid cell.~Fig.~\ref{fig:DLM_DFIBM} shows how the number of markers changes with the $d/h$ for a single particle scenario.~It is seen that as $d/h$ increases, the present needs much fewer markers compared with the DLM method.~Third, the multi-direct forcing algorithm includes an outer loop with $m$ ranging from 1 to $N_{s}$, which controls the degree of coupling between Eulerian and Lagrangian variables.~The
original method of Uhlmann~\cite{uhlmann2005immersed} corresponds to the case of $N_{s} = 0$.~Increasing $N_{s}$ can enhance their coupling but will also increase the computational load.~Based on the experience in the previous work~\cite{kempe2012improved,breugem2012second} and tests presented in this paper, it is sufficient to set $N_{s}$ to 2-3 for all cases in Section~\ref{sec:Result}.~Finally, if a system has multiple particles, each particle goes into 1 to $N_{s}$ loop sequentially.~The corrected Eulerian forces we employ take into account the effects of all particles.~This consideration also applies to the calculation of the particle volume fraction (PVF) field in Section~\ref{sec:R-pvf}.

\begin{algorithm}[H]
    \caption {Multidirect forcing method for fluid-particle interaction}\label{mlr}
    \begin{algorithmic}[1]
      \State $\mathbf{u}^{(0)} = \widetilde{\mathbf{u}}^{*, n+1}$
      \For{$m=1$ \textbf{to} $N_{s}$}
        \State Interpolate Lagrangian Velocity,
        \State $\mathbf{U}^{m-1}\left(\mathbf{X}^{n}_l\right)=\sum_{i=1}^{N_x} \sum_{j=1}^{N_y} \sum_{k=1}^{N_z} \mathbf{u}^{(m-1)}\left(\mathbf{x}_{i, j, k}\right) \delta_h\left(\mathbf{x}_{i, j, k}-\mathbf{X}^{n}_l\right) h^3$
        \State Calculate Lagrangian Force,
        \State $\mathbf{F}^{m}=\mathbf{F}^{m-1}+\left(\mathbf{U}^d\left(\mathbf{X}^{n}_l\right)-\mathbf{U}^{m-1}\left(\mathbf{X}^{n}_l\right)\right) / \Delta t$
        \State Spreading Lagrangian Force onto Eulerian Force,
        \State $\mathbf{f}^{m}\left(\mathbf{x}_{i, j, k}\right)=\sum_{l=1}^{N_l} \mathbf{F}^{m}\left(\mathbf{X}_l\right) \delta_h\left(\mathbf{x}_{i, j, k}-\mathbf{X}_l\right) \Delta V_l$
        \State Correct Eulerian Velocity, 
        \State $\mathbf{u}^{(m)}=\mathbf{u}^{(0)}+\Delta t \mathbf{f}^{m}\left(\mathbf{x}_{i, j, k}\right)$
      \EndFor
      \State ${\mathbf{u}}^{*, n+1} = \mathbf{u}^{(m)}$
    \end{algorithmic}
\end{algorithm}

\textbf{Step 3}: With the updated intermediate velocity ${\mathbf{u}}^{*, n+1}$ in \textbf{Step 2}, a level projection operator is applied to obtain the updated pressure $p^{n+1/2}$ and velocity ${\mathbf{u}}^{n+1}$fields.~An auxiliary variable $\boldsymbol{V}$ is first calculated by
\begin{linenomath*}
\begin{linenomath*} \begin{equation}\label{eq:ns_lp1}
\boldsymbol{V} =  \frac{{\mathbf{u}}^{*,n+1}}{\Delta t} + \frac{1}{\rho_{f}} \bm{\nabla} p^{n-\frac{1}{2}}.
\end{equation}\end{linenomath*}
\end{linenomath*}
Then, $\boldsymbol{V}$ is projected onto the divergence-free velocity field to obtain the updated pressure $p^{n+1/2}$ via

\begin{linenomath*}
\begin{linenomath*} \begin{equation}\label{eq:ns_lp2}
L^{cc,l}_{\rho_{f}} p^{n+1/2} =  \bm{\nabla} \cdot \boldsymbol{V},  
\end{equation}\end{linenomath*}
\end{linenomath*}
where $L^{cc}_{\rho_{f}}p^{n+1/2}$ is the density-weighted Laplacian operator to $\bm{\nabla} \cdot (1/\rho_{f}\bm{\nabla} p^{n+1/2})$~\cite{almgren1998conservative,zeng2022aparallel}.~Finally, the divergence-free velocity ${\mathbf{u}}^{n+1}$ on level $l$ is obtained as

\begin{linenomath*}
\begin{linenomath*} \begin{equation}\label{eq:ns_lp3}
{\mathbf{u}}^{n+1} = \Delta t \left(\boldsymbol{V} - \frac{1}{\rho_{f}} \bm{\nabla} p^{n+1/2}\right).
\end{equation}\end{linenomath*}
\end{linenomath*}
The projection is stable and appears to be well-behaved in various numerical tests~\cite{almgren1996numerical,rider1995approximate} and practical applications~\cite{sussman1999adaptive,martin2000cell}.

\textbf{Step 4}: After completing \textbf{Step 3}, we obtain the divergence-free fluid velocity at $t^{n+1}$.~The particle-related information also needs to be updated from $t^{n}$ to $t^{n+1}$.~Depending on different kinematic constraints, the particle motion is categorized into prescribed motion and free motion.~The specific updates are detailed in Section~\ref{S:typesofsolidconstraints}.

\subsection{Types of kinematic constraints}\label{S:typesofsolidconstraints}

\subsubsection{Prescribed motion}\label{S:typesofsolidconstraints_one}
If the motion of the particle is prescribed, then its velocity and position are known \emph{a priori} and not influenced by the surrounding fluid.~Thus, the centroid position $\mathbf{X}_{r}^{n}$, centroid velocity $\mathbf{U}_{r}^{n}$ at $t^{n}$, centroid velocity $\mathbf{U}_{r}^{n+1}$ at $t^{n+1}$, and angular velocity $\mathbf{W}_{r}^{n}$ of the body are given.~The desired velocity $\mathbf{U}^d\left(\mathbf{X}^{n}_l\right)$ of the markers in Algorithm~\ref{mlr} is calculated as
\begin{linenomath*}
\begin{equation}\label{eq:pr1}
\begin{aligned}
\mathbf{U}^d\left(\mathbf{X}^{n}_l\right)=\mathbf{U}_{r}^{n}+\mathbf{W}_{r}^{n} \times \mathbf{R}_l^n,
\end{aligned}
\end{equation} \end{linenomath*}
where $\mathbf{R}_l^n = \left(\mathbf{X}^{n}_{l} - \mathbf{X}_{r}^{n}\right)$.~The new position of the centroid of the particle $\mathbf{X}_{r}^{n+1}$  is updated using the midpoint scheme as
\begin{linenomath*}
\begin{equation}\label{eq:pr2}
\begin{aligned}
\mathbf{X}_{r}^{n+1}=\mathbf{X}_{r}^{n}+\frac{\Delta t}{2}(\mathbf{U}_{r}^{n+1}+\mathbf{U}_{r}^{n}). 
\end{aligned}
\end{equation} \end{linenomath*}

\subsubsection{Free motion}\label{S:typesofsolidconstraints_two}
In contrast to the prescribed kinematics case, the motion of a freely moving particle is influenced by the surrounding fluid.~To account for this two-way interaction, the following governing equations of the particle systems are solved~\cite{kempe2012improved,breugem2012second}.

\begin{linenomath*}
\begin{equation}\label{eq:particle1}
\begin{aligned}
\rho_p V_p \frac{d \mathbf{U}_r}{d t} \approx-\rho_f \sum_{l=1}^{N_L} \mathbf{F}_l^{n+1/2} \Delta V_l+\rho_f \frac{d}{d t}\left(\int_{V_p} \mathbf{u} d V\right) \\ +\left(\rho_p-\rho_f\right) V_p \mathbf{g}+\mathbf{F}_c^{n+1 / 2},
\end{aligned}
\end{equation} \end{linenomath*}

\begin{linenomath*}
\begin{equation}\label{eq:particle2}
\begin{aligned}
I_p \frac{d \mathbf{W}_r}{d t} \approx-\rho_f \sum_{l=1}^{N_L} \mathbf{R}_l^n \times \mathbf{F}_l^{n+1 / 2} \Delta V_l \\ +\rho_f \frac{d}{d t}\left(\int_{V_p} \mathbf{r} \times \mathbf{u} d V\right)+\mathbf{T}_c^{n+1 / 2}
\end{aligned}
\end{equation} \end{linenomath*}

In the right-hand side of Eqs.~\ref{eq:particle1}-~\ref{eq:particle2}, the term $\mathbf{F}_l^{n+1/2}$ refers to the Lagrangian Force, coming from the final value of $\mathbf{F}^{m}$ in Algorithm $\ref{mlr}$.~The time derivatives of momentum integration $\rho_f \frac{d}{d t}\left(\int_{V_p} \mathbf{u} d V\right)$ and angular momentum integration $\rho_f \frac{d}{d t}\left(\int_{V_p} \mathbf{r} \times \mathbf{u} d V\right)$ within the particle are also included.~These two integrated terms account for flow unsteadiness by using the PVF field (Section~\ref{sec:R-pvf}).~The term $\left(\rho_p-\rho_f\right) V_p \mathbf{g}$ considers the buoyancy effects.~The terms $\mathbf{F}_c^{n+1 / 2}$ and $\mathbf{T}_c^{n+1 / 2}$ refer to the induced force and torque generated by the particle collision, respectively.~If there is only one single particle in the system, both $\mathbf{F}_c^{n+1 / 2}$ and $\mathbf{T}_c^{n+1 / 2}$ are set to be zero.~In the left-hand side of Eqs.~\ref{eq:particle1}-~\ref{eq:particle2}, we use the second-order mid-point scheme to integrate particle motions~\cite{zhu2022particle}.~After updating the particle centroid velocity $\mathbf{U}_{r}^{n+1}$ and angular velocity $\mathbf{W}_{r}^{n+1}$ at $t^{n+1}$, we go back to Eq.~\ref{eq:pr2} to update new position of the particle centroid $\mathbf{X}_{r}^{n+1}$.

The time advancement scheme in this work is not fully implicit~\cite{zhu2022particle}, yet it can deal with the free motion applies to particles either with a large density ratio (i.e., $\frac{\rho_p}{\rho_f} \geq 10$) or a small density ratio (i.e., $\frac{\rho_p}{\rho_f} \approx 1.5 - 2$)~\cite{kempe2012improved,breugem2012second}.~We found it is robust and fast enough to handle all the testing cases in Section~\ref{sec:Result}.~Before ending this Section, we also emphasize that our method is similar to the "weak coupling" method used in the sharp-interfaced immersed boundary method, which requires only one solution for fluid and solid solver during each time step and no iterations are needed between these two solvers~\cite{balaras2009adaptive,cui2018sharp}.~It is easier to extend the current portable solver to the "strong coupling" method, which then re-projects the flow part, re-updates the solid particle, and performs a convergence checking between the fluid solver and the solid solver during each sub-iteration~\cite{he2022numerical}.

\subsection{Multi-level advancement} \label{S:multileveladvancement}
In this work, we use a level-by-level time advancement method~\cite{zeng2022aparallel,zeng2023consistent} to advance the fluid and particle solution on the multi-level grid within the BSAMR framework.~Specifically, we introduce both the subcycling method and the non-subcycling method (Session~\ref{S:52}).~Finally, we describe several synchronization operations to better achieve the composite solution (Session~\ref{S:sync}).  

\subsubsection{Subcycling and non-subcycling methods} \label{S:52}

\begin{figure*}
    \centering
    \includegraphics[width=0.7\textwidth]{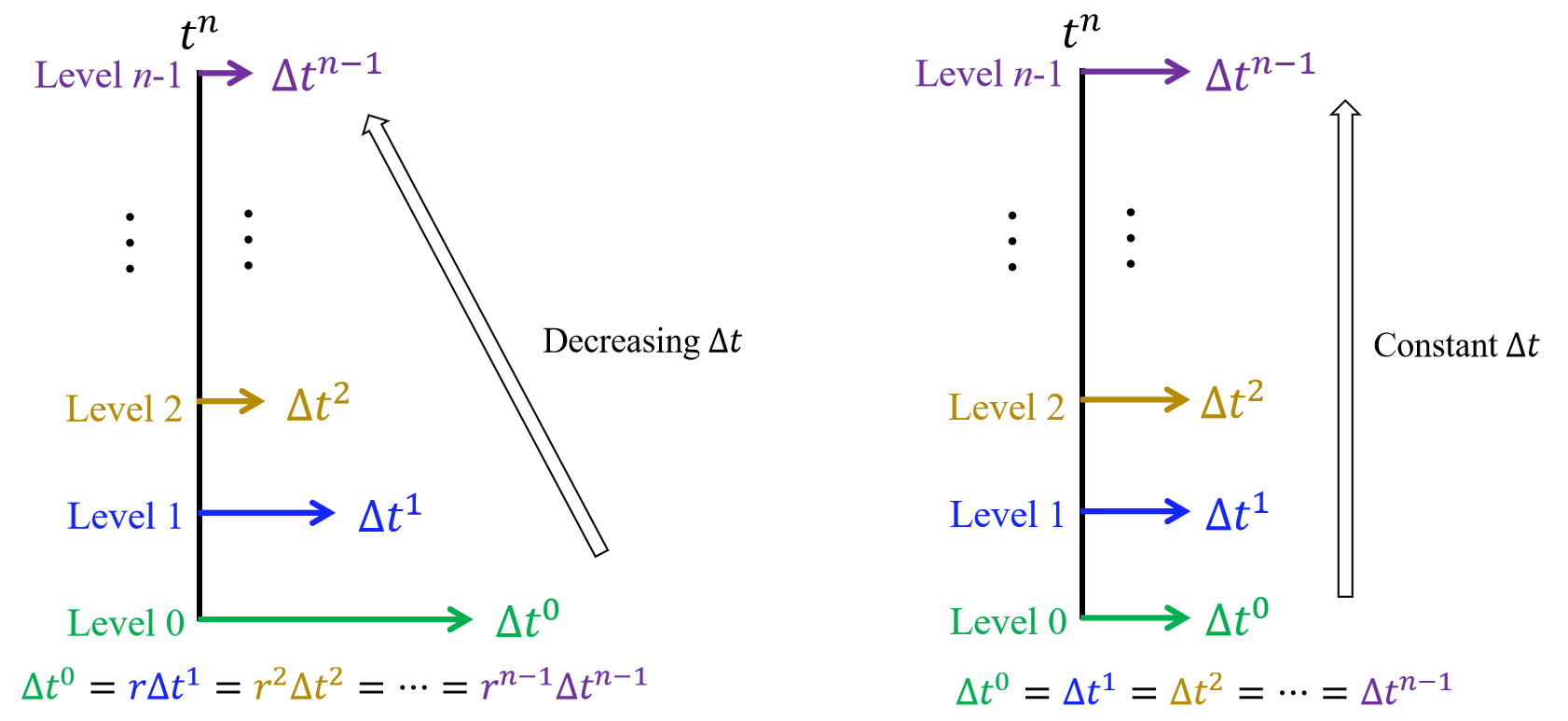}
    \caption{Schematic of the substeps in the level-by-level advancement method for a $n$-level grid.~Left: the subcycling method.~Right:the non-subcycling method.~The parameter $r=2$ is the refining ratio between two consecutive levels.}
    \label{fig:subnosub}
\end{figure*}

To advance variables on a multi-level grid, we utilize both the subcycling and non-subcycling methods with a level-by-level approach.~In the subcycling method, variables on different levels progress with distinct time step sizes.~The primary advantage of this approach is that maintaining the Courant–Friedrichs–Lewy (CFL) number constant across grid levels allows for larger time steps on coarser grids due to their larger spacing~\cite{almgren1998conservative,almgren1996numerical}.~For instance, with a refinement ratio of two between adjacent levels, the time step size on the coarser level, $\Delta{t^{l}}$, can be twice as large as that on the finer level, $\Delta{t^{l+1}}$.~Conversely, in the non-subcycling method, variables on all levels advance with the same time step size determined by the finest level $l_{\rm max}$.Fig.~\ref{fig:subnosub} schematically shows how the subcycling and non-subcycling methods are used to advance the variables on a multi-level grid with $n$ levels.~It should be noted that both of these methods produce consistent and accurate results for the single particles and multiple particle cases in Session~\ref{sec:Result}. Within the level-by-level framework, the non-subcycling method is relatively easier to implement and no temporal recursive procedure is involved. The subcycling method, on the other hand, allows large time steps on the coarser levels and thus reduces the overall computational cost.

\subsubsection{Synchronization} \label{S:sync}

The synchronization operations are used to make the solution data consistent across all levels~\cite{almgren1998conservative,martin2000cell,martin2008cell}.~There are three substeps of synchronization operations after the level advancement.
First, the flow velocity ${\mathbf{u}}$ and pressure $p$ on coarser levels are replaced by the corresponding averaging value on the finer levels.~There is no need to average the particle-related variables since they only exist on the finest level.~Second, we use a refluxing operation to account for an imbalance of the momentum and scalar fluxes at the coarse fine (CF) boundary~\cite{zeng2022aparallel,zeng2022subcycling}.~Our previous work~\cite{zeng2022aparallel} has validated that the refluxing operation can help add the mass and momentum conservation for tracer advection and double shear layer problems.~Last, a composite grid projection is applied to enforce the divergence-free condition on the velocity field across the entire hierarchy~\cite{almgren1998conservative,martin2000cell}.

\subsubsection{Summary of the multi-level advancement}\label{S:smla}

The synchronization operations are used to make the solution data consistent across all levels~\cite{almgren1998conservative,martin2000cell,martin2008cell}.~There are three substeps of synchronization operations after the level advancement.~First, the flow velocity ${\mathbf{u}}$ and pressure $p$ on coarser levels are replaced by the corresponding averaging value on the finer levels.~There is no need to average the particle-related variables since they only exist on the finest level.~Second, AMR requires numerical methods to deal with the coarse fine (CF) boundary interface where cells in different refinement levels meet. We thus use a refluxing operation to account for an imbalance of the momentum and scalar fluxes at the CF boundary~\cite{zeng2022aparallel,zeng2022subcycling}.~Our previous work~\cite{zeng2022aparallel} has validated that the refluxing operation can help add the mass and momentum conservation for tracer advection during vortex merging and double shear layer problems.~Last, a composite grid projection is applied to enforce the divergence-free condition on the velocity field across the entire hierarchy~\cite{almgren1998conservative,martin2000cell}.

Algorithm~\ref{MLAG} summarizes the adaptive multi-level advancement framework using both subcycling and non-subcycling methods. After initializing flow-related variables on all levels and particle-related variables on the finest level, time advancement can proceed using either method. Synchronization occurs when a coarser level catches up with a finer level.

As a final remark, we emphasize that our multi-level advancement algorithm employs a level-by-level approach, distinct from the composite advancement method~\cite{bhalla2013unified,sussman1999adaptive,griffith2007adaptive}.~In the level-by-level method, each level's variables advance independently until synchronization, reducing time step constraints on coarser levels.~Conversely, the composite advancement method uses composite variables for time advancement, relying only on variables in non-overlapping regions.~This makes it less flexible to integrate both subcycling and non-subcycling methods.~Our level-by-level approach, however, handles both methods with ease.

\begin{figure}[!t]
\begin{algorithm}[H]
    \caption{Multi-level advancement}\label{MLAG}
    \label{alg:solver}
    \begin{algorithmic} [1] 
        \State Initialize $\mathbf{X}^{0}_{\mathrm{r}}$, $\mathbf{U}^{0}_{\mathrm{r}}$, $\mathbf{W}^{0}_{\mathrm{r}}$, $\mathbf{u}^0$, and $p^0$ on level $0$
        \State $l \gets 0$
        \While{refinement criteria are satisfied on level $l$ and $l<l_{\rm max}$}
          \State Regrid the grid to level $l+1$
          \State Initialize $\mathbf{X}^{0}_{\mathrm{r}}$, $\mathbf{U}^{0}_{\mathrm{r}}$, $\mathbf{W}^{0}_{\mathrm{r}}$, $\mathbf{u}^0$, and $p^0$ on level $l+1$
          \State $l \gets l+1$
        \EndWhile
        \State Initialize $\mathbf{X}_{l}^{0}$, $\mathbf{U}_{l}^{0}$ and $\mathbf{R}_{l}^{0}$ for all Lagrangian markers on level $l_{\rm max}$
        \If{subcycling method is used}
          \State $\Delta t^l = 2^{l_{\rm max}-l} \Delta t^{l_{\rm max}}$ for all $0\leq l < l_{\rm max}$
        \Else
          \State $\Delta t^l = \Delta t^{l_{\rm max}}$ for all $0\leq l < l_{\rm max}$
        \EndIf
        \For{$n=1,n_{\rm max}$}\Comment{$n_{\rm max}$ is the number of time steps in the simulation}
          \State \Call{Single-level-advancement}{$0$, $t_n^0$, $t_n^0+\Delta t^0$, $\Delta t^0$}
          \State Apply the synchronization projection~\cite{zeng2022aparallel,zeng2022subcycling,zeng2023consistent}
          \State Refine the grid and interpolate $\mathbf{u}$ and $p$ onto new levels
        \EndFor
        \State		
        \Procedure{Single-level-advancement}{$l$, $t^l$, $t^l_{\rm max}$, $\Delta t^l$}
        \While{$t^l<t^l_{\rm max}$}
            \State Solve momentum Eq.~\eqref{eq:31} to obtain $\widetilde{\mathbf{u}}^{*,n+1}$
            \State Apply the Algorithm~\ref{mlr} for the fluid-particle coupling and obtain the intermediate velocity ${\mathbf{u}}^{*, n+1}$
            \State Apply the level projection using Eq.~\eqref{eq:ns_lp1}-~\eqref{eq:ns_lp3} to obtain the updated pressure $p^{n+1/2}$ and velocity ${\mathbf{u}}^{n+1}$fields
            \State Update particle-related information from $t^{n}$ to $t^{n+1}$ based on different constraints
          
          \If{$l<l_{\rm max}$}
            \State \Call{Single-level-advancement}{$l+1$, $t^l$, $t^l+\Delta t^l$, $\Delta t^{l+1}$}
          \EndIf
          \State $t^l \gets t^l+\Delta t^l$
        \EndWhile
        \If{$l>0$}
        \State Average all data from finer levels onto coarser levels~\cite{zeng2022aparallel,zeng2022subcycling,zeng2023consistent}
        \EndIf
        \If{$l<l_{\rm max}$}
        \State Perform the refluxing operation~\cite{zeng2022aparallel,zeng2022subcycling,zeng2023consistent}
        \EndIf
        \EndProcedure
    \end{algorithmic}
\end{algorithm}
\end{figure}

\subsection{IAMReX framework} \label{S:IAMReX}
We extend the AMReX-based~\cite{zhang2019amrex,zhang2020amrex} application IAMR~\cite{almgren1996numerical,almgren1998conservative} to a much more powerful framework IAMReX.~In the IAMReX, the Navier-Stokes euqations are solved on a semi-staggered multi-level grid using the projection method~\cite{almgren1998conservative}.~The gas-liquid interface is captured using either the level set (LS) method~\cite{zeng2022aparallel,zeng2023consistent}.~And the fluid-particle interface is resolved using the multidirect forcing immersed boundary method~\cite{zhu2022particle,breugem2012second}.~IAMReX is a publicly accessible platform designed specifically for developing massively parallel block-structured adaptive mesh refinement (BSAMR) applications.~The code now supports hybrid parallelization using either pure MPI or MPI+OpenMP for multicore machines~\cite{zhang2019amrex}.~The source code for IAMReX, testing cases used in this work can be accessed at~\url{https://github.com/ruohai0925/IAMR/tree/development}.~The scalability of AMReX-based apps has been thoroughly validated in the previous works~\cite{zhang2020amrex,min2022towards,yao2022massively}.~All input scripts and raw postprocessing data are uploaded into~\url{https://pan.baidu.com/s/1bZRoDunjBv7bqYL8CI3ASA?pwd=i5c2} for interested readers to reproduce the results in Session~\ref{sec:Result}.

\section{\label{sec:Result}Results}

This section presents several canonical fluid-particle interaction problems to validate the capabilities and robustness of the proposed IAMReX framework.~For each case, $\Delta t_{0}$ refers to the time step on level 0, and $\Delta x_{0}$, $\Delta y_{0}$, and $\Delta z_{0}$ are the grid spacings in the $x$-direction, $y$-direction, and $z$-direction, respectively, on level 0.

\subsection{\label{sec:R-pvf}PVF}
\begin{figure}[h]
	\centering
	\includegraphics[trim=0 0 0 0,clip,width=0.8\linewidth]{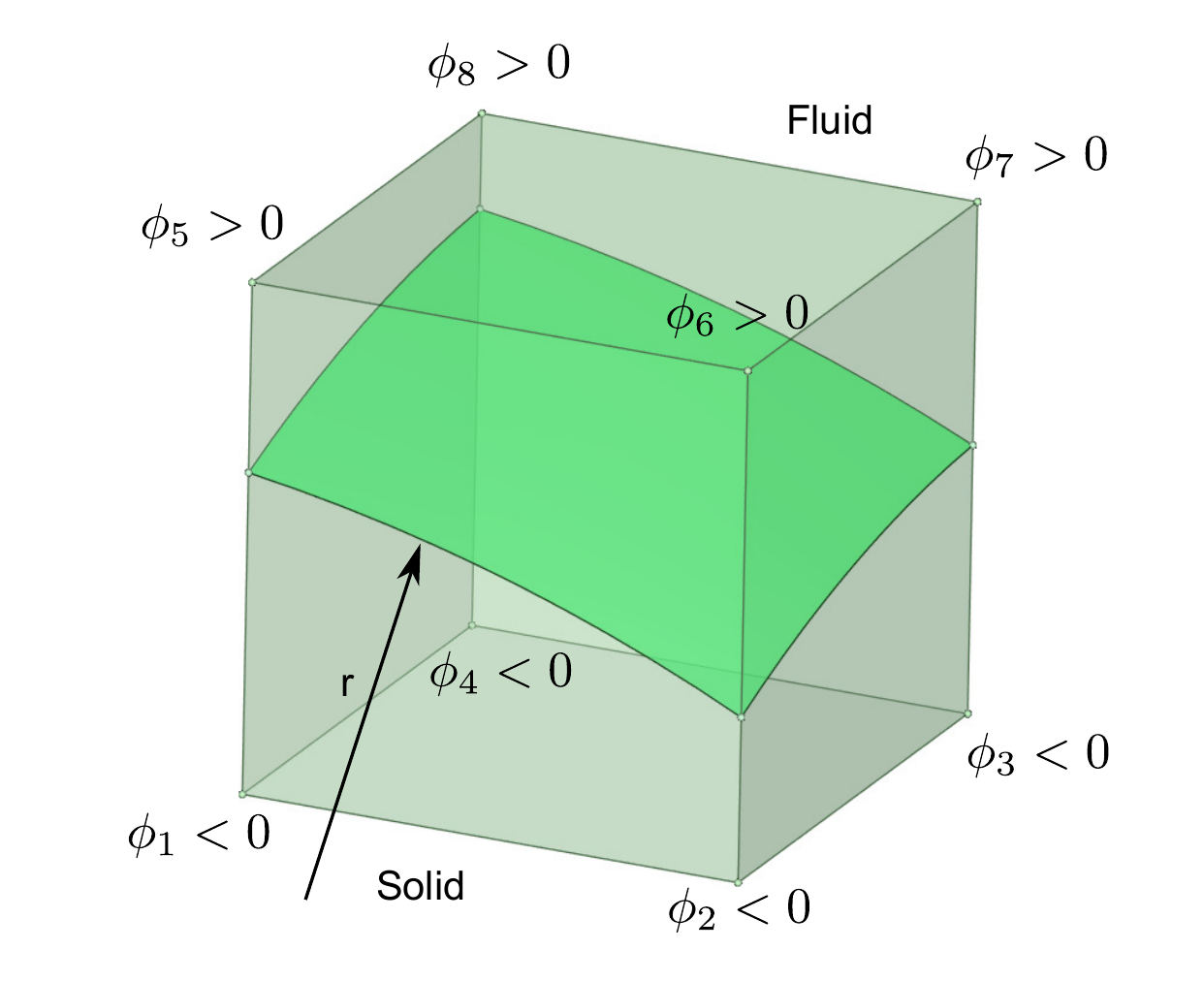}
	\caption{Sketch of $\phi$ at each corner of a grid cell with interface.}
	\label{fig:pvf_level_set_function}
\end{figure}

The particle volume fraction (PVF), which joins the calculation of free motion updates, is introduced and validated in this session.~The PVF is approximated by the signed-distance level-set function $\phi$ of the fluid-particle interface.~The level-set function $\phi$ is located at cell nodes and is calculated at the eight corners of each cell.~The symbol of $\phi$ as well as the intersected interface is shown in Fig.~\ref{fig:pvf_level_set_function}.~Here, $\phi$ is negative inside the particle and positive outside the particle.

Based on the level-set function $\phi$, an approximation of PVF can be obtained by the following equation,  

\begin{equation}
    \alpha_{i,j,k} = \frac{\sum_{m=1}^{8}-\phi_mH(-\phi_m)}{\sum_{m=1}^8|\phi_m|},
    \label{equ:level-set function}
\end{equation}
where $H$ is the Heaviside function, defined by
\begin{equation}
    H(\phi) = 
    \left\{
    \begin{aligned}
    0, \phi \le 0\\
    1, \phi > 0\\
    \end{aligned}
    \right.
    \label{equ:Heaviside function}
\end{equation}

\begin{figure}[ht]
	\centering
	\includegraphics[trim=0 0 0 0,clip,width=0.9\linewidth]{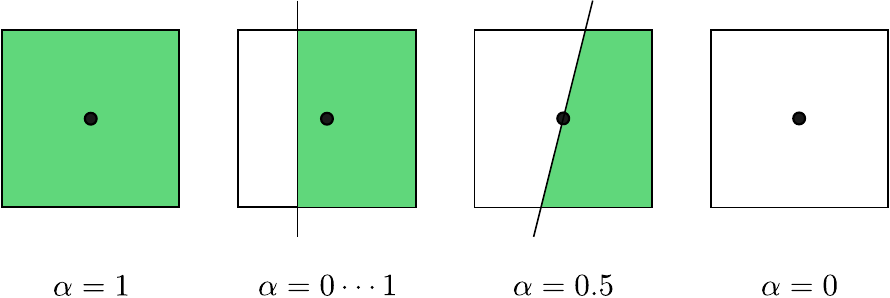}
	\caption{Sketch of PVF with different values.~The white and green colors refer to the cell areas within the fluid and particle, respectively.}
	\label{fig:pvf_cases_of_exact_value}
\end{figure}

In the right side of Eq.~\ref{equ:level-set function}, the value of $\phi$ for each cell corner depends on the location of the fluid-particle interface.~When the shape of the particle surface is analytically given, the $\phi$ value can be determined by calculating the Euclidean distance from the corner point of the cell to the particle surface.~The calculation of PVF is then transformed from an exact integral to a numerical integral.~As shown in Fig.~\ref{fig:pvf_cases_of_exact_value}, the cell value of PVF varies from $0$ to $1$, depending on the relative position between the cell center and the fluid-particle interface.


\begin{figure}[ht]
	\centering
	\includegraphics[trim=0 0 0 0,clip,width=0.9\linewidth]{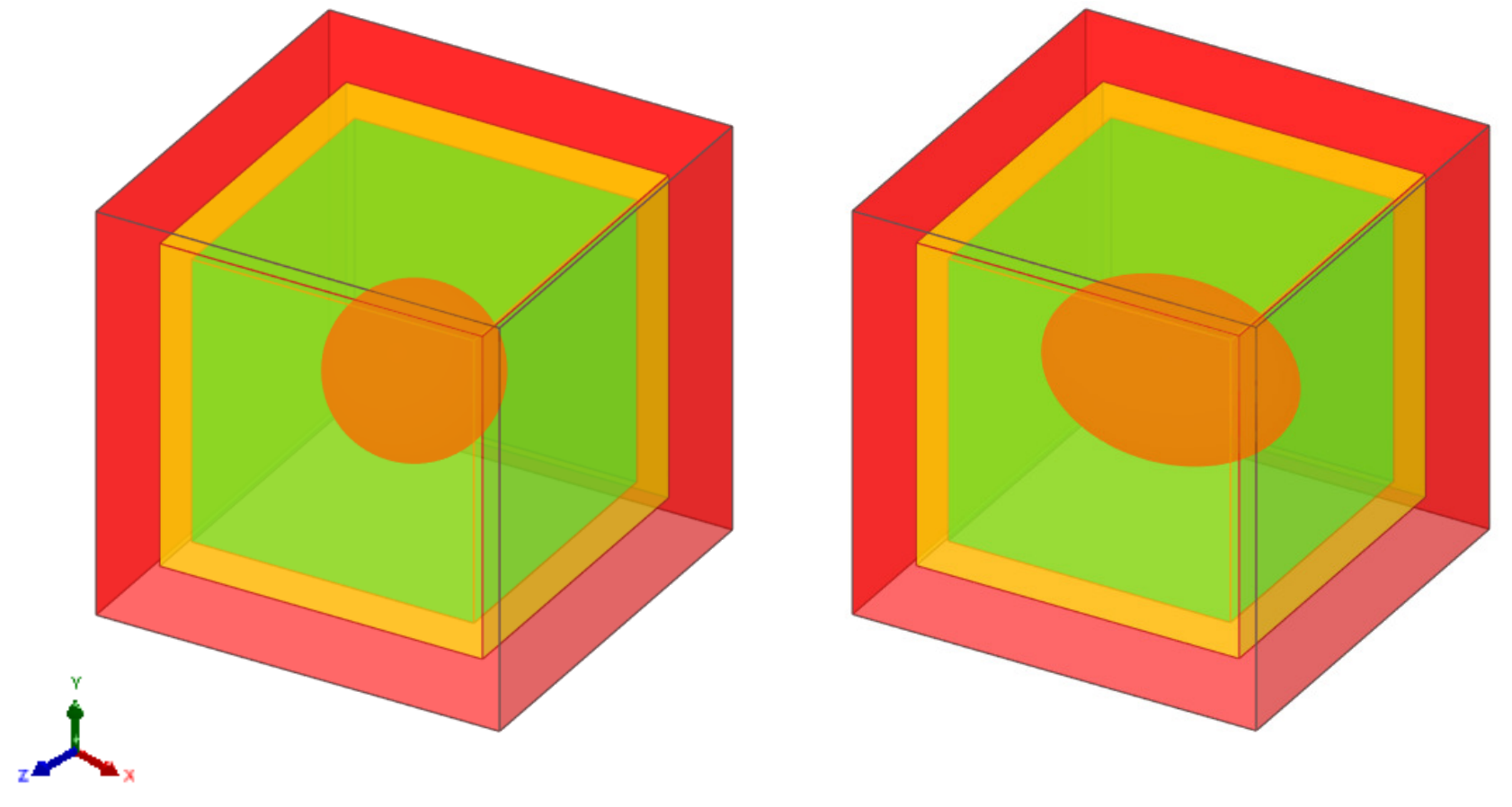}
	\caption{Results of a three-level AMR grid. The red, orange, and green color represents the grid on levels 0, 1, and 2, respectively.~Left: a spherical particle; Right: an ellipsoidal particle.}
	\label{fig:pvf_amr_block}
\end{figure}

We validate the correctness and convergence of the above PVF approximation by calculating the volumes of spherical and ellipsoidal surfaces on the Cartesian grid.~The exact solutions for the sphere and ellipsoid are given by the following formula,

\begin{equation}
    V_{exact} = 4\pi \frac{abc}{3}
    \label{equ:exact solutions of the sphere and ellipsoid}
\end{equation}
where a, b, and c are the semi-axes of the ellipsoid.~For a sphere case, we have $a=b=c$.~, the signed-distance level-set function is 
\begin{equation}
    \phi_{i,j,k}= \\ 
    \sqrt{\frac{(x_{i,j,k}-x_{p})^2}{a^2} + 
    \frac{(x_{i,j,k}-y_{p})^2}{b^2} + 
    \frac{(x_{i,j,k}-z_{p})^2}{c^2}}-1,
    \label{equ:level-set function of spheroidal geometries}
\end{equation}
for any Eulerian cell $(i,j,k)$.~The computational domain is $L_x \times L_y \times L_z = 2 \times 2 \times 2$, the sphere diameter is $D = 0.8$, and the semi-axes of the ellipsoid are set to be $a = 0.4$, $b = 0.6$, $c = 0.4$.~The centers of both two particles are $(x_p, y_p, z_p)=(1,1,1)$.~As shown in Fig.~\ref{fig:pvf_amr_block}, three levels of AMR grid are used during the PVF calculation, and the particles are enclosed by the finest level. 

Table.~\ref{tab:pvf result of sphere} and ~\ref{tab:pvf result of ellipsoid} show the calculation results of the sphere and ellipsoid.~The numerical errors decrease with the increase of the $D/h$, where $h$ is the Cartesian grid spacing on level $0$.~If the resolution on the finest level keeps unchanged, we validated that the results of a three-level grid are the same as those of the corresponding single-level grid. In addition, our results show the second-order convergence and agree well with the results in ~\citet{kempe2012improved}. It also matches the overall second-order accuracy of the basic fluid solver.

\begin{table}[h]
\caption{Calculate sphere volume fraction by using a three-level AMR grid}
\begin{ruledtabular}
\begin{tabular}{cccc}
$D/h$ & Volume & $\epsilon_{sphere}[\%]$ & $s_{sphere}$\\
\hline
16 & 0.2667230796 & $5.071 \cdot 10^{-1}$ & \\
32 & 0.2677639589 & $1.188 \cdot 10^{-1}$ & 2.09374 \\
64 & 0.2679990393 & $3.116 \cdot 10^{-2}$ & 1.93077 \\
128 & 0.2680627154 & $7.407 \cdot 10^{-3}$ & 2.07273 \\
\end{tabular}
\end{ruledtabular}
\label{tab:pvf result of sphere}
\end{table}

\begin{table}[ht]
\caption{\label{tab:pvf result of ellipsoid} Calculate ellipsoid volume fraction by using a three-level AMR grid}
\begin{ruledtabular}
\begin{tabular}{cccc}
$D/h$ & Volume & $\epsilon_{ellispoid}[\%]$ & $s_{ellispoid}$ \\
\hline
16 & 0.4004903567 & $4.062 \cdot 10^{-1}$ & \\
32 & 0.4016884973 & $1.083 \cdot 10^{-1}$ & 1.90716 \\
64 & 0.4020166444 & $2.666 \cdot 10^{-2}$ & 2.02228 \\
128 & 0.40209828964 & $6.359 \cdot 10^{-3}$ & 2.0678 \\
\end{tabular}
\end{ruledtabular}
\end{table}

Lastly, it is noted that this method is also applicable when multiple particles are close to each other or their surfaces are in direct contact. Because the PVF calculation is a separate operation for each particle, the total volume fraction is not needed as long as the Eulerian force considers the effects of all particles~\cite{breugem2012second}.

\subsection{\label{sec:lid-driven-sphere} 3D Lid-driven cavity with fixed spherical particle}

\begin{figure}[h]
    \includegraphics[width=0.35\textwidth]{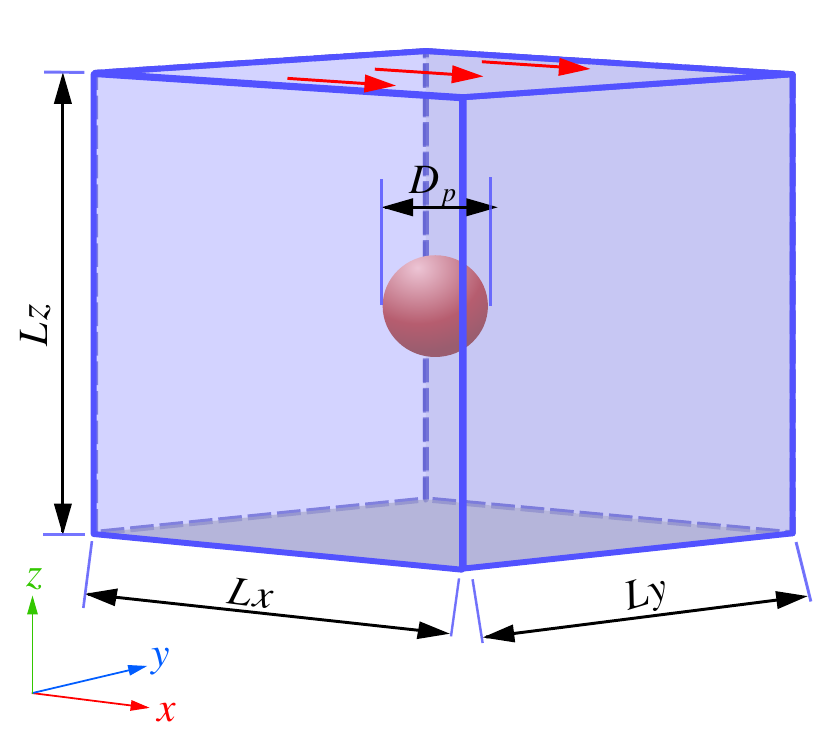}
    \caption{The schematic of 3D lid-driven cavity flow with fixed spherical particle}
    \label{fig:scheme_LDC}
\end{figure}

We start to validate the convergence and accuracy of our solver using a lid-driven cavity over a spherical particle case (Fig.~\ref{fig:scheme_LDC}).~The particle's diameter $D_p = 0.25$ and the computation domain $L_x \times L_y \times L_z = 4D_p \times 4D_p \times 4D_p$.~The top wall has a constant velocity $U = 1$ at $x$ direction and no-slip stationary conditions are applied on the other wall of the cube and surface of particle.~The Reynolds number of the flow is $Re = \rho_f U L_x / \mu_f$. Nine cases are considered in Table~\ref{tab:lid_driven_cavity_problem_set}, which includes the single-level, three-level non-subcycling AMR, and three-level subcycling AMRsimulations.~Each case also includes three different Reynolds numbers, i.e., $Re=$1,100, and 400. For the AMR cases, the particle is always refined to the finest level.

\begin{table}[h]
\caption{Parameters of the lid-driven cavity with spherical particle problem.}
    \begin{ruledtabular}
    \begin{tabular}{cccc}
        Case no. & Grids number on level 0  & $ l_{max} $ & Cycling method\\
        \hline
        1 & $16 \times 16 \times 16$ & 0 & - \\
        2 & $32 \times 32 \times 32$ & 0 & - \\
        3 & $64 \times 64 \times 64$ & 0 & - \\
        4 & $4  \times 4  \times 4 $ & 2 & None \\
        5 & $8  \times 8  \times 8 $ & 2 & None \\
        6 & $16 \times 16 \times 16$ & 2 & None \\
        7 & $4  \times 4  \times 4 $ & 2 & Auto \\
        8 & $8  \times 8  \times 8 $ & 2 & Auto \\
        9 & $16 \times 16 \times 16$ & 2 & Auto \\
    \end{tabular}
    \end{ruledtabular}
    \label{tab:lid_driven_cavity_problem_set} 
\end{table}

\begin{figure*}
    \includegraphics[width=0.9\textwidth]{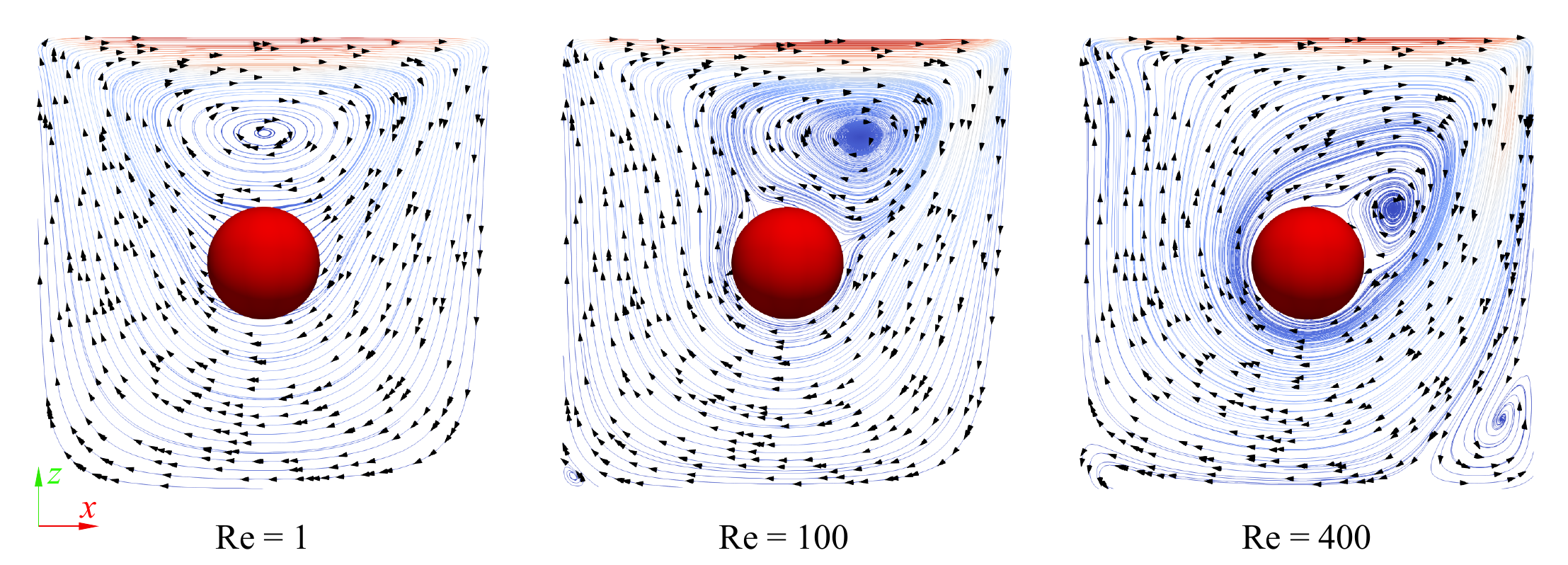}
    \caption{Streamline contours in the $x-z$ plane at $y = 0.5$ for Case 2 in Table~\ref{tab:lid_driven_cavity_problem_set}. (a) $Re=1$; (b) $Re=100$; (c) $Re=400$.}
    \label{fig:lid_driven_stream_line}
\end{figure*}

\begin{figure*}
    \includegraphics[width=1.0\textwidth]{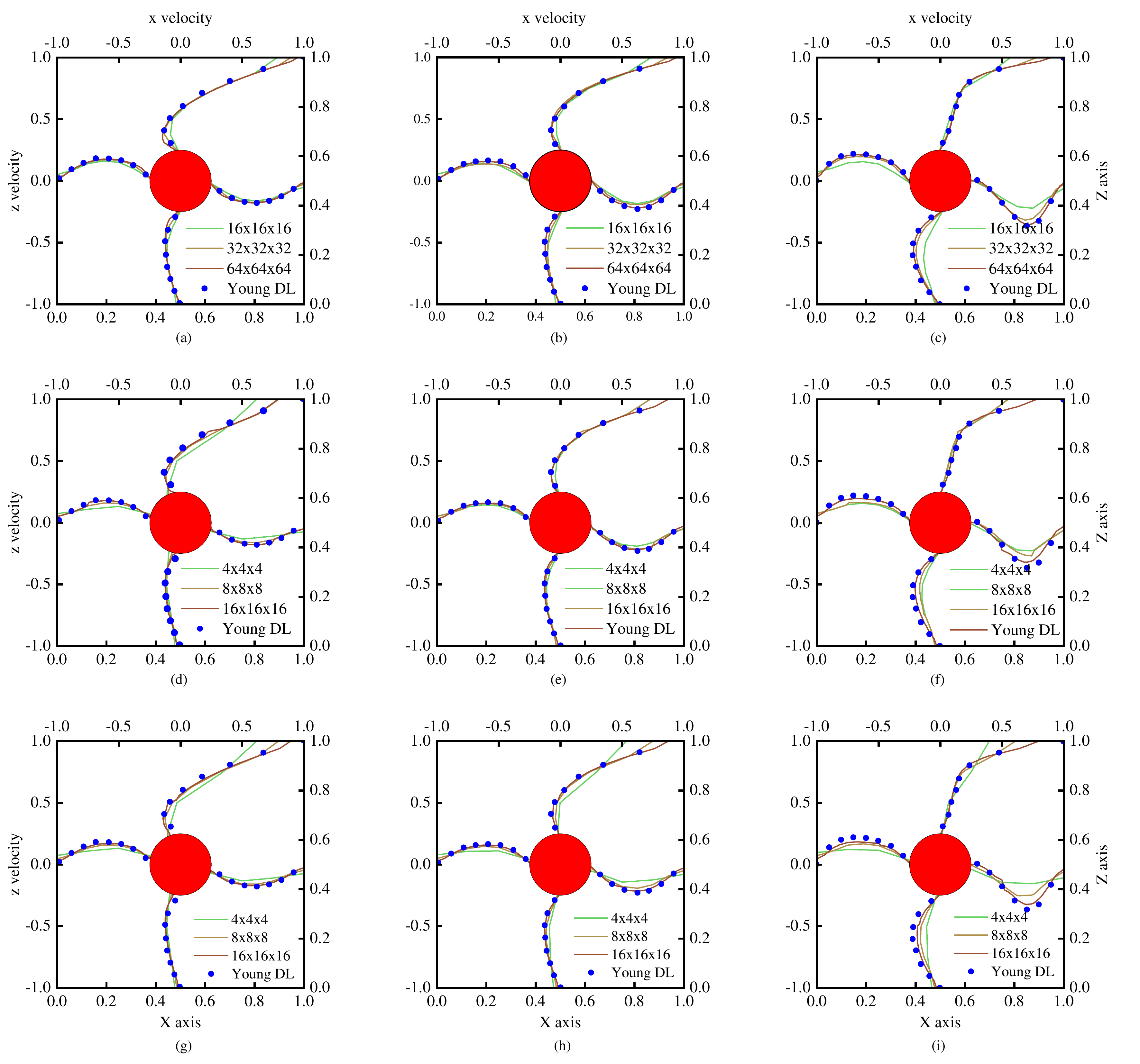}
    \caption{Velocity profiles at horizontal and vertical line of the lid-driven cavity over a spherical particle problem.~Here, (a), (b), and (c) represent the velocity on a single-level grid (Cases 1-3) with $Re=1,100$, and $400$, respectively;~(d), (e), and (f) represent $Re=1,100$, and $400$ of non-subcycling method (Cases 4-6) on a three-level AMR grid;~(g), (h) and (i) represent $Re=1,100$, and $400$ of subcycling method (Cases 7-9) on a three-level AMR grid.}
    \label{fig:lid_driven_result}
\end{figure*}

We first plot the contour results of Case 2 with different Reynolds numbers, which are varied by maintaining a constant flow rate driven by the top wall and changing the viscosity.~As shown in Fig.~\ref{fig:lid_driven_stream_line}, the flow passes around the spherical particle and generates the clockwise vortex. As $Re$ increases, the secondary vortex appears near the particle surface and the bottom-right corner. 

Fig.\ref{fig:lid_driven_result} shows velocity distribution on the $x$ and $z$ direction in the $x-z$ plane at $y = 0.5$. The present results converge as the grid number increases under all three $Re$ scenarios. The results of single level, non-subcycling method, and subcycling method produce consistent results, and all of them closely match~\citet{Young2009TheMO}.


\subsection{\label{sec:R-fps}Flow Past Fixed Sphere}

\begin{figure}[H]
    \includegraphics[width=0.45\textwidth]{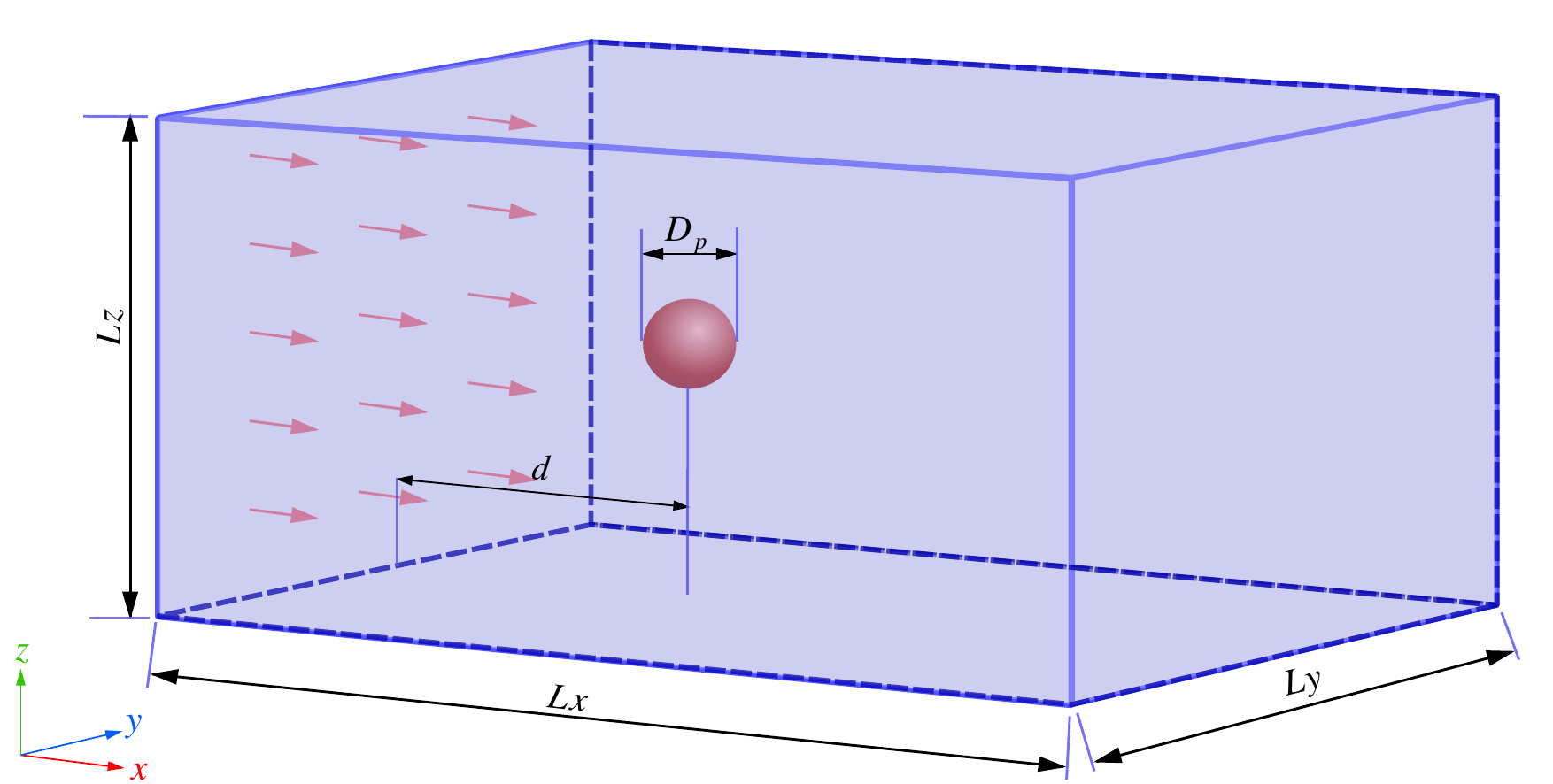}
    \caption{The schematic of the flow passing through the spherical particles}
    \label{fig:scheme_FPS}
\end{figure}

In this session, we validate the accuracy and efficacy of our adaptive solver by simulating a spherical particle in uniform flow with different particle Reynolds numbers~\cite{schiller1933uber,zhu2022particle}.~The sketch of the fluid flow passing through a spherical particle is shown in Fig.~\ref{fig:scheme_FPS}, the diameter of the particle is $D_p = 1$, the computational domain is $L_x \times L_y \times L_z = 20D_p \times 10D_p \times 10D_p$, the distance of the particle from the inlet is $d = 5D_p$ and located in the center of the $y-z$ plane.~The inlet and outlet boundaries are applied in the $x$ direction and the inlet velocity $U$ is $1m/s$.~Both $y$ and $z$ directions are periodic boundaries. 

\begin{figure}[H] 
    \includegraphics[width=0.45\textwidth]{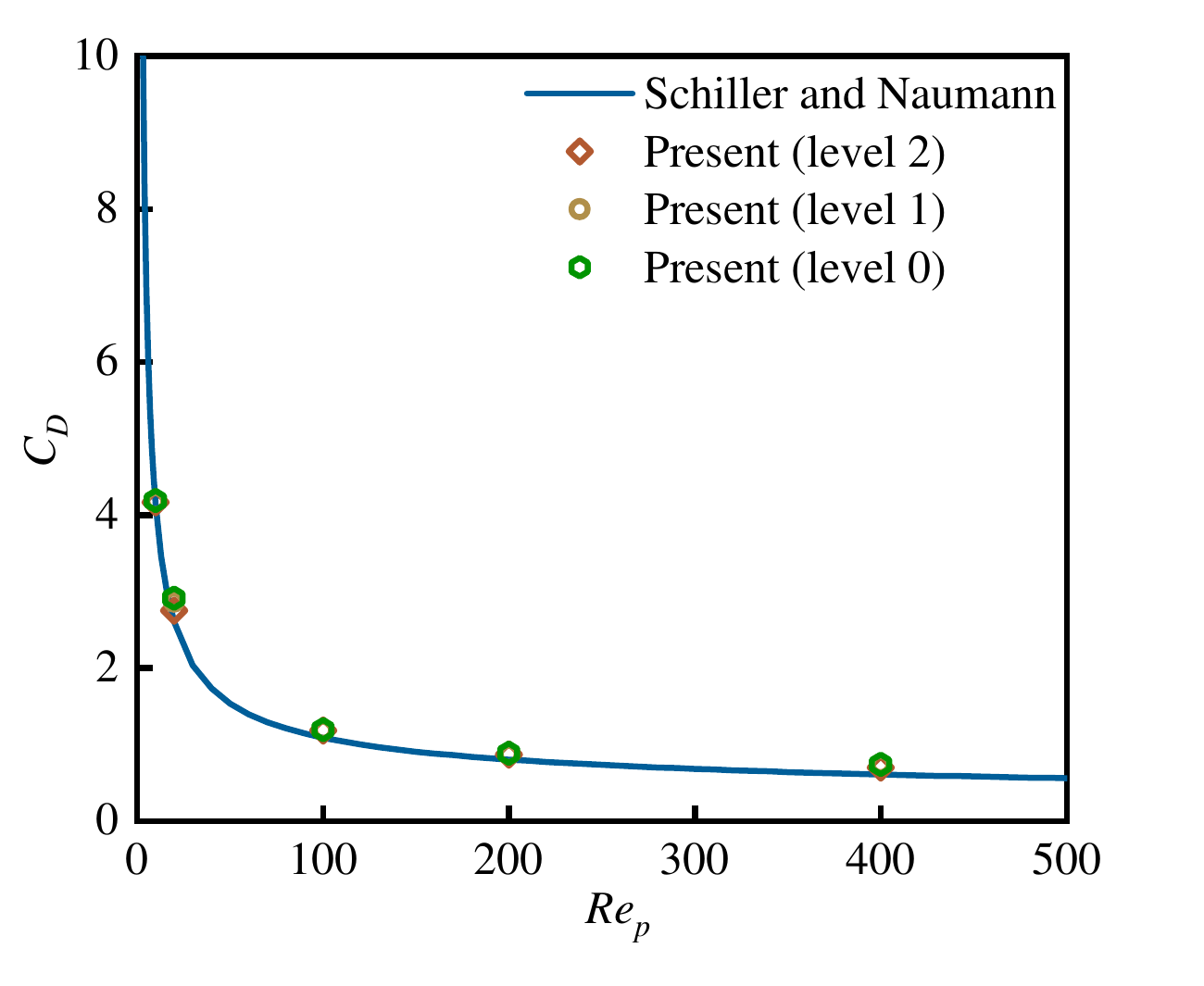}
    \caption{The drag coefficient of the particle under uniform flows varies with Particle Reynolds number at different AMR levels.}
    \label{fig:result_FPS}
\end{figure}

\begin{figure}[H]
    \includegraphics[width=0.45\textwidth]{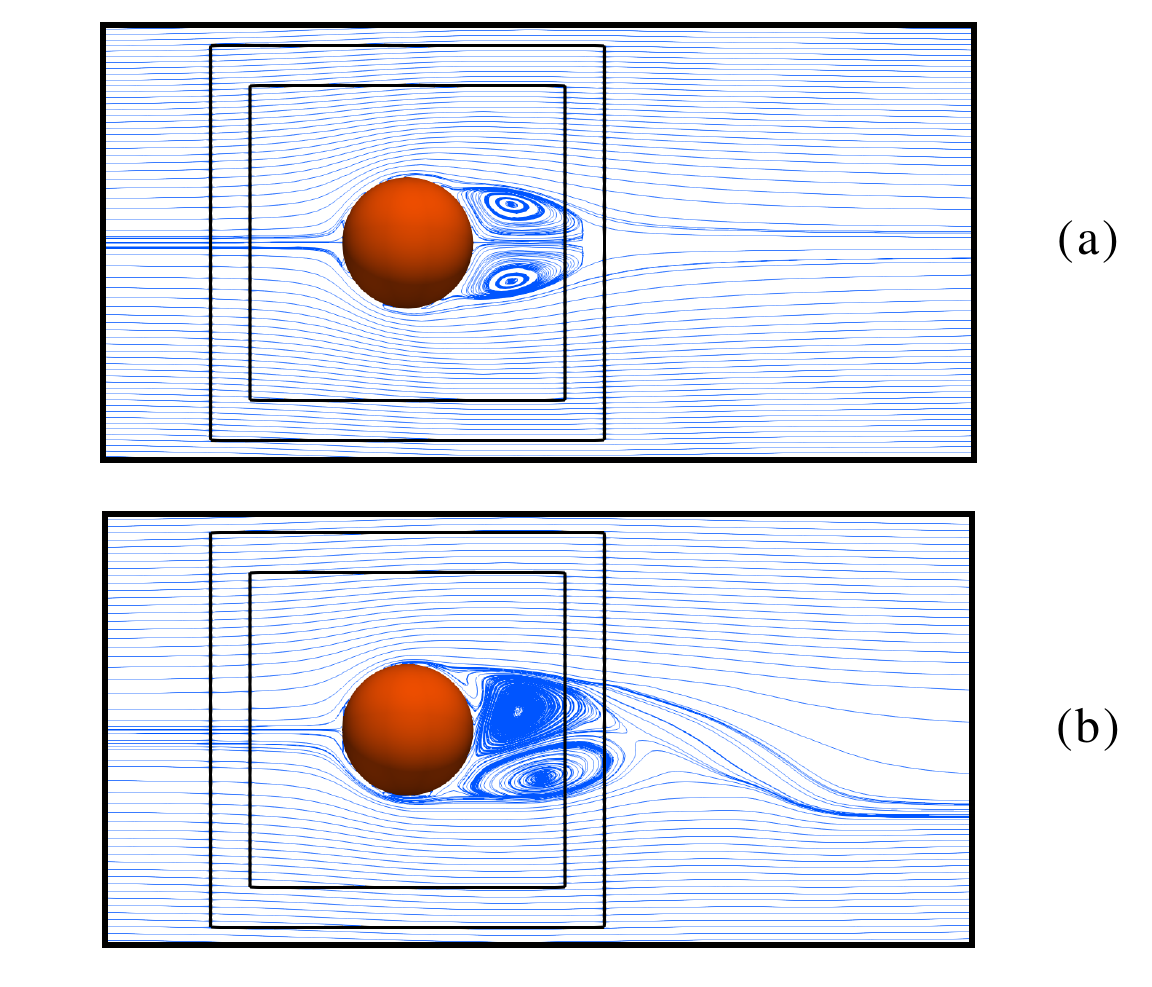}
    \caption{Streamlines on fluid Eulerian grids around the particle at different particle Reynolds numbers.(a) $Re_p=100$; (b) $Re_p=400$.}
    \label{fig:stream_line_FPS}
\end{figure}

The influence of AMR on the simulation results is investigated by using the subcycling method with different levels. As shown in Fig.~\ref{fig:result_FPS}, three different types of grid were selected: level 0 indicates the single-level grid containing no AMR, level 1 indicates the two-level grid, and level 2 indicates the three-level grid.~For all three types of grid, the ratio of the diameter of the particles to the grid spacing on the finest level is 16.~The drag force, including the contributions of Lagrangian force and the PVF function, is calculated by, 

\begin{equation}\label{eq:FD}
\mathbf{F}_D=-\rho_f \sum_{l=1}^{N_L} \mathbf{F}_l^{n+1/2} \Delta V_l+\rho_f \frac{d}{d t}\left(\int_{V_p} \mathbf{u} d V\right),
\end{equation}
The theoretical S-N law for calculating the drag coefficient of the shaped particles is,

\begin{equation}\label{eq:SN}
C_D = (24/Re_p)(1+0.15Re_p^{0.687})
\end{equation}
which is proposed by~\citet{schiller1933uber}, and $Re_p = UD_p/\nu$ represents the particle Reynolds number.~From Fig.~\ref{fig:result_FPS}, and it can be seen that the present results under different particle Reynolds numbers are in good agreement with S-N law. The fact that different levels of grid produce the nearly identical results validated the accuracy of our solver on the adaptive grid. Fig.~\ref{fig:stream_line_FPS} illustrates the streamlines of the flow field and vortex form under different $Re_p$.~The streamline is continuous over the coarse-fine boundaries with the help of synchronization operations in Session~\ref{S:sync}.~As the Reynolds number increases, more unsteadiness appears and the vortex behind the particles gradually becomes asymmetrical as expected~\cite{gong2023cp3d}.

\subsection{\label{sec:R-sfs}Flow Past Rotating and Moving Sphere}
\begin{figure*}[ht]
    \centering
    \subfigure[]{
        \includegraphics[width=0.45\textwidth]{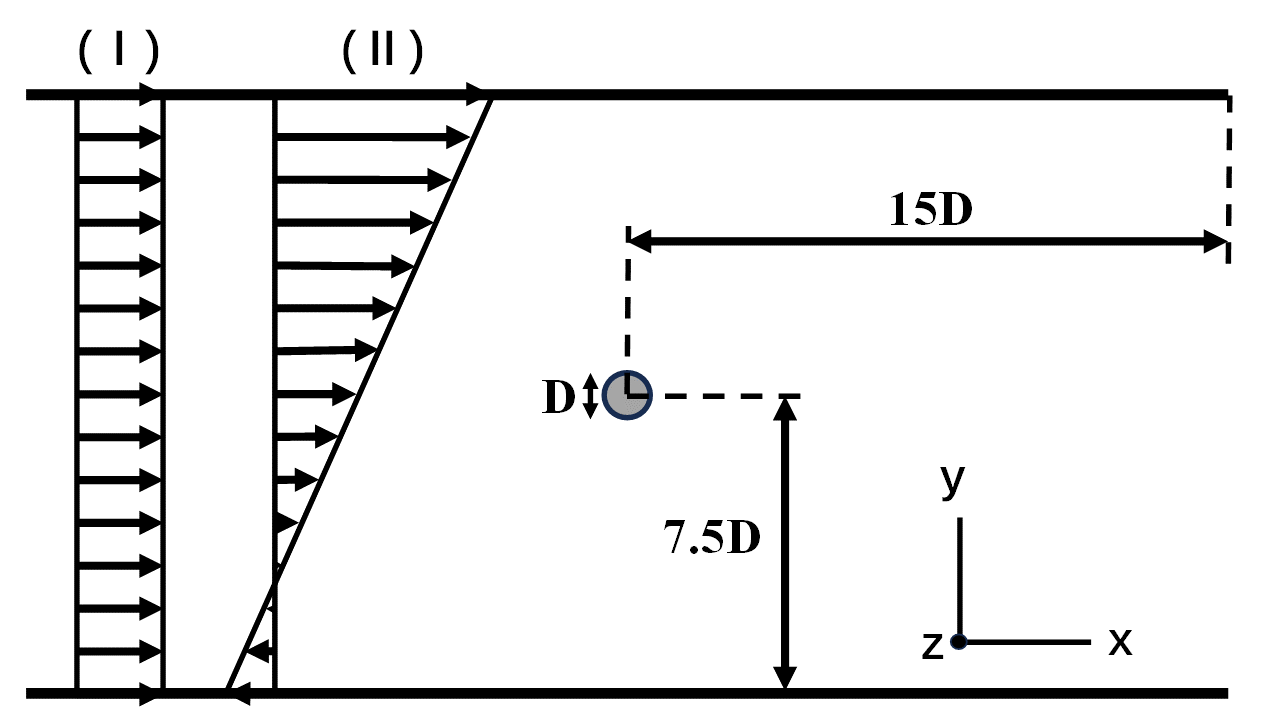}
        \label{fig:Sketch_Map}
    }
    \subfigure[]{
        \includegraphics[width=0.45\textwidth]{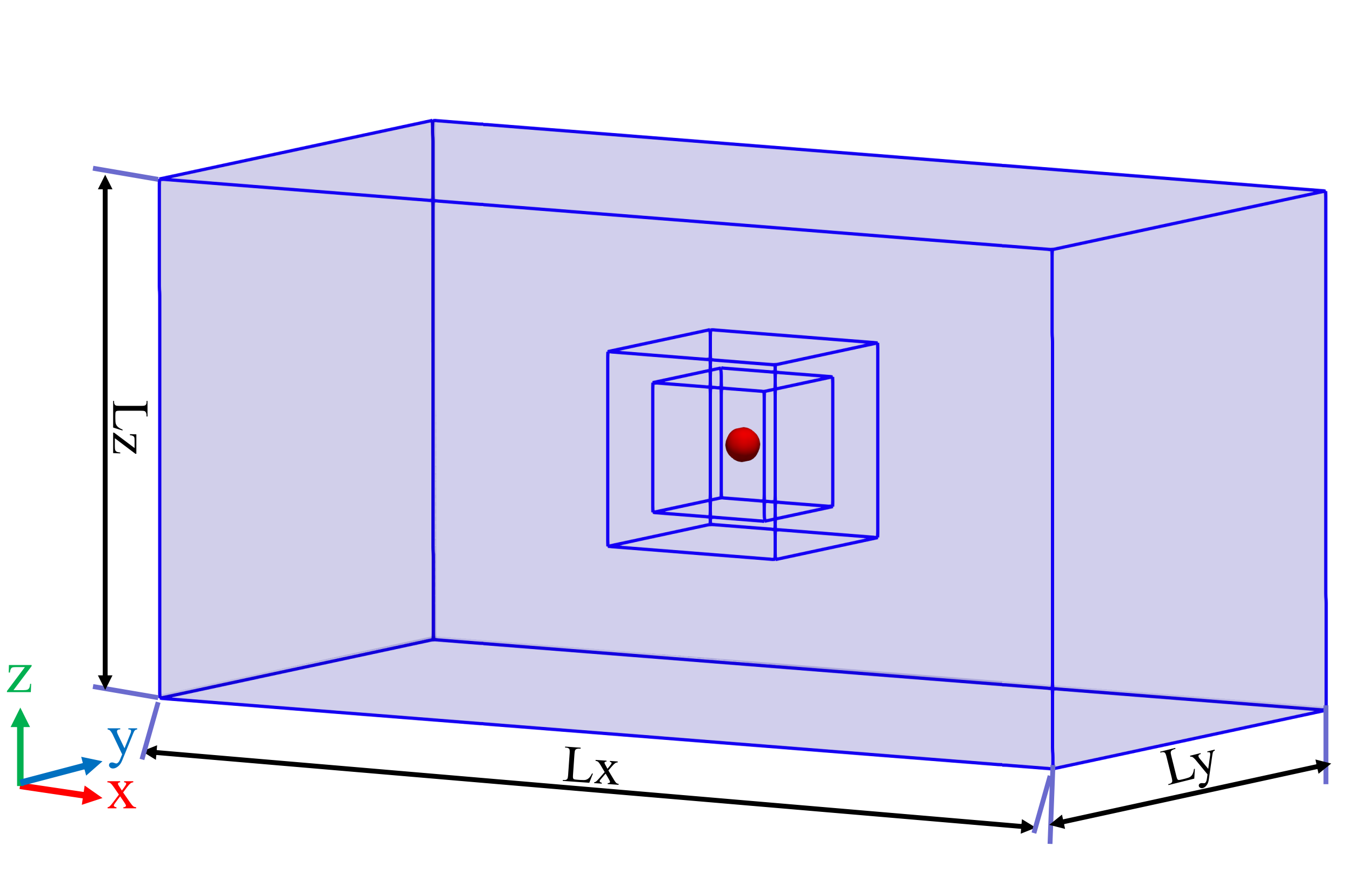}
        \label{fig:AMR_Sketch_Map}
    }
    \caption{Flow Past Rotating and Moving Sphere example, where (a) is a sketch with (\MakeUppercase{\romannumeral 1}) representing the uniform flow and (\MakeUppercase{\romannumeral 2}) representing the shear flow, and (b) is a three-level AMR grid.}
    \label{fig:Example_Schematic_Diagram}
\end{figure*}

The particle is stationary in the previous validation case.~In this session, we primarily validate the accuracy and effectiveness of our adaptive solver in handling particles under translational and rotational motion. We consider two different cases: the first case involves a particle with translational degrees of freedom in the x-direction under uniform flow, and the second case involves a particle with rotational degrees of freedom in the z-direction under shear flow~\cite{tschisgale2017non}.~The two cases are shown in Fig.~\ref{fig:Sketch_Map}, where (\MakeUppercase{\romannumeral 1}) and (\MakeUppercase{\romannumeral 2}) represent the uniform flow and shear flow, respectively.~The size of the computational domain is \(L_x \times L_y \times L_z = 30 \times 15 \times 15\), and the particle diameter is \(D_p = 1\).~The initial position of the particles is at the center of the flow field.~The density ratio of particles to the flow field \((R = \rho_p / \rho_f)\) is set as \(R = 1.05\) and \(R = 5\)~\cite{kempe2012improved,tschisgale2017non}.

\begin{figure*}[ht]
    \centering
    \subfigure[]{
        \includegraphics[width=0.47\textwidth]{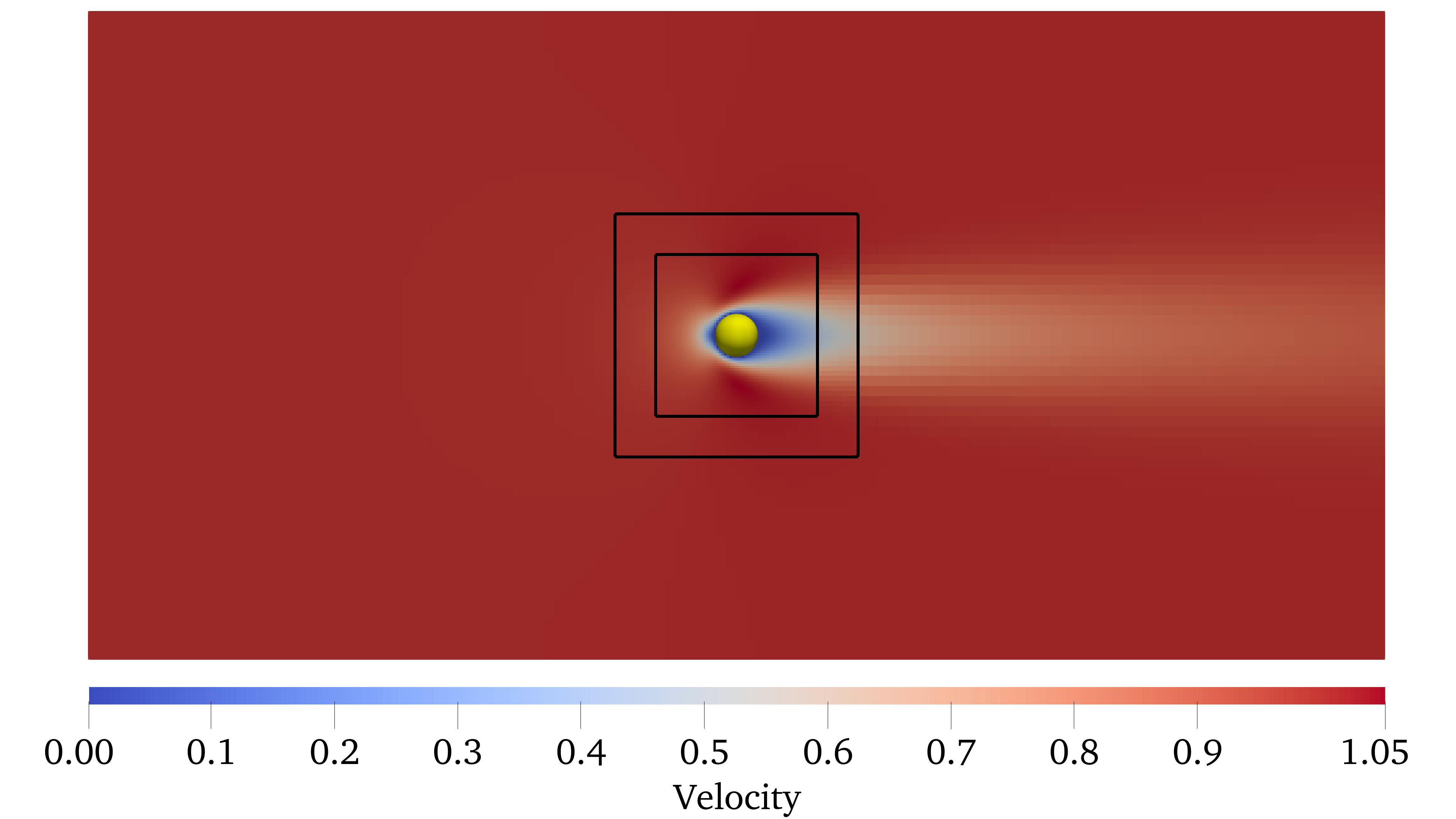}
        \label{fig:angular_velocity}
    }
    \subfigure[]{
        \includegraphics[width=0.47\textwidth]{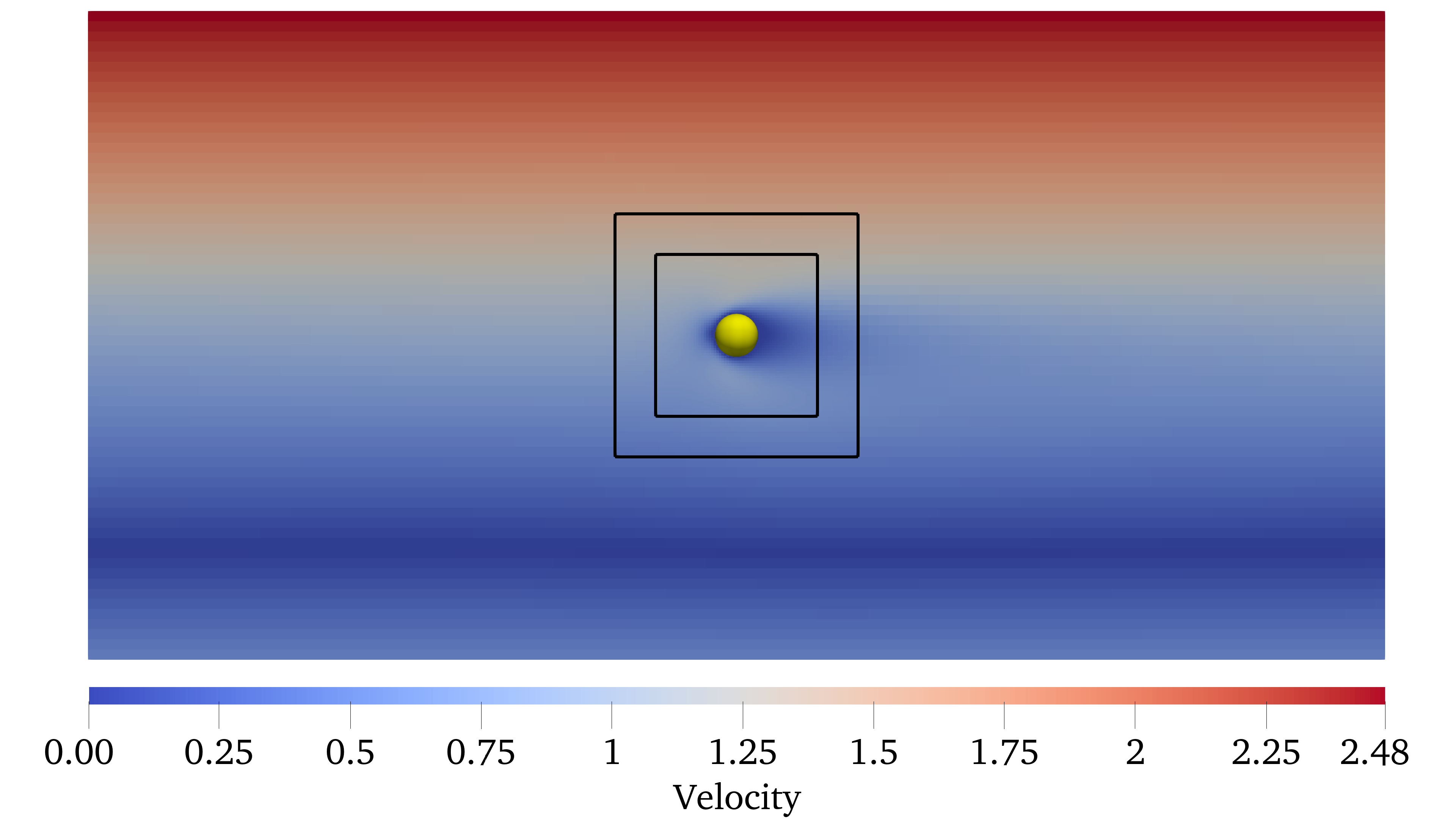}
        \label{fig:linear_velocity}
    }
    \caption{Steady-state flow field velocity contours in (a) uniform flow and (b) shear flow. The black line represents the finer-level grid.}
    \label{fig:Flow_Field_Steady_State_Diagram}
\end{figure*}

\begin{figure*}
    \includegraphics[width=0.9\textwidth]{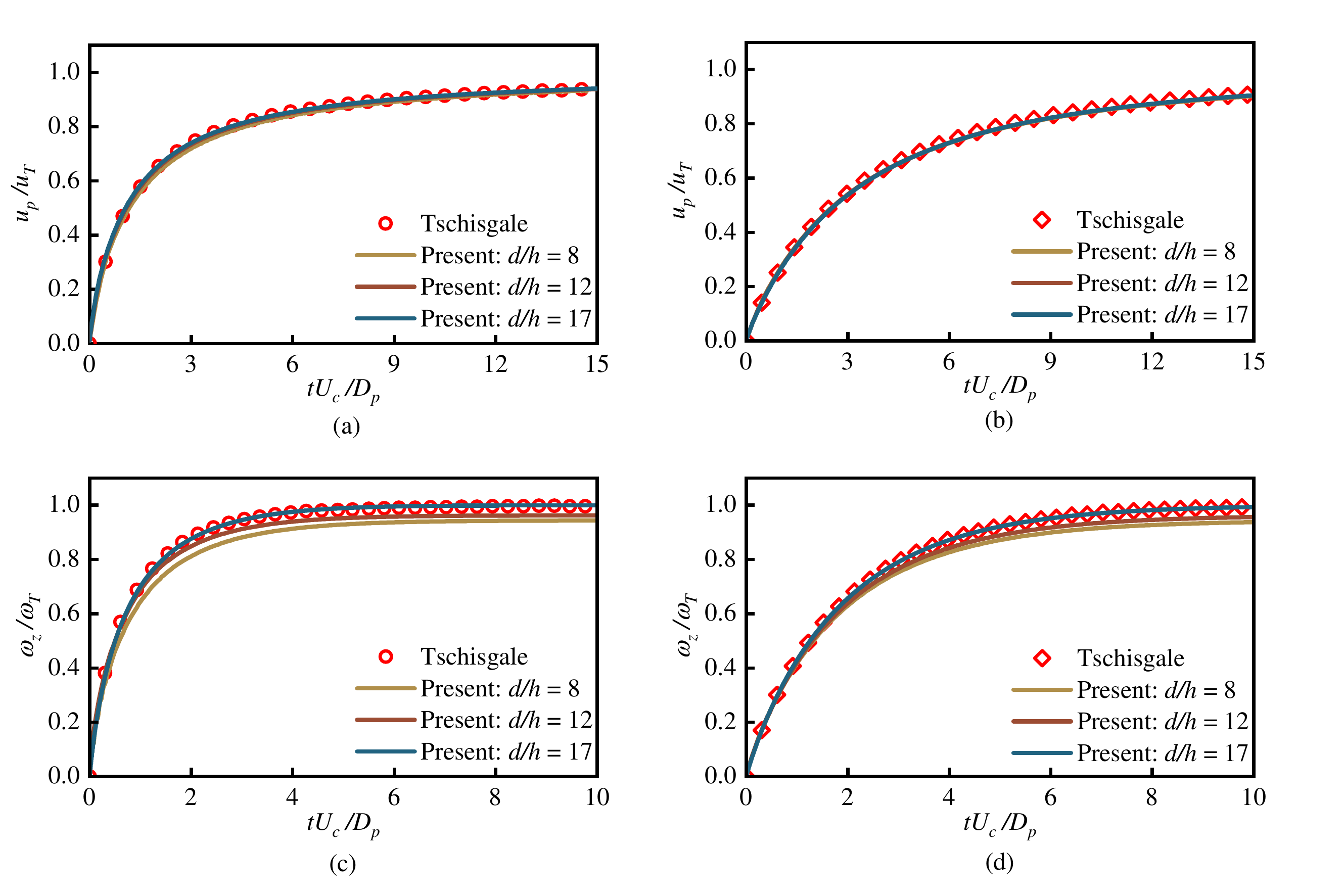}
    \caption{Comparison of the particle movement between the present results and Tschisgale et al~\citep{tschisgale2017non}; (a) and (b) show the evolution of the particle's linear velocity over time in uniform flows with density ratios of 1.05 and 5, while (c) and (d) show the evolution of the particle's angular velocity over time in shear flows with density ratios of 1.05 and 5.}
    \label{fig:sfs}
\end{figure*}


For the first case, inlet-outlet boundary conditions were applied in the \(x\) direction, while no-slip boundary conditions were used in the \(y\) and \(z\) directions.~For the second case, periodic boundary conditions were applied in the \(x\) and \(z\) directions.~No-slip boundary conditions were applied in the \(y\) direction.~To simulate different flow field conditions, we adjust the shear rate of the background flow at the inlet by changing the velocities of the upper and lower boundaries.~The shear flow is given by
\begin{equation}
    \mathbf{u}(y) = U_0 + Sy,
    \label{Shear_flow}
\end{equation}
where \( \mathbf{u}(y) \) represents the velocity at different heights, \( U_0 \) represents the velocity of the lower surface, and \( S \) is the shear rate.~In the uniform flow case, ~The inlet velocity was set to 1 to simulate the flow field conditions.~In the shear flow case, the upper boundary velocity \(u_{\text{top}}\) is set to be 2.5 and the lower boundary velocity \(u_{\text{bottom}}\) is set to be -0.5.~This velocity setting creates a velocity gradient in the \(y\) direction, thus forming a shear effect in the flow field.

In these two cases, a three-level AMR grid is used to save the computational cost without compromising the simulation accuracy. The sketch is shown in Fig.~\ref{fig:AMR_Sketch_Map}, the nested cell consists of levels from the outermost to the innermost, designated as level 0, level 1, and level 2, respectively.~The level 0, 1, and 2 grid are set to be \(N_x \times N_y \times N_z = 128 \times 64 \times 64\), \(N_x \times N_y \times N_z = 48 \times 48 \times 48\), and \(N_x \times N_y \times N_z = 64 \times 64 \times 64\), respectively.~The cell resolution for resolving the particle is measured by the ratio of \(D_p / h\), where \(D_p\) represents the particle diameter and \(h\) represents the cell spacing on the finest level.~We consider three different ratios to validate the convergence of our algorithm in this session, i.e.,~$D_p / h$=8, 12, and 17.~The definition of the Reynolds number is then defined as
\begin{equation}
    Re = \frac{u_p D_p}{\nu},
    \label{Shear_flow_Re}
\end{equation}
where \( u_p \) is the velocity of the flow field at the height of the center of the particle, and \( \nu \) is the kinematic viscosity of the flow field. The Reynolds number is chosen as \( Re = 20 \) for all validation cases in this session.

During the simulation, we first fix the particle and simulate the fluid motion alone.~Fig.~\ref{fig:angular_velocity} and Fig.~\ref{fig:linear_velocity} show the steady flow field under the uniform inflow and the shear inflow, respectively flow field is continuous and smooth across the coarse-fine boundaries with the help of synchronization operations in Session~\ref{S:sync}.~Once the fluid reaches a steady state, we release the particle and record the variation of the particle's translational velocity in the x-direction over time under uniform flow, as well as the variation of the particle's angular velocity in the z-direction over time under shear flow.~As shown in Fig.~\ref{fig:sfs}, increasing $D_p / h$ achieves better convergence. The results of $D_p / h=17$ are in good agreement with data in Tschisgale et al~\citep{tschisgale2017non}, which demonstrates the accuracy of our adaptive solver in dealing with particle motions with different degrees of freedom.

\subsection{\label{sec:R-fs}Falling Sphere}

\begin{figure}[h]
    \centering
    \includegraphics[width=0.3\textwidth]{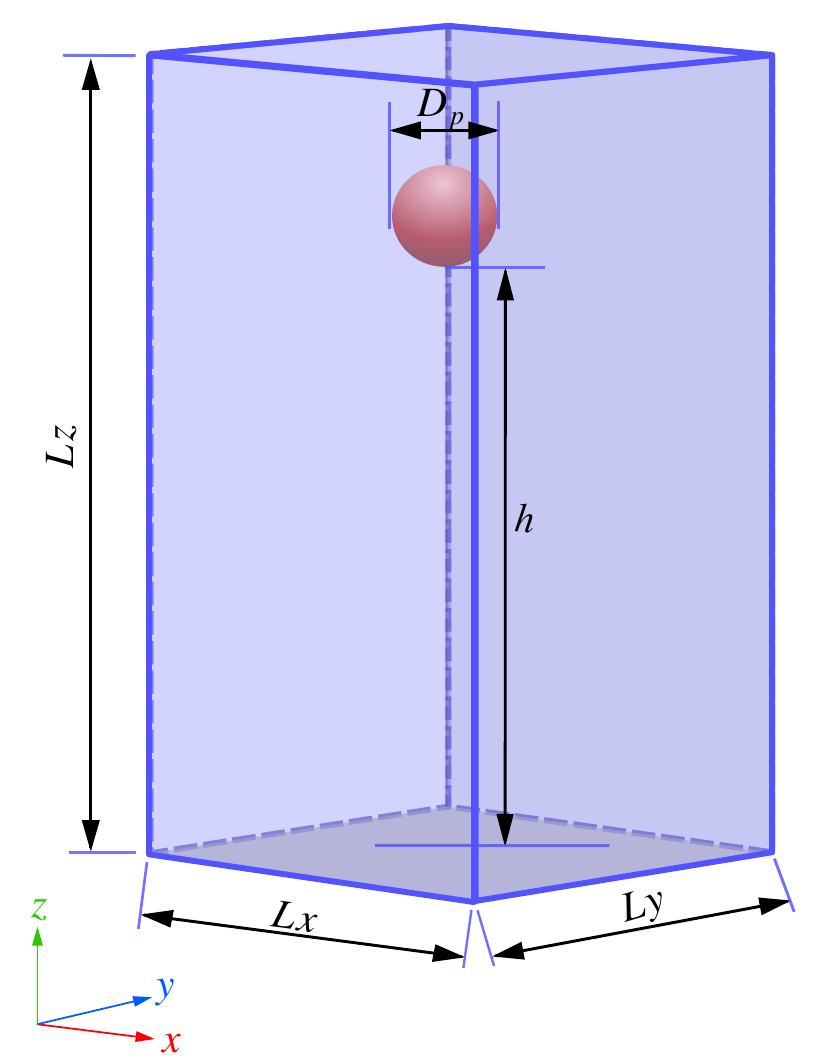}
    \caption{Sketch of a falling sphere case}
    \label{fig:scheme_of_FS}
\end{figure}

The falling sphere example is used to compare the subcycling and non-subcycling methods for particles with free motions~\cite{uhlmann2005immersed,kempe2012improved}.~We start with the computational setup~\citet{uhlmann2005immersed} and consider two particle density ratios $\rho_p/\rho_f$ in Table~\ref{tab:falling-rigid}. As the sketch in Fig.~\ref{fig:scheme_of_FS}, the computational domain is $L_x \times L_y \times L_z = 15D_p\times 15D_p\times L_z$, in which the particle diameter is $D_p=1/6$.~The initial position of the particles is $(x,y,z) = (12.5D_p, 12.5D_p, 114D_p)$, and the gravitational acceleration $g=-9.8 m/s$ is in the z-axis.~For all of following cases, the ratio of the particle diameter to the spacing of the finest grid is set to be $27$.

\begin{figure*}
    \includegraphics[width=0.9\textwidth]{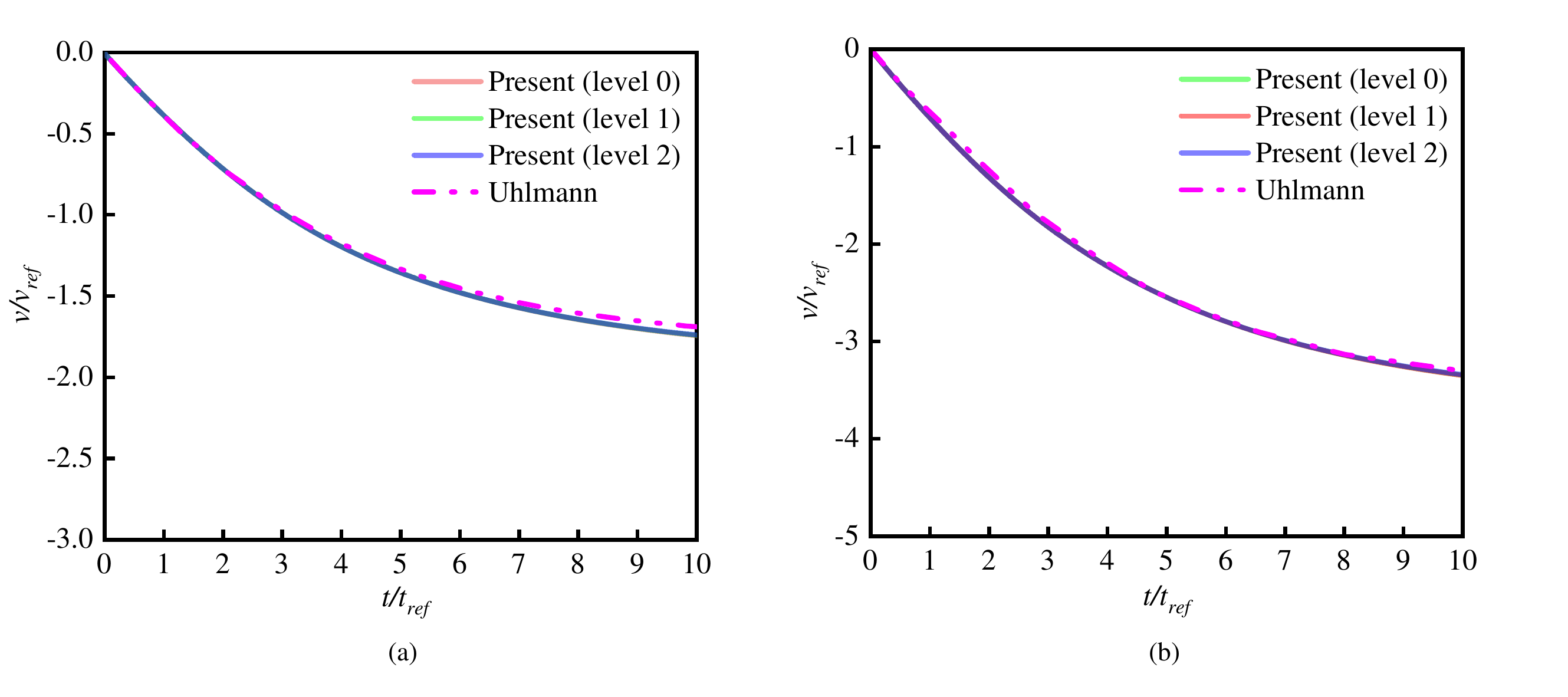}
    \caption{ Comparison of the settling velocity for different density ratios between the present results and~\citet{uhlmann2005immersed}: (a) $\rho_p/\rho_f = 2.56$; (b) $\rho_p/\rho_f = 7.71$.}
    \label{fig:falling-uhlmann}
\end{figure*}

\begin{table}[h]
\caption{\label{tab:falling-rigid}Parameters of the falling sphere case with two different density ratios}
\begin{ruledtabular}
\begin{tabular}{ccc}
    $\rho_p/\rho_f$ & $L_z(m)$ & $\nu_f(m^2/s)$\\
    \hline
    2.56 & 120$D_p$ & 0.00104238\\
    7.71 & $180D_p$ & 0.00267626\\
\end{tabular}
\end{ruledtabular}
\end{table}

Fig.~\ref{fig:falling-uhlmann} shows the present results on different levels using the non-subcycling method, in which $u_{ref}=\sqrt{|g|D_p}$ and $t_{ref}=\sqrt{D_p/|g|}$ are the reference velocity and time, respectively.~The present results validate the consistency of our solver on different levels and show good agreement with~\citet{uhlmann2005immersed} on both high-density ratio (i.e., $\rho_p/\rho_f = 7.71$) and relatively low-density ratio (i.e., $\rho_p/\rho_f = 2.56$).

We further test our solver on another canonical setup~\cite{tenCate2002}. The size of the entire calculation domain is $Lx \times Ly \times L_z = 6.67D_p\times 6.67D_p\times 13.34D_p$, the diameter of the particle $Dp = 0.015$, and the initial position of the particle is $(x,y,z) = (3.33D_p, 3.33D_p, 8.7D_p)$.~The particle is released from the initial position until it touches the bottom boundary.~The gravity $g=-9.81m/s^2$ is vertically downward and the density of the particles is $\rho_p = 1120 kg/m^3$.~For the bottom and side boundaries, both of them are no-slip boundaries, while the top is a free-slip boundary.~The gird number on level 0 is $32\times 32\times 64$.~Other parameters are listed in Fig.~\ref{tab:falling-pvf}, in which four cases with AMR are set up and each of them utilizes both the subcycling and non-subcycling methods.

\begin{table}[h]
    \caption{\label{tab:falling-pvf}Parameters of the same falling sphere case as~\citet{tenCate2002}}
    \begin{ruledtabular}
    \begin{tabular}{ccccc}
        Case No. & $\rho_f(kg/m^3)$ & $\nu_f(m^2/s)$ & Re & $l_{max}$\\
        \hline
        1 & 970 & $3.845\times 10^{-4}$ & 1.5  & 2\\
        2 & 965 & $2.197\times 10^{-4}$ & 4.1  & 2\\
        3 & 962 & $1.175\times 10^{-4}$ & 11.6 & 2\\
        4 & 960 & $6.042\times 10^{-5}$ & 32.2 & 2\\
    \end{tabular}
    \end{ruledtabular}
\end{table}



\begin{figure*}
    \includegraphics[width=0.9\textwidth]{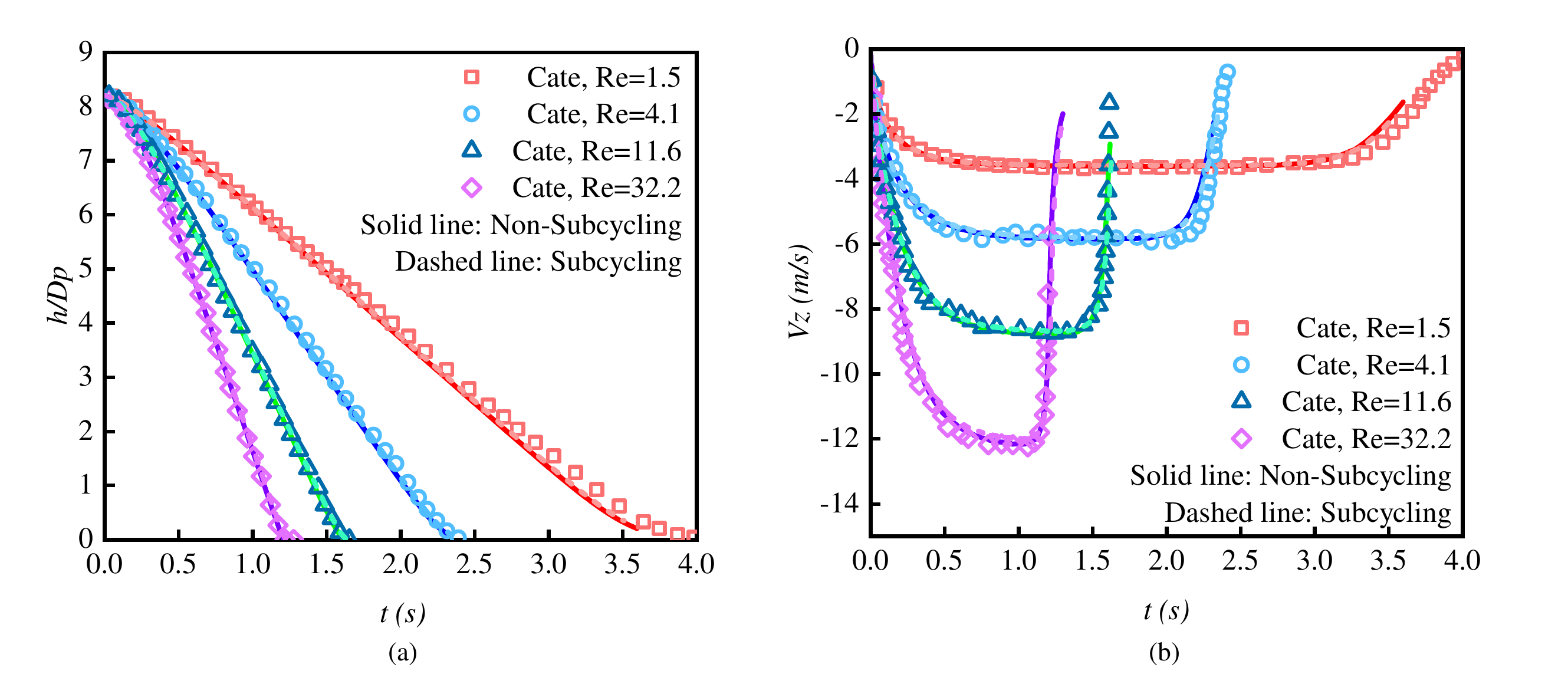}
    \caption{Comparison of the subcycling and non-subcycling results with~\citet{tenCate2002}: (a) the time series of the dimensionless height; (b) the time series of the dimensional z-velocity.}
    \label{fig:falling-pvf}
\end{figure*}

\begin{figure}[H]
    \includegraphics[width=0.45\textwidth]{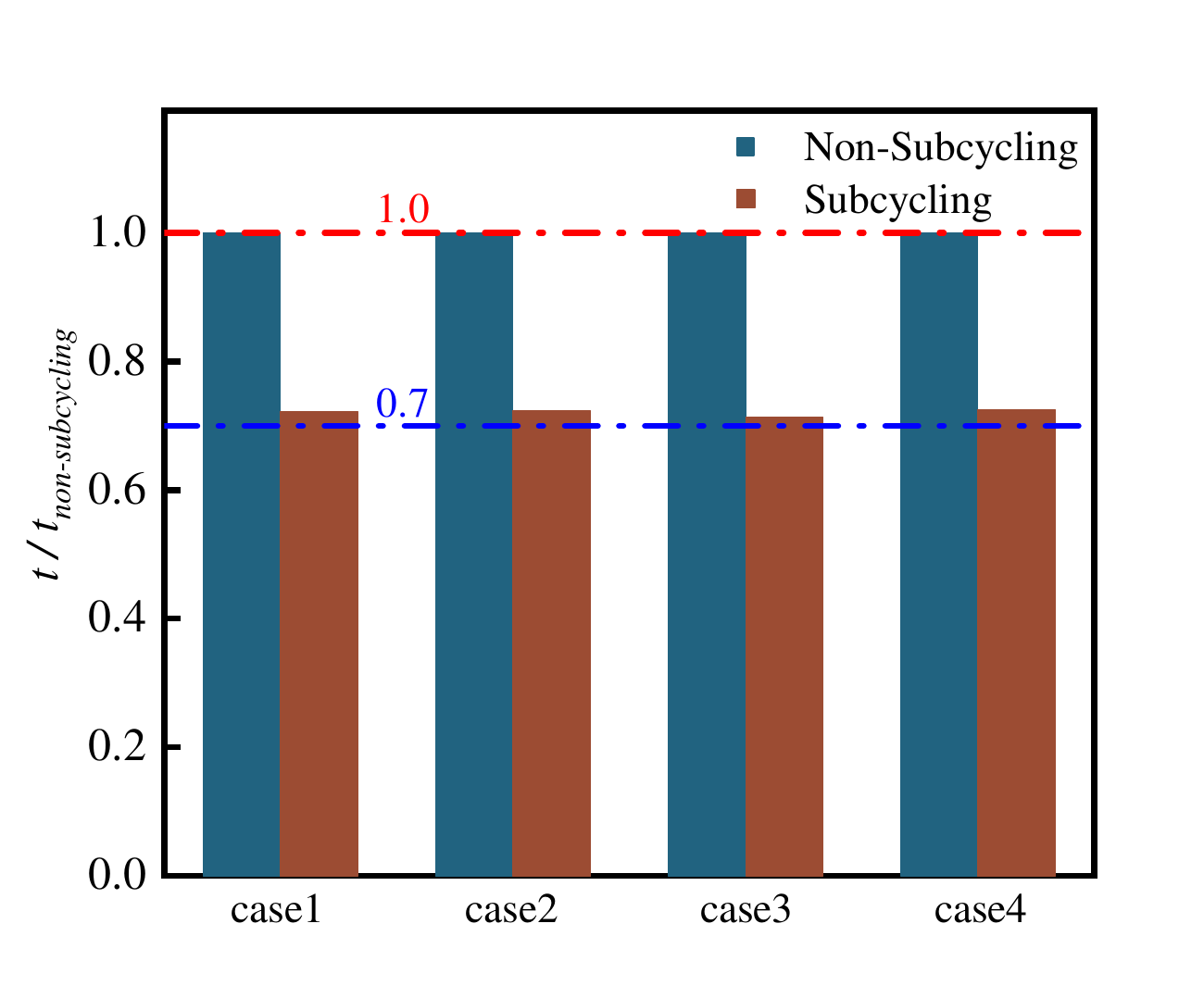}
    \caption{Comparison of the computational cost between the subcycling and non-subcycling method for all falling sphere cases in Table~\ref{tab:falling-pvf}}
    \label{fig:falling-pvf-cycling-time}
\end{figure}

As shown in Fig.~\ref{fig:falling-pvf}, our present numerical results achieve good agreement with the experimental results of~\citet{tenCate2002}.~For cases with different density ratios, the particle accelerates at the initial stage, reaches the steady state, and then touches the wall.~The overlapping between the solid line and the dashed line also validates the accuracy and consistency of our subcyling and non-subcycling methods for particles with free motions.~In Fig.\ref{fig:falling-pvf-cycling-time}, the computational cost between the subcycling method is compared with the non-subcycling method for all four cases in Table~\ref{tab:falling-pvf}. During the comparison, the running time, without including the IO process, is added and normalized by the total time of the non-subcycling method. It is seen that the subcycling method is more efficient in the falling sphere simulation, which takes around $30$ percent less time compared with the non-subcycling time. 



\subsection{\label{sec:R-dkt}Drafting-Kissing-Tumbling}
In this section, we study the drafting, kissing, and tumbling (DKT) phenomenon of a pair of particles.~This phenomenon is frequently observed in particle sedimentations and has been studied in previous work~\cite{uhlmann2005immersed,zhu2022particle,gong2023cp3d}.~The objective here is to validate the correctness and accuracy of our adaptive solver when a collision model of two particles is combined with AMR. Following the previous work of~\citet{breugem2012second}, a collision model between the particles is added as follows:
\begin{eqnarray}
    F_{c,ij} = \left\{ \begin{array}{cc}
        -\frac{m_p ||{\bf g}||}{\epsilon_c}\left(\frac{||\delta_{i,j}|| - D_c}{d_c}\right)^2\frac{\delta_{i,j}}{||\delta_{i,j}||} & |\delta_{i,j}| < D_c\\
        0&|\delta_{i,j}| > D_c
    \end{array}\right.
    \label{eq:DKT_model}
\end{eqnarray}

In Eq.~\ref{eq:DKT_model}, $F_c$ refers to the repulsive force, $m$ denotes the mass of the particles, $\bf g$ represents the gravitational acceleration, $\delta$ represents the distance between the two particles, and $D_c$ represents the sum of the radius of the two particles and the grid size. The $d_c$ represents the grid spacing on the finest level and the dimensionless constant $\epsilon_c$ is set to be $10^{-4}$.

In this DKT case, the diameters of the two particles are the same, $D_p=1.67mm$, and the computational domain is $Lx \times Ly \times L_z = 6D_p\times6D_p\times24D_p$.~The initial position of the particle at the higher location is $(x_h, y_h, z_h) = (3.03D_p,3.03D_p,21.0D_p)$, while the particle at the lower location starts at $(x_l, y_l, z_l) = (2.97D_p,2.97D_p,18.96D_p)$.~In addition, the physical parameters of the fluid flow and particles are as follows: $\rho_p = 1140 kg/m^3$, $\rho_f=1000 kg/m^3$, and $\nu_f = 10^{-6} m^2/s$.~The gravity $g=-9.81m/s^2$ is vertically downward.~All boundaries of the computational domain have no-slip conditions, and the ratio of the particle diameter to the grid spacing on the finest level is consistent in all directions, i.e., $D_p/d_c = 16$.

\begin{figure}[h]
    \includegraphics[width=0.45\textwidth]{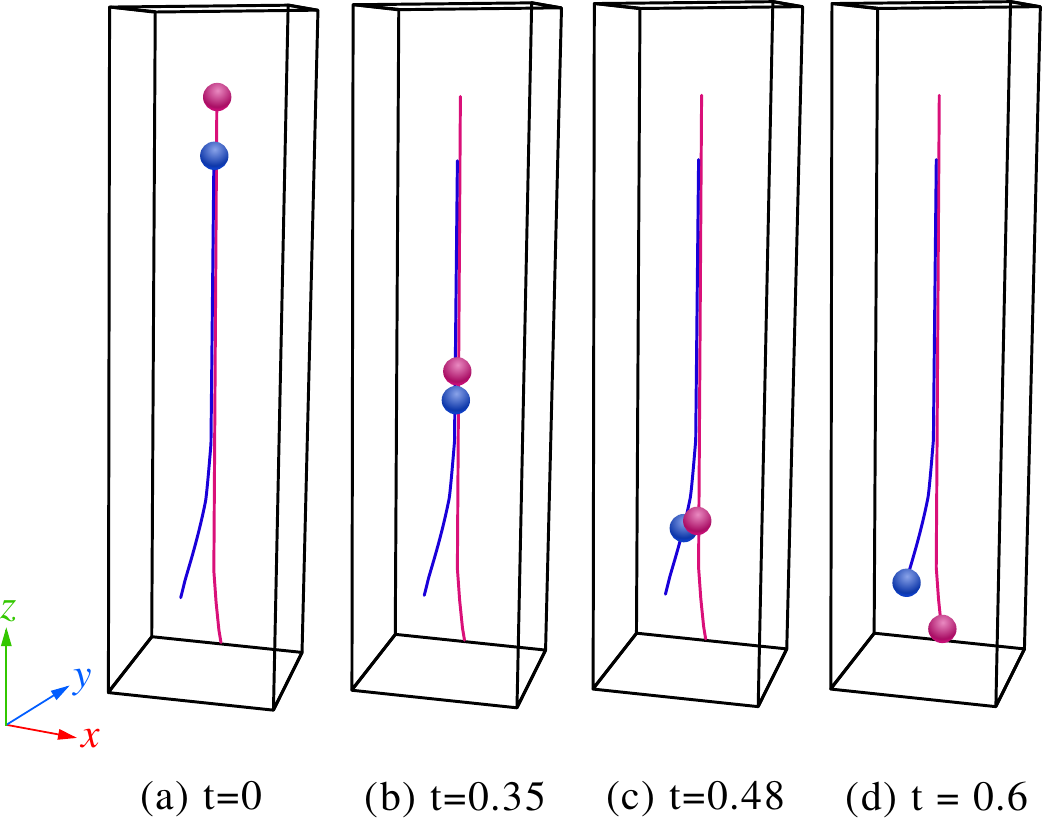}
    \caption{Trajectory of two particles and their positions at different time steps}
    \label{fig:dkt-particle-path}
\end{figure}

Fig.~\ref{fig:dkt-particle-path} shows the motion trajectories of the particles and their positions at different time steps.~At the initial stage of the sedimentation process, the upper particle is slightly higher than the lower particle.~From $t=0$ to $t=0.35$,  the lower pressure in the wake of
the lower particle results in the larger velocity of the upper particle, which helps it to gradually catch up with the lower particle.~This process is known as "drafting".~Then the distance between the two particles gradually decreases until they eventually collide.~From $t=0.35$ to $t=0.48$ in Fig.~\ref{fig:dkt-particle-path}(b) and (c), the distance between the two particles remains almost constant, and during this period, the upper particle gradually shifts to the side of the lower particle.~This process is called "kissing".~After the "kissing" stage, the particle that was originally higher flips to the side of the lower particle, and the unstable vertical alignment makes the upper particle push the lower one aside and take the lead. This process is called "tumbling". Fig.~\ref{fig:dkt-particle-path} shows that our present results can qualitatively and reasonably reproduce the DKT process~\cite{gong2023cp3d,breugem2012second}.

\begin{figure*}
    \includegraphics[width=0.9\textwidth]{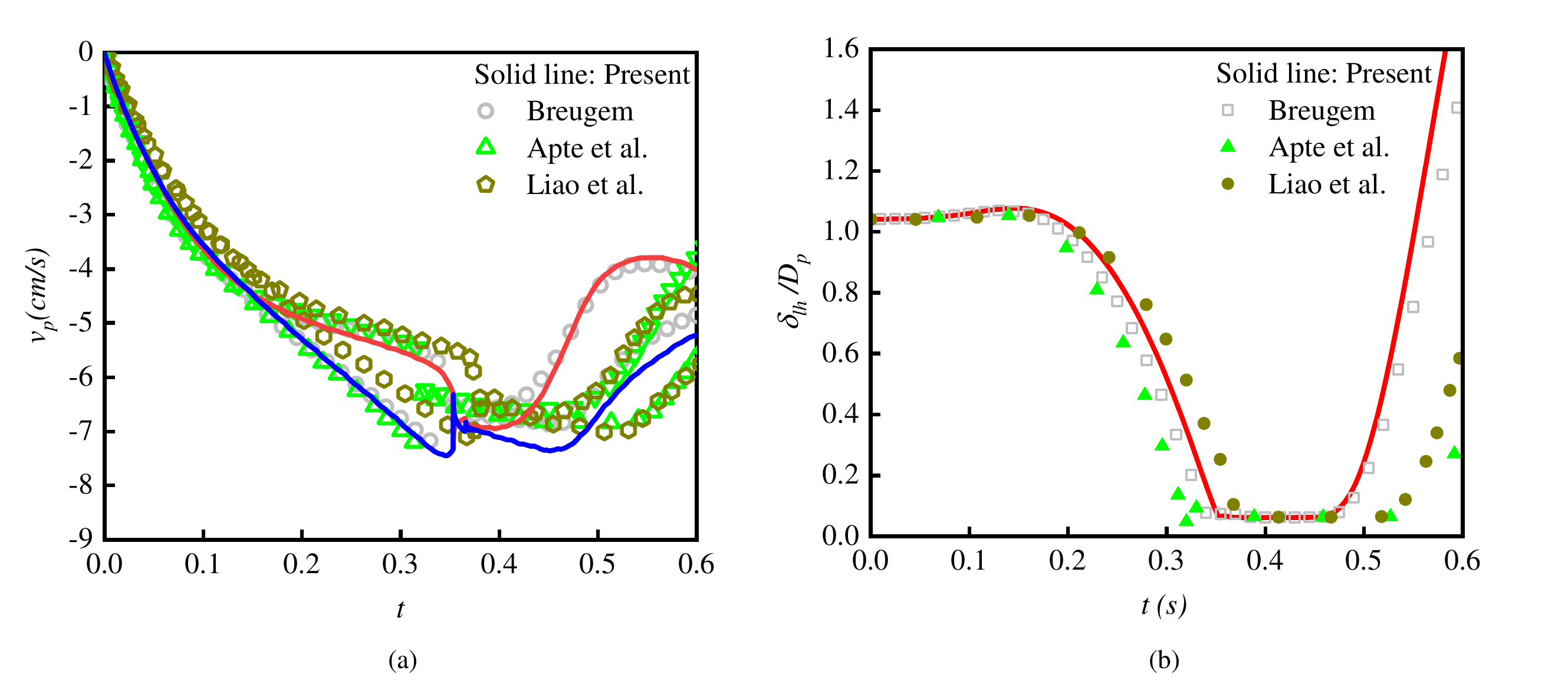}
    \caption{(a)~Time series of the vertical velocities of the two particles; (b)~Time series of the distance of two particles.~The present results are compared with~\citet{breugem2012second},~\citet{Apte2009ANM}, and~\citet{Liao2015SimulationsOT}.}
    \label{fig:dkt-level2}
\end{figure*}

The time series of the distance and the vertical velocities of these two
particles are shown in Fig.~\ref{fig:dkt-level2}.~Our adaptive results with the non-subcycling method agree well with the data of~\citet{breugem2012second},~\citet{Apte2009ANM}, and~\citet{Liao2015SimulationsOT}.


\subsection{\label{sec:R-mp}Cluster of monodisperse particles}

\begin{figure*}
    \includegraphics[width=0.9\textwidth]{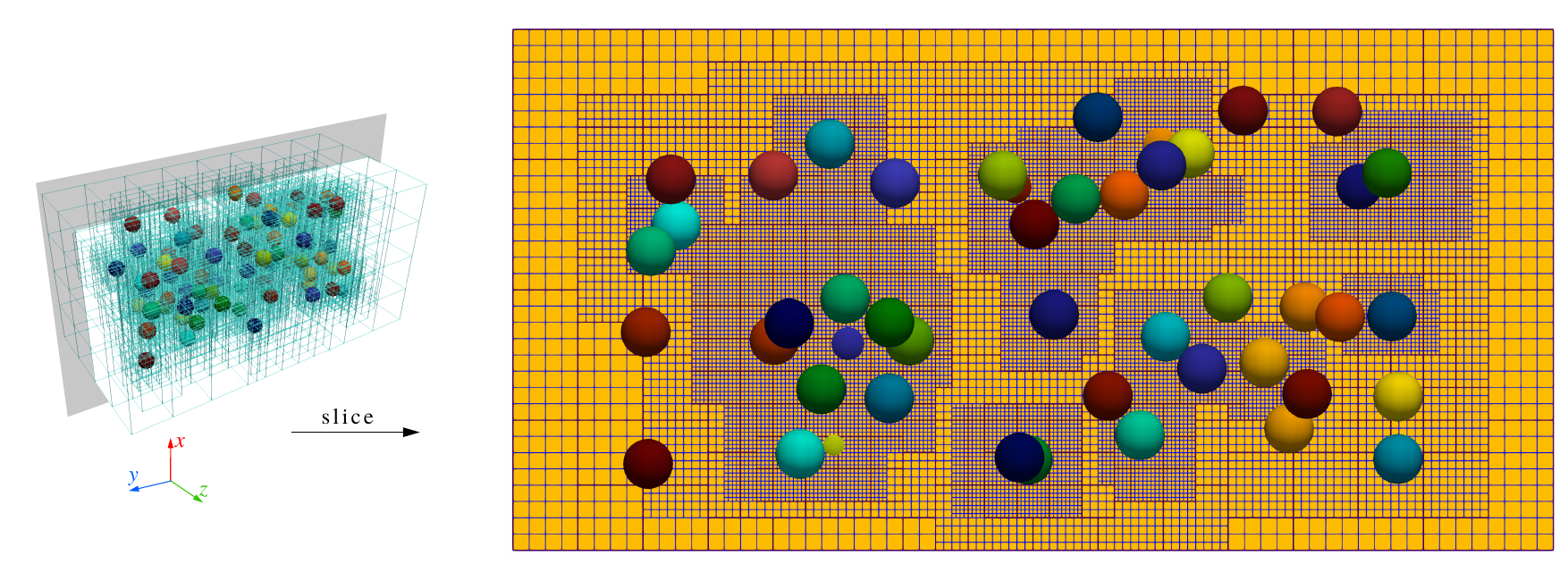}
    \caption{Monodisperse particles on a three-level AMR grid}
    \label{mono-particle-level2}
\end{figure*}

As the last example, we demonstrate the accuracy and efficacy of our codes for simulating clusters of particles on the multi-level grid.~As shown in Fig.~\ref{mono-particle-level2}, 80 particles of diameter D = 1 are randomly distributed in a channel of size $L_x\times L_y \times L_z = 10\times 20 \times 10$, and the fluid flow is driven by applying a pressure gradient of 1.0 in the $z$ direction.~This case can represent a porous medium with a volume fraction of 0.02.~Three levels of the AMR grid is applied and the $D/h=16$ is used on the finest level.~As shown in Table~\ref{tab:gridCellsInMonodisperse}, we found that the total number of cells in the AMR grid is 2,256,320, which is a 72.46\% reduction compared with the single-level simulation without AMR.

\begin{table}[h]
\caption{\label{tab:gridCellsInMonodisperse}Number of grid cells for monodisperse particle cases}
\begin{ruledtabular}
\begin{tabular}{ccccc}
Case no. & Level 0 cells & Level 1 cells & Level 2 cells & Total cells\\
\hline
1 & 8,192,000 & - & - & 8,192,000  \\
2 & 128,000  & 534,656 & 1,593,664 &  2,256,320 \\
\end{tabular}
\end{ruledtabular}
\end{table}

\begin{figure}[H]
    \centering
    \includegraphics[width=0.45\textwidth]{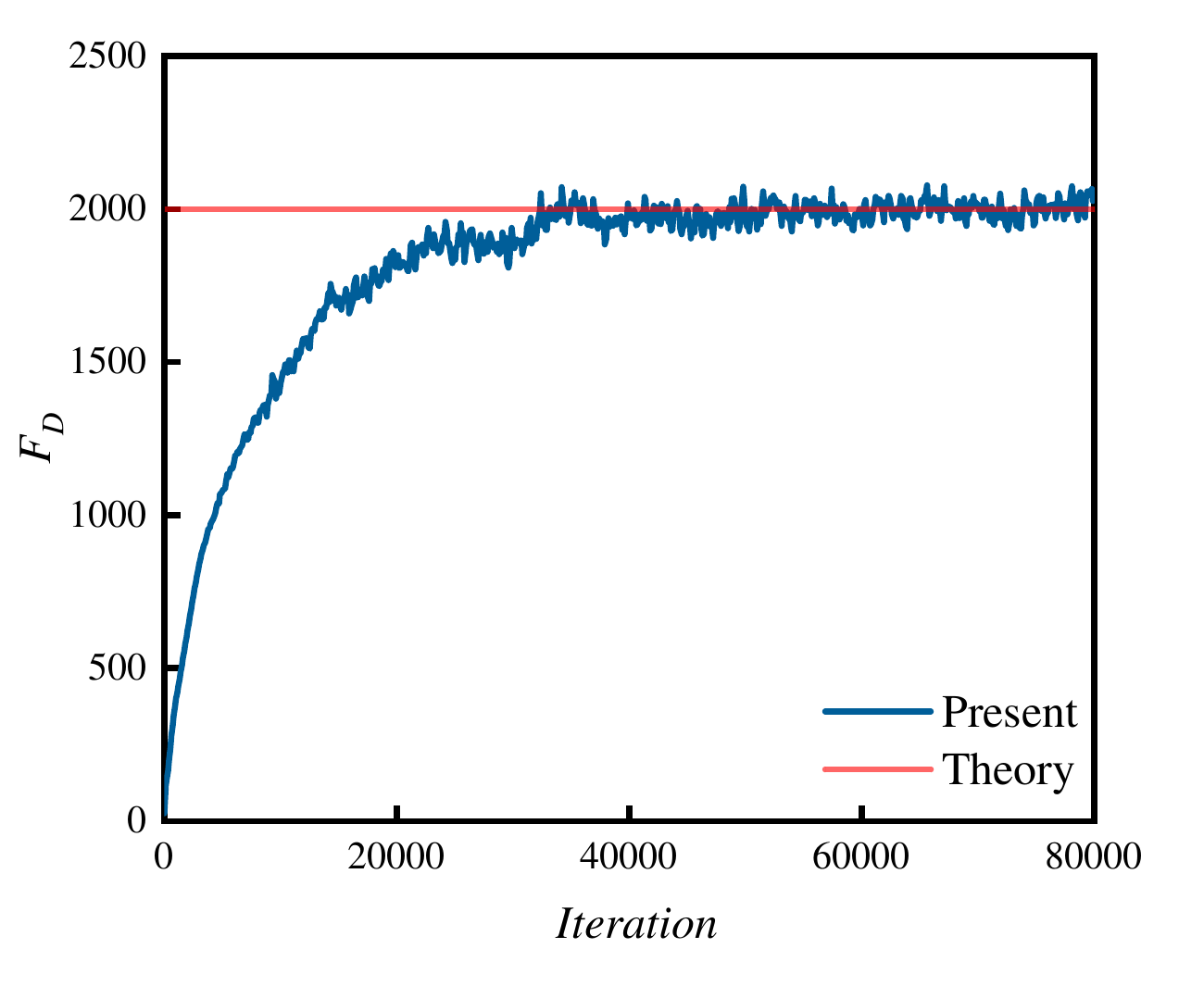}
    \caption{Comparison of total drag force between the theory and present results.}
    \label{fig:monodisperse_fz}
\end{figure}

\begin{figure}[h]
    \includegraphics[width=0.5\textwidth]{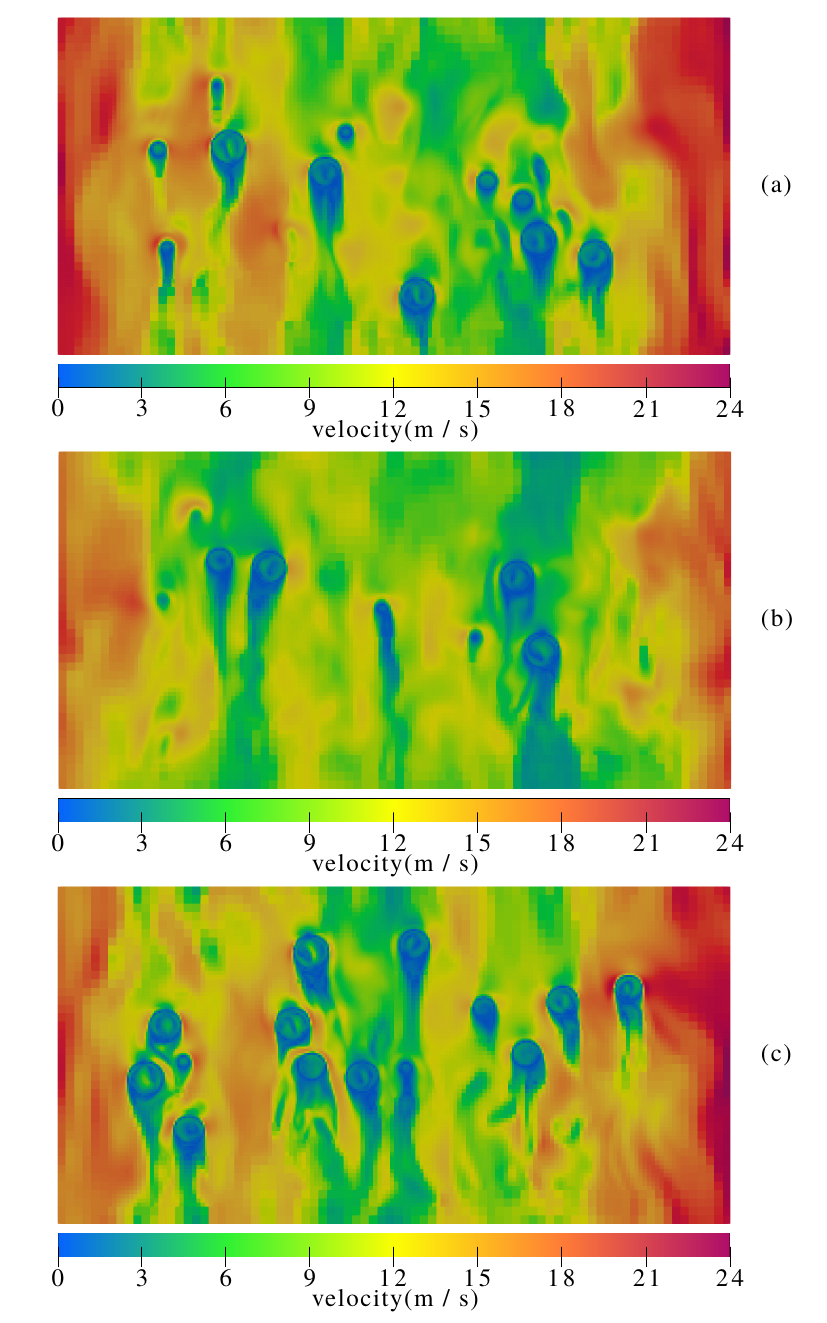}
    \caption{Contours of velocity magnitude in $y-z$ plane at the different $x$ position:~(a) $x=2.5$; (b) $x=5.0$; (c) $x=7.5$.}
    \label{fig:monodispers_v}
\end{figure}

When the simulation reaches the steady state, the total pressure drop balances the IB force generated by all particles in the streamwise $z$ direction.~Following the equation in~\cite{akiki2016immersed}, the theoretical drag force would be
\begin{equation}
    F_{theory} = (\frac{\Delta p}{\Delta z}L_z)L_xL_y.
    \label{equ:total_pressure_drop}
\end{equation}
Fig.~\ref{fig:monodispers_v} represents the velocity contour of three different interfaces in the x direction at the steady state. The flow passes around the particles and generates the wakes behind. Fig.~\ref{fig:monodisperse_fz} shows the time series of total IB force for all particles.~The resistance gradually reaches a steady state after $40,000$ steps.~In this case, the theoretical value of drag force given by Eq.~\ref{equ:total_pressure_drop} is 2000, while the present value at steady state is around 2002.~This close agreement validates the accuracy of our proposed framework in dealing with large amounts of particles in the fluid system. 

\section{\label{sec:Conclusion}Conclusion}

In this work, we established a novel adaptive solver with subcycling and non-subcycling time advancement methods for simulating fluid-particle interaction.~The proposed multi-level advancement algorithm uses the level-by-level advancement technique for time-marching the variables in the adaptive meshes and decouples the time advancement at different levels. When the subcycling method is applied, the time step constraint on the coarser levels is relaxed compared to the finer levels.~The accuracy and efficacy of both subcycling and non-subcycling methods are validated by the classic flow past sphere and falling sphere cases.

We also implemented different types of constraints for particles, including the prescribed and free motions.~The numerical approximations of the PVF variable are shown to be second-order accurate and match our time advancement algorithms~\cite{zeng2022aparallel}.~The particle motions are also validated using the sphere in uniform and shear flow cases. Besides, the collision model with repulsive force is correctly added to account for particle-particle collision and validated by the DKT case on the adaptive meshes.~More advanced collision models, such as the solid sphere model (SSM) and Adaptive Collision Time Model (ACTM)~\cite{gong2023cp3d,zhu2023multiple}, will be included in future work.

We brought two memory optimization techniques of our adaptive solver.~First, the Lagrangian markers associated with particles only exist on the finest level of the adaptive grid.~Since the Eulerian-Lagrangian information exchange information only happens on the finest level, we do not need to define any auxiliary variables and particle information on the coarser levels.~This helps to save memory compared with our previous study~\cite{zeng2022subcycling}, in which coarser levels also store Eulerian forces for using the force averaging schemes.~Second, there is only one set of Lagrangian markers when many particles are simulated. Since we loop over particles, memory associated with Lagrangian markers can be reused.


At last, the cluster of monodispersed particles case shows the accuracy and robustness of the computational framework while simulating large amounts of particles.~This capability enabled us to simulate a bunch of practical engineering problems, including aeolian sand and dust movement, sediment transport, and fluidized-bed processes.~The codes are openly available in the GitHub and the raw postprocessing data is also attached for reproducing all results in Session~\ref{sec:Result}.

Regarding ongoing work,~we will extend our adaptive solver and add some new features, including GPU running~\cite{liu2024investigate,yao2022massively}, fully implicit schemes of fluid-particle coupling~\cite{wu2024implicit}, non-Newtonain fluids~\cite{sverdrup2018highly,sverdrup2019embedded}, and more particle collision models~\cite{costa2015collision,zhu2022particle}.~Considering different shapes of particles~\cite{gan2016cfd} and running the simulations on the non-uniform grid~\cite{akiki2016immersed,pinelli2010immersed} are also two promising directions for practical applications.

\begin{acknowledgments}
X.L., Y.Z., and Z.Z. are grateful to Ann Almgren, Andy Nonaka, Andrew Myers, Axel Huebl, and Weiqun Zhang in the Lawrence Berkeley National Laboratory (LBNL) for their discussions related to AMReX and IAMR.~Y.Z. and Z.Z. also thank Prof.~Lian Shen, Prof.~Ruifeng Hu and Prof.~Xiaojing Zheng during their Ph.D. studies. 
\end{acknowledgments}

\section*{Data Availability Statement}

The codes that support the findings of this study are openly available in~\url{https://github.com/ruohai0925/IAMR/tree/development}.~Please help to submit a pulling request (PR) in the GitHub if needed.~Raw postprocessing data of all cases can be downloaded from~\url{https://pan.baidu.com/s/1bZRoDunjBv7bqYL8CI3ASA?pwd=i5c2}.

\FloatBarrier
\bibliography{aipsamp}

\begin{thebibliography}{97}%
\makeatletter
\providecommand \@ifxundefined [1]{%
 \@ifx{#1\undefined}
}%
\providecommand \@ifnum [1]{%
 \ifnum #1\expandafter \@firstoftwo
 \else \expandafter \@secondoftwo
 \fi
}%
\providecommand \@ifx [1]{%
 \ifx #1\expandafter \@firstoftwo
 \else \expandafter \@secondoftwo
 \fi
}%
\providecommand \natexlab [1]{#1}%
\providecommand \enquote  [1]{``#1''}%
\providecommand \bibnamefont  [1]{#1}%
\providecommand \bibfnamefont [1]{#1}%
\providecommand \citenamefont [1]{#1}%
\providecommand \href@noop [0]{\@secondoftwo}%
\providecommand \href [0]{\begingroup \@sanitize@url \@href}%
\providecommand \@href[1]{\@@startlink{#1}\@@href}%
\providecommand \@@href[1]{\endgroup#1\@@endlink}%
\providecommand \@sanitize@url [0]{\catcode `\\12\catcode `\$12\catcode `\&12\catcode `\#12\catcode `\^12\catcode `\_12\catcode `\%12\relax}%
\providecommand \@@startlink[1]{}%
\providecommand \@@endlink[0]{}%
\providecommand \url  [0]{\begingroup\@sanitize@url \@url }%
\providecommand \@url [1]{\endgroup\@href {#1}{\urlprefix }}%
\providecommand \urlprefix  [0]{URL }%
\providecommand \Eprint [0]{\href }%
\providecommand \doibase [0]{http://dx.doi.org/}%
\providecommand \selectlanguage [0]{\@gobble}%
\providecommand \bibinfo  [0]{\@secondoftwo}%
\providecommand \bibfield  [0]{\@secondoftwo}%
\providecommand \translation [1]{[#1]}%
\providecommand \BibitemOpen [0]{}%
\providecommand \bibitemStop [0]{}%
\providecommand \bibitemNoStop [0]{.\EOS\space}%
\providecommand \EOS [0]{\spacefactor3000\relax}%
\providecommand \BibitemShut  [1]{\csname bibitem#1\endcsname}%
\let\auto@bib@innerbib\@empty
\bibitem [{\citenamefont {Balachandar}\ and\ \citenamefont {Eaton}(2010)}]{balachandar2010turbulent}%
  \BibitemOpen
  \bibfield  {author} {\bibinfo {author} {\bibfnamefont {S.}~\bibnamefont {Balachandar}}\ and\ \bibinfo {author} {\bibfnamefont {J.~K.}\ \bibnamefont {Eaton}},\ }\bibfield  {title} {\enquote {\bibinfo {title} {Turbulent dispersed multiphase flow},}\ }\href@noop {} {\bibfield  {journal} {\bibinfo  {journal} {Annual review of fluid mechanics}\ }\textbf {\bibinfo {volume} {42}},\ \bibinfo {pages} {111--133} (\bibinfo {year} {2010})}\BibitemShut {NoStop}%
\bibitem [{\citenamefont {Brandt}\ and\ \citenamefont {Coletti}(2022)}]{brandt2022particle}%
  \BibitemOpen
  \bibfield  {author} {\bibinfo {author} {\bibfnamefont {L.}~\bibnamefont {Brandt}}\ and\ \bibinfo {author} {\bibfnamefont {F.}~\bibnamefont {Coletti}},\ }\bibfield  {title} {\enquote {\bibinfo {title} {Particle-laden turbulence: progress and perspectives},}\ }\href@noop {} {\bibfield  {journal} {\bibinfo  {journal} {Annual Review of Fluid Mechanics}\ }\textbf {\bibinfo {volume} {54}},\ \bibinfo {pages} {159--189} (\bibinfo {year} {2022})}\BibitemShut {NoStop}%
\bibitem [{\citenamefont {Squires}\ and\ \citenamefont {Eaton}(1990)}]{squires1990particle}%
  \BibitemOpen
  \bibfield  {author} {\bibinfo {author} {\bibfnamefont {K.~D.}\ \bibnamefont {Squires}}\ and\ \bibinfo {author} {\bibfnamefont {J.~K.}\ \bibnamefont {Eaton}},\ }\bibfield  {title} {\enquote {\bibinfo {title} {Particle response and turbulence modification in isotropic turbulence},}\ }\href@noop {} {\bibfield  {journal} {\bibinfo  {journal} {Physics of Fluids A: Fluid Dynamics}\ }\textbf {\bibinfo {volume} {2}},\ \bibinfo {pages} {1191--1203} (\bibinfo {year} {1990})}\BibitemShut {NoStop}%
\bibitem [{\citenamefont {Wang}\ and\ \citenamefont {Maxey}(1993)}]{wang1993settling}%
  \BibitemOpen
  \bibfield  {author} {\bibinfo {author} {\bibfnamefont {L.-P.}\ \bibnamefont {Wang}}\ and\ \bibinfo {author} {\bibfnamefont {M.~R.}\ \bibnamefont {Maxey}},\ }\bibfield  {title} {\enquote {\bibinfo {title} {Settling velocity and concentration distribution of heavy particles in homogeneous isotropic turbulence},}\ }\href@noop {} {\bibfield  {journal} {\bibinfo  {journal} {Journal of fluid mechanics}\ }\textbf {\bibinfo {volume} {256}},\ \bibinfo {pages} {27--68} (\bibinfo {year} {1993})}\BibitemShut {NoStop}%
\bibitem [{\citenamefont {Ferrante}\ and\ \citenamefont {Elghobashi}(2003)}]{ferrante2003physical}%
  \BibitemOpen
  \bibfield  {author} {\bibinfo {author} {\bibfnamefont {A.}~\bibnamefont {Ferrante}}\ and\ \bibinfo {author} {\bibfnamefont {S.}~\bibnamefont {Elghobashi}},\ }\bibfield  {title} {\enquote {\bibinfo {title} {On the physical mechanisms of two-way coupling in particle-laden isotropic turbulence},}\ }\href@noop {} {\bibfield  {journal} {\bibinfo  {journal} {Physics of fluids}\ }\textbf {\bibinfo {volume} {15}},\ \bibinfo {pages} {315--329} (\bibinfo {year} {2003})}\BibitemShut {NoStop}%
\bibitem [{\citenamefont {Vance}, \citenamefont {Squires},\ and\ \citenamefont {Simonin}(2006)}]{vance2006properties}%
  \BibitemOpen
  \bibfield  {author} {\bibinfo {author} {\bibfnamefont {M.~W.}\ \bibnamefont {Vance}}, \bibinfo {author} {\bibfnamefont {K.~D.}\ \bibnamefont {Squires}}, \ and\ \bibinfo {author} {\bibfnamefont {O.}~\bibnamefont {Simonin}},\ }\bibfield  {title} {\enquote {\bibinfo {title} {Properties of the particle velocity field in gas-solid turbulent channel flow},}\ }\href@noop {} {\bibfield  {journal} {\bibinfo  {journal} {Physics of Fluids}\ }\textbf {\bibinfo {volume} {18}} (\bibinfo {year} {2006})}\BibitemShut {NoStop}%
\bibitem [{\citenamefont {Zhao}, \citenamefont {Andersson},\ and\ \citenamefont {Gillissen}(2010)}]{zhao2010turbulence}%
  \BibitemOpen
  \bibfield  {author} {\bibinfo {author} {\bibfnamefont {L.}~\bibnamefont {Zhao}}, \bibinfo {author} {\bibfnamefont {H.~I.}\ \bibnamefont {Andersson}}, \ and\ \bibinfo {author} {\bibfnamefont {J.}~\bibnamefont {Gillissen}},\ }\bibfield  {title} {\enquote {\bibinfo {title} {Turbulence modulation and drag reduction by spherical particles},}\ }\href@noop {} {\bibfield  {journal} {\bibinfo  {journal} {Physics of Fluids}\ }\textbf {\bibinfo {volume} {22}} (\bibinfo {year} {2010})}\BibitemShut {NoStop}%
\bibitem [{\citenamefont {Lee}\ and\ \citenamefont {Lee}(2015)}]{lee2015modification}%
  \BibitemOpen
  \bibfield  {author} {\bibinfo {author} {\bibfnamefont {J.}~\bibnamefont {Lee}}\ and\ \bibinfo {author} {\bibfnamefont {C.}~\bibnamefont {Lee}},\ }\bibfield  {title} {\enquote {\bibinfo {title} {Modification of particle-laden near-wall turbulence: Effect of stokes number},}\ }\href@noop {} {\bibfield  {journal} {\bibinfo  {journal} {Physics of Fluids}\ }\textbf {\bibinfo {volume} {27}} (\bibinfo {year} {2015})}\BibitemShut {NoStop}%
\bibitem [{\citenamefont {Li}, \citenamefont {Luo},\ and\ \citenamefont {Fan}(2016)}]{li2016modulation}%
  \BibitemOpen
  \bibfield  {author} {\bibinfo {author} {\bibfnamefont {D.}~\bibnamefont {Li}}, \bibinfo {author} {\bibfnamefont {K.}~\bibnamefont {Luo}}, \ and\ \bibinfo {author} {\bibfnamefont {J.}~\bibnamefont {Fan}},\ }\bibfield  {title} {\enquote {\bibinfo {title} {Modulation of turbulence by dispersed solid particles in a spatially developing flat-plate boundary layer},}\ }\href@noop {} {\bibfield  {journal} {\bibinfo  {journal} {Journal of Fluid Mechanics}\ }\textbf {\bibinfo {volume} {802}},\ \bibinfo {pages} {359--394} (\bibinfo {year} {2016})}\BibitemShut {NoStop}%
\bibitem [{\citenamefont {Wang}\ and\ \citenamefont {Richter}(2019)}]{wang2019two}%
  \BibitemOpen
  \bibfield  {author} {\bibinfo {author} {\bibfnamefont {G.}~\bibnamefont {Wang}}\ and\ \bibinfo {author} {\bibfnamefont {D.}~\bibnamefont {Richter}},\ }\bibfield  {title} {\enquote {\bibinfo {title} {Two mechanisms of modulation of very-large-scale motions by inertial particles in open channel flow},}\ }\href@noop {} {\bibfield  {journal} {\bibinfo  {journal} {Journal of Fluid Mechanics}\ }\textbf {\bibinfo {volume} {868}},\ \bibinfo {pages} {538--559} (\bibinfo {year} {2019})}\BibitemShut {NoStop}%
\bibitem [{\citenamefont {Zheng}, \citenamefont {Feng},\ and\ \citenamefont {Wang}(2021)}]{zheng2021modulation}%
  \BibitemOpen
  \bibfield  {author} {\bibinfo {author} {\bibfnamefont {X.}~\bibnamefont {Zheng}}, \bibinfo {author} {\bibfnamefont {S.}~\bibnamefont {Feng}}, \ and\ \bibinfo {author} {\bibfnamefont {P.}~\bibnamefont {Wang}},\ }\bibfield  {title} {\enquote {\bibinfo {title} {Modulation of turbulence by saltating particles on erodible bed surface},}\ }\href@noop {} {\bibfield  {journal} {\bibinfo  {journal} {Journal of Fluid Mechanics}\ }\textbf {\bibinfo {volume} {918}},\ \bibinfo {pages} {A16} (\bibinfo {year} {2021})}\BibitemShut {NoStop}%
\bibitem [{\citenamefont {Pan}\ and\ \citenamefont {Banerjee}(1997)}]{pan1997numerical}%
  \BibitemOpen
  \bibfield  {author} {\bibinfo {author} {\bibfnamefont {Y.}~\bibnamefont {Pan}}\ and\ \bibinfo {author} {\bibfnamefont {S.}~\bibnamefont {Banerjee}},\ }\bibfield  {title} {\enquote {\bibinfo {title} {Numerical investigation of the effects of large particles on wall-turbulence},}\ }\href@noop {} {\bibfield  {journal} {\bibinfo  {journal} {Physics of Fluids}\ }\textbf {\bibinfo {volume} {9}},\ \bibinfo {pages} {3786--3807} (\bibinfo {year} {1997})}\BibitemShut {NoStop}%
\bibitem [{\citenamefont {Bagchi}\ and\ \citenamefont {Balachandar}(2003)}]{bagchi2003effect}%
  \BibitemOpen
  \bibfield  {author} {\bibinfo {author} {\bibfnamefont {P.}~\bibnamefont {Bagchi}}\ and\ \bibinfo {author} {\bibfnamefont {S.}~\bibnamefont {Balachandar}},\ }\bibfield  {title} {\enquote {\bibinfo {title} {Effect of turbulence on the drag and lift of a particle},}\ }\href@noop {} {\bibfield  {journal} {\bibinfo  {journal} {Physics of fluids}\ }\textbf {\bibinfo {volume} {15}},\ \bibinfo {pages} {3496--3513} (\bibinfo {year} {2003})}\BibitemShut {NoStop}%
\bibitem [{\citenamefont {Burton}\ and\ \citenamefont {Eaton}(2005)}]{burton2005fully}%
  \BibitemOpen
  \bibfield  {author} {\bibinfo {author} {\bibfnamefont {T.~M.}\ \bibnamefont {Burton}}\ and\ \bibinfo {author} {\bibfnamefont {J.~K.}\ \bibnamefont {Eaton}},\ }\bibfield  {title} {\enquote {\bibinfo {title} {Fully resolved simulations of particle-turbulence interaction},}\ }\href@noop {} {\bibfield  {journal} {\bibinfo  {journal} {Journal of Fluid Mechanics}\ }\textbf {\bibinfo {volume} {545}},\ \bibinfo {pages} {67--111} (\bibinfo {year} {2005})}\BibitemShut {NoStop}%
\bibitem [{\citenamefont {Shao}, \citenamefont {Wu},\ and\ \citenamefont {Yu}(2012)}]{shao2012fully}%
  \BibitemOpen
  \bibfield  {author} {\bibinfo {author} {\bibfnamefont {X.}~\bibnamefont {Shao}}, \bibinfo {author} {\bibfnamefont {T.}~\bibnamefont {Wu}}, \ and\ \bibinfo {author} {\bibfnamefont {Z.}~\bibnamefont {Yu}},\ }\bibfield  {title} {\enquote {\bibinfo {title} {Fully resolved numerical simulation of particle-laden turbulent flow in a horizontal channel at a low reynolds number},}\ }\href@noop {} {\bibfield  {journal} {\bibinfo  {journal} {Journal of Fluid Mechanics}\ }\textbf {\bibinfo {volume} {693}},\ \bibinfo {pages} {319--344} (\bibinfo {year} {2012})}\BibitemShut {NoStop}%
\bibitem [{\citenamefont {Picano}, \citenamefont {Breugem},\ and\ \citenamefont {Brandt}(2015)}]{picano2015turbulent}%
  \BibitemOpen
  \bibfield  {author} {\bibinfo {author} {\bibfnamefont {F.}~\bibnamefont {Picano}}, \bibinfo {author} {\bibfnamefont {W.-P.}\ \bibnamefont {Breugem}}, \ and\ \bibinfo {author} {\bibfnamefont {L.}~\bibnamefont {Brandt}},\ }\bibfield  {title} {\enquote {\bibinfo {title} {Turbulent channel flow of dense suspensions of neutrally buoyant spheres},}\ }\href@noop {} {\bibfield  {journal} {\bibinfo  {journal} {Journal of Fluid Mechanics}\ }\textbf {\bibinfo {volume} {764}},\ \bibinfo {pages} {463--487} (\bibinfo {year} {2015})}\BibitemShut {NoStop}%
\bibitem [{\citenamefont {Wang}\ \emph {et~al.}(2016)\citenamefont {Wang}, \citenamefont {Peng}, \citenamefont {Guo},\ and\ \citenamefont {Yu}}]{wang2016flow}%
  \BibitemOpen
  \bibfield  {author} {\bibinfo {author} {\bibfnamefont {L.-P.}\ \bibnamefont {Wang}}, \bibinfo {author} {\bibfnamefont {C.}~\bibnamefont {Peng}}, \bibinfo {author} {\bibfnamefont {Z.}~\bibnamefont {Guo}}, \ and\ \bibinfo {author} {\bibfnamefont {Z.}~\bibnamefont {Yu}},\ }\bibfield  {title} {\enquote {\bibinfo {title} {Flow modulation by finite-size neutrally buoyant particles in a turbulent channel flow},}\ }\href@noop {} {\bibfield  {journal} {\bibinfo  {journal} {Journal of Fluids Engineering}\ }\textbf {\bibinfo {volume} {138}},\ \bibinfo {pages} {041306} (\bibinfo {year} {2016})}\BibitemShut {NoStop}%
\bibitem [{\citenamefont {Wang}\ \emph {et~al.}(2022)\citenamefont {Wang}, \citenamefont {Zhu}, \citenamefont {Hu},\ and\ \citenamefont {Shen}}]{wang2022direct}%
  \BibitemOpen
  \bibfield  {author} {\bibinfo {author} {\bibfnamefont {Y.}~\bibnamefont {Wang}}, \bibinfo {author} {\bibfnamefont {Z.}~\bibnamefont {Zhu}}, \bibinfo {author} {\bibfnamefont {R.}~\bibnamefont {Hu}}, \ and\ \bibinfo {author} {\bibfnamefont {L.}~\bibnamefont {Shen}},\ }\bibfield  {title} {\enquote {\bibinfo {title} {Direct numerical simulation of a stationary spherical particle in fluctuating inflows},}\ }\href@noop {} {\bibfield  {journal} {\bibinfo  {journal} {AIP Advances}\ }\textbf {\bibinfo {volume} {12}} (\bibinfo {year} {2022})}\BibitemShut {NoStop}%
\bibitem [{\citenamefont {Wang}\ \emph {et~al.}(2023)\citenamefont {Wang}, \citenamefont {Lei}, \citenamefont {Zhu},\ and\ \citenamefont {Zheng}}]{wang2023drag}%
  \BibitemOpen
  \bibfield  {author} {\bibinfo {author} {\bibfnamefont {P.}~\bibnamefont {Wang}}, \bibinfo {author} {\bibfnamefont {Y.}~\bibnamefont {Lei}}, \bibinfo {author} {\bibfnamefont {Z.}~\bibnamefont {Zhu}}, \ and\ \bibinfo {author} {\bibfnamefont {X.}~\bibnamefont {Zheng}},\ }\bibfield  {title} {\enquote {\bibinfo {title} {{Drag model of finite-sized particle in turbulent wall-bound flow over sediment bed}},}\ }\href@noop {} {\bibfield  {journal} {\bibinfo  {journal} {Journal of Fluid Mechanics}\ }\textbf {\bibinfo {volume} {964}},\ \bibinfo {pages} {A9} (\bibinfo {year} {2023})}\BibitemShut {NoStop}%
\bibitem [{\citenamefont {Mittal}\ and\ \citenamefont {Iaccarino}(2005)}]{mittal2005immersed}%
  \BibitemOpen
  \bibfield  {author} {\bibinfo {author} {\bibfnamefont {R.}~\bibnamefont {Mittal}}\ and\ \bibinfo {author} {\bibfnamefont {G.}~\bibnamefont {Iaccarino}},\ }\bibfield  {title} {\enquote {\bibinfo {title} {Immersed boundary methods},}\ }\href {\doibase https://doi.org/10.1146/annurev.fluid.37.061903.175743} {\bibfield  {journal} {\bibinfo  {journal} {Annual Review of Fluid Mechanics}\ }\textbf {\bibinfo {volume} {37}},\ \bibinfo {pages} {239--261} (\bibinfo {year} {2005})}\BibitemShut {NoStop}%
\bibitem [{\citenamefont {Sotiropoulos}\ and\ \citenamefont {Yang}(2014)}]{sotiropoulos2014immersed}%
  \BibitemOpen
  \bibfield  {author} {\bibinfo {author} {\bibfnamefont {F.}~\bibnamefont {Sotiropoulos}}\ and\ \bibinfo {author} {\bibfnamefont {X.}~\bibnamefont {Yang}},\ }\bibfield  {title} {\enquote {\bibinfo {title} {Immersed boundary methods for simulating fluid--structure interaction},}\ }\href@noop {} {\bibfield  {journal} {\bibinfo  {journal} {Progress in Aerospace Sciences}\ }\textbf {\bibinfo {volume} {65}},\ \bibinfo {pages} {1--21} (\bibinfo {year} {2014})}\BibitemShut {NoStop}%
\bibitem [{\citenamefont {Griffith}\ and\ \citenamefont {Patankar}(2020)}]{griffith2020immersed}%
  \BibitemOpen
  \bibfield  {author} {\bibinfo {author} {\bibfnamefont {B.~E.}\ \bibnamefont {Griffith}}\ and\ \bibinfo {author} {\bibfnamefont {N.~A.}\ \bibnamefont {Patankar}},\ }\bibfield  {title} {\enquote {\bibinfo {title} {Immersed methods for fluid--structure interaction},}\ }\href@noop {} {\bibfield  {journal} {\bibinfo  {journal} {Annual review of fluid mechanics}\ }\textbf {\bibinfo {volume} {52}},\ \bibinfo {pages} {421--448} (\bibinfo {year} {2020})}\BibitemShut {NoStop}%
\bibitem [{\citenamefont {Verzicco}(2023)}]{verzicco2023immersed}%
  \BibitemOpen
  \bibfield  {author} {\bibinfo {author} {\bibfnamefont {R.}~\bibnamefont {Verzicco}},\ }\bibfield  {title} {\enquote {\bibinfo {title} {Immersed boundary methods: Historical perspective and future outlook},}\ }\href@noop {} {\bibfield  {journal} {\bibinfo  {journal} {Annual Review of Fluid Mechanics}\ }\textbf {\bibinfo {volume} {55}},\ \bibinfo {pages} {129--155} (\bibinfo {year} {2023})}\BibitemShut {NoStop}%
\bibitem [{\citenamefont {Goldstein}, \citenamefont {Handler},\ and\ \citenamefont {Sirovich}(1993)}]{goldstein1993modeling}%
  \BibitemOpen
  \bibfield  {author} {\bibinfo {author} {\bibfnamefont {D.}~\bibnamefont {Goldstein}}, \bibinfo {author} {\bibfnamefont {R.}~\bibnamefont {Handler}}, \ and\ \bibinfo {author} {\bibfnamefont {L.}~\bibnamefont {Sirovich}},\ }\bibfield  {title} {\enquote {\bibinfo {title} {Modeling a no-slip flow boundary with an external force field},}\ }\href@noop {} {\bibfield  {journal} {\bibinfo  {journal} {Journal of computational physics}\ }\textbf {\bibinfo {volume} {105}},\ \bibinfo {pages} {354--366} (\bibinfo {year} {1993})}\BibitemShut {NoStop}%
\bibitem [{\citenamefont {Saiki}\ and\ \citenamefont {Biringen}(1996)}]{saiki1996numerical}%
  \BibitemOpen
  \bibfield  {author} {\bibinfo {author} {\bibfnamefont {E.~M.}\ \bibnamefont {Saiki}}\ and\ \bibinfo {author} {\bibfnamefont {S.}~\bibnamefont {Biringen}},\ }\bibfield  {title} {\enquote {\bibinfo {title} {Numerical simulation of a cylinder in uniform flow: application of a virtual boundary method},}\ }\href@noop {} {\bibfield  {journal} {\bibinfo  {journal} {Journal of computational physics}\ }\textbf {\bibinfo {volume} {123}},\ \bibinfo {pages} {450--465} (\bibinfo {year} {1996})}\BibitemShut {NoStop}%
\bibitem [{\citenamefont {Angot}, \citenamefont {Bruneau},\ and\ \citenamefont {Fabrie}(1999)}]{angot1999penalization}%
  \BibitemOpen
  \bibfield  {author} {\bibinfo {author} {\bibfnamefont {P.}~\bibnamefont {Angot}}, \bibinfo {author} {\bibfnamefont {C.-H.}\ \bibnamefont {Bruneau}}, \ and\ \bibinfo {author} {\bibfnamefont {P.}~\bibnamefont {Fabrie}},\ }\bibfield  {title} {\enquote {\bibinfo {title} {A penalization method to take into account obstacles in incompressible viscous flows},}\ }\href@noop {} {\bibfield  {journal} {\bibinfo  {journal} {Numerische Mathematik}\ }\textbf {\bibinfo {volume} {81}},\ \bibinfo {pages} {497--520} (\bibinfo {year} {1999})}\BibitemShut {NoStop}%
\bibitem [{\citenamefont {Specklin}\ and\ \citenamefont {Delaur{\'e}}(2018)}]{specklin2018sharp}%
  \BibitemOpen
  \bibfield  {author} {\bibinfo {author} {\bibfnamefont {M.}~\bibnamefont {Specklin}}\ and\ \bibinfo {author} {\bibfnamefont {Y.}~\bibnamefont {Delaur{\'e}}},\ }\bibfield  {title} {\enquote {\bibinfo {title} {A sharp immersed boundary method based on penalization and its application to moving boundaries and turbulent rotating flows},}\ }\href@noop {} {\bibfield  {journal} {\bibinfo  {journal} {European Journal of Mechanics-B/Fluids}\ }\textbf {\bibinfo {volume} {70}},\ \bibinfo {pages} {130--147} (\bibinfo {year} {2018})}\BibitemShut {NoStop}%
\bibitem [{\citenamefont {Lai}\ and\ \citenamefont {Peskin}(2000)}]{lai2000immersed}%
  \BibitemOpen
  \bibfield  {author} {\bibinfo {author} {\bibfnamefont {M.-C.}\ \bibnamefont {Lai}}\ and\ \bibinfo {author} {\bibfnamefont {C.~S.}\ \bibnamefont {Peskin}},\ }\bibfield  {title} {\enquote {\bibinfo {title} {An immersed boundary method with formal second-order accuracy and reduced numerical viscosity},}\ }\href@noop {} {\bibfield  {journal} {\bibinfo  {journal} {Journal of computational Physics}\ }\textbf {\bibinfo {volume} {160}},\ \bibinfo {pages} {705--719} (\bibinfo {year} {2000})}\BibitemShut {NoStop}%
\bibitem [{\citenamefont {Lee}(2003)}]{lee2003stability}%
  \BibitemOpen
  \bibfield  {author} {\bibinfo {author} {\bibfnamefont {C.}~\bibnamefont {Lee}},\ }\bibfield  {title} {\enquote {\bibinfo {title} {Stability characteristics of the virtual boundary method in three-dimensional applications},}\ }\href@noop {} {\bibfield  {journal} {\bibinfo  {journal} {Journal of Computational Physics}\ }\textbf {\bibinfo {volume} {184}},\ \bibinfo {pages} {559--591} (\bibinfo {year} {2003})}\BibitemShut {NoStop}%
\bibitem [{\citenamefont {Fadlun}\ \emph {et~al.}(2000)\citenamefont {Fadlun}, \citenamefont {Verzicco}, \citenamefont {Orlandi},\ and\ \citenamefont {Mohd-Yusof}}]{fadlun2000combined}%
  \BibitemOpen
  \bibfield  {author} {\bibinfo {author} {\bibfnamefont {E.~A.}\ \bibnamefont {Fadlun}}, \bibinfo {author} {\bibfnamefont {R.}~\bibnamefont {Verzicco}}, \bibinfo {author} {\bibfnamefont {P.}~\bibnamefont {Orlandi}}, \ and\ \bibinfo {author} {\bibfnamefont {J.}~\bibnamefont {Mohd-Yusof}},\ }\bibfield  {title} {\enquote {\bibinfo {title} {Combined immersed-boundary finite-difference methods for three-dimensional complex flow simulations},}\ }\href@noop {} {\bibfield  {journal} {\bibinfo  {journal} {Journal of computational physics}\ }\textbf {\bibinfo {volume} {161}},\ \bibinfo {pages} {35--60} (\bibinfo {year} {2000})}\BibitemShut {NoStop}%
\bibitem [{\citenamefont {Griffith}\ and\ \citenamefont {Peskin}(2005)}]{griffith2005order}%
  \BibitemOpen
  \bibfield  {author} {\bibinfo {author} {\bibfnamefont {B.~E.}\ \bibnamefont {Griffith}}\ and\ \bibinfo {author} {\bibfnamefont {C.~S.}\ \bibnamefont {Peskin}},\ }\bibfield  {title} {\enquote {\bibinfo {title} {On the order of accuracy of the immersed boundary method: Higher order convergence rates for sufficiently smooth problems},}\ }\href@noop {} {\bibfield  {journal} {\bibinfo  {journal} {Journal of Computational Physics}\ }\textbf {\bibinfo {volume} {208}},\ \bibinfo {pages} {75--105} (\bibinfo {year} {2005})}\BibitemShut {NoStop}%
\bibitem [{\citenamefont {Uhlmann}(2005)}]{uhlmann2005immersed}%
  \BibitemOpen
  \bibfield  {author} {\bibinfo {author} {\bibfnamefont {M.}~\bibnamefont {Uhlmann}},\ }\bibfield  {title} {\enquote {\bibinfo {title} {An immersed boundary method with direct forcing for the simulation of particulate flows},}\ }\href@noop {} {\bibfield  {journal} {\bibinfo  {journal} {J. Comput. Phys.}\ }\textbf {\bibinfo {volume} {209}},\ \bibinfo {pages} {448--476} (\bibinfo {year} {2005})}\BibitemShut {NoStop}%
\bibitem [{\citenamefont {Luo}\ \emph {et~al.}(2007)\citenamefont {Luo}, \citenamefont {Wang}, \citenamefont {Fan},\ and\ \citenamefont {Cen}}]{luo2007full}%
  \BibitemOpen
  \bibfield  {author} {\bibinfo {author} {\bibfnamefont {K.}~\bibnamefont {Luo}}, \bibinfo {author} {\bibfnamefont {Z.}~\bibnamefont {Wang}}, \bibinfo {author} {\bibfnamefont {J.}~\bibnamefont {Fan}}, \ and\ \bibinfo {author} {\bibfnamefont {K.}~\bibnamefont {Cen}},\ }\bibfield  {title} {\enquote {\bibinfo {title} {Full-scale solutions to particle-laden flows: Multidirect forcing and immersed boundary method},}\ }\href@noop {} {\bibfield  {journal} {\bibinfo  {journal} {Physical Review E—Statistical, Nonlinear, and Soft Matter Physics}\ }\textbf {\bibinfo {volume} {76}},\ \bibinfo {pages} {066709} (\bibinfo {year} {2007})}\BibitemShut {NoStop}%
\bibitem [{\citenamefont {Kempe}\ and\ \citenamefont {Fr{\"o}hlich}(2012)}]{kempe2012improved}%
  \BibitemOpen
  \bibfield  {author} {\bibinfo {author} {\bibfnamefont {T.}~\bibnamefont {Kempe}}\ and\ \bibinfo {author} {\bibfnamefont {J.}~\bibnamefont {Fr{\"o}hlich}},\ }\bibfield  {title} {\enquote {\bibinfo {title} {{An improved immersed boundary method with direct forcing for the simulation of particle laden flows}},}\ }\href@noop {} {\bibfield  {journal} {\bibinfo  {journal} {Journal of Computational Physics}\ }\textbf {\bibinfo {volume} {231}},\ \bibinfo {pages} {3663--3684} (\bibinfo {year} {2012})}\BibitemShut {NoStop}%
\bibitem [{\citenamefont {Breugem}(2012)}]{breugem2012second}%
  \BibitemOpen
  \bibfield  {author} {\bibinfo {author} {\bibfnamefont {W.-P.}\ \bibnamefont {Breugem}},\ }\bibfield  {title} {\enquote {\bibinfo {title} {{A second-order accurate immersed boundary method for fully resolved simulations of particle-laden flows}},}\ }\href@noop {} {\bibfield  {journal} {\bibinfo  {journal} {Journal of Computational Physics}\ }\textbf {\bibinfo {volume} {231}},\ \bibinfo {pages} {4469--4498} (\bibinfo {year} {2012})}\BibitemShut {NoStop}%
\bibitem [{\citenamefont {Martin}\ and\ \citenamefont {Colella}(2000)}]{martin2000cell}%
  \BibitemOpen
  \bibfield  {author} {\bibinfo {author} {\bibfnamefont {D.~F.}\ \bibnamefont {Martin}}\ and\ \bibinfo {author} {\bibfnamefont {P.}~\bibnamefont {Colella}},\ }\bibfield  {title} {\enquote {\bibinfo {title} {{A cell-centered adaptive projection method for the incompressible Euler equations}},}\ }\href@noop {} {\bibfield  {journal} {\bibinfo  {journal} {J. Comput. Phys.}\ }\textbf {\bibinfo {volume} {163}},\ \bibinfo {pages} {271--312} (\bibinfo {year} {2000})}\BibitemShut {NoStop}%
\bibitem [{\citenamefont {Martin}, \citenamefont {Colella},\ and\ \citenamefont {Graves}(2008)}]{martin2008cell}%
  \BibitemOpen
  \bibfield  {author} {\bibinfo {author} {\bibfnamefont {D.~F.}\ \bibnamefont {Martin}}, \bibinfo {author} {\bibfnamefont {P.}~\bibnamefont {Colella}}, \ and\ \bibinfo {author} {\bibfnamefont {D.}~\bibnamefont {Graves}},\ }\bibfield  {title} {\enquote {\bibinfo {title} {{A cell-centered adaptive projection method for the incompressible Navier--Stokes equations in three dimensions}},}\ }\href@noop {} {\bibfield  {journal} {\bibinfo  {journal} {J. Comput. Phys.}\ }\textbf {\bibinfo {volume} {227}},\ \bibinfo {pages} {1863--1886} (\bibinfo {year} {2008})}\BibitemShut {NoStop}%
\bibitem [{\citenamefont {Almgren}\ \emph {et~al.}(1998)\citenamefont {Almgren}, \citenamefont {Bell}, \citenamefont {Colella}, \citenamefont {Howell},\ and\ \citenamefont {Welcome}}]{almgren1998conservative}%
  \BibitemOpen
  \bibfield  {author} {\bibinfo {author} {\bibfnamefont {A.~S.}\ \bibnamefont {Almgren}}, \bibinfo {author} {\bibfnamefont {J.~B.}\ \bibnamefont {Bell}}, \bibinfo {author} {\bibfnamefont {P.}~\bibnamefont {Colella}}, \bibinfo {author} {\bibfnamefont {L.~H.}\ \bibnamefont {Howell}}, \ and\ \bibinfo {author} {\bibfnamefont {M.~L.}\ \bibnamefont {Welcome}},\ }\bibfield  {title} {\enquote {\bibinfo {title} {{A conservative adaptive projection method for the variable density incompressible Navier--Stokes equations}},}\ }\href@noop {} {\bibfield  {journal} {\bibinfo  {journal} {J. Comput. Phys.}\ }\textbf {\bibinfo {volume} {142}},\ \bibinfo {pages} {1--46} (\bibinfo {year} {1998})}\BibitemShut {NoStop}%
\bibitem [{\citenamefont {Ji}\ \emph {et~al.}(2013)\citenamefont {Ji}, \citenamefont {Munjiza}, \citenamefont {Avital}, \citenamefont {Ma},\ and\ \citenamefont {Williams}}]{ji2013direct}%
  \BibitemOpen
  \bibfield  {author} {\bibinfo {author} {\bibfnamefont {C.}~\bibnamefont {Ji}}, \bibinfo {author} {\bibfnamefont {A.}~\bibnamefont {Munjiza}}, \bibinfo {author} {\bibfnamefont {E.}~\bibnamefont {Avital}}, \bibinfo {author} {\bibfnamefont {J.}~\bibnamefont {Ma}}, \ and\ \bibinfo {author} {\bibfnamefont {J.}~\bibnamefont {Williams}},\ }\bibfield  {title} {\enquote {\bibinfo {title} {Direct numerical simulation of sediment entrainment in turbulent channel flow},}\ }\href@noop {} {\bibfield  {journal} {\bibinfo  {journal} {Physics of Fluids}\ }\textbf {\bibinfo {volume} {25}} (\bibinfo {year} {2013})}\BibitemShut {NoStop}%
\bibitem [{\citenamefont {Ji}\ \emph {et~al.}(2014)\citenamefont {Ji}, \citenamefont {Munjiza}, \citenamefont {Avital}, \citenamefont {Xu},\ and\ \citenamefont {Williams}}]{ji2014saltation}%
  \BibitemOpen
  \bibfield  {author} {\bibinfo {author} {\bibfnamefont {C.}~\bibnamefont {Ji}}, \bibinfo {author} {\bibfnamefont {A.}~\bibnamefont {Munjiza}}, \bibinfo {author} {\bibfnamefont {E.}~\bibnamefont {Avital}}, \bibinfo {author} {\bibfnamefont {D.}~\bibnamefont {Xu}}, \ and\ \bibinfo {author} {\bibfnamefont {J.}~\bibnamefont {Williams}},\ }\bibfield  {title} {\enquote {\bibinfo {title} {Saltation of particles in turbulent channel flow},}\ }\href@noop {} {\bibfield  {journal} {\bibinfo  {journal} {Physical Review E}\ }\textbf {\bibinfo {volume} {89}},\ \bibinfo {pages} {052202} (\bibinfo {year} {2014})}\BibitemShut {NoStop}%
\bibitem [{\citenamefont {Kidanemariam}\ and\ \citenamefont {Uhlmann}(2014)}]{kidanemariam2014direct}%
  \BibitemOpen
  \bibfield  {author} {\bibinfo {author} {\bibfnamefont {A.~G.}\ \bibnamefont {Kidanemariam}}\ and\ \bibinfo {author} {\bibfnamefont {M.}~\bibnamefont {Uhlmann}},\ }\bibfield  {title} {\enquote {\bibinfo {title} {Direct numerical simulation of pattern formation in subaqueous sediment},}\ }\href@noop {} {\bibfield  {journal} {\bibinfo  {journal} {Journal of Fluid Mechanics}\ }\textbf {\bibinfo {volume} {750}},\ \bibinfo {pages} {R2} (\bibinfo {year} {2014})}\BibitemShut {NoStop}%
\bibitem [{\citenamefont {Kidanemariam}\ and\ \citenamefont {Uhlmann}(2017)}]{kidanemariam2017formation}%
  \BibitemOpen
  \bibfield  {author} {\bibinfo {author} {\bibfnamefont {A.~G.}\ \bibnamefont {Kidanemariam}}\ and\ \bibinfo {author} {\bibfnamefont {M.}~\bibnamefont {Uhlmann}},\ }\bibfield  {title} {\enquote {\bibinfo {title} {Formation of sediment patterns in channel flow: minimal unstable systems and their temporal evolution},}\ }\href@noop {} {\bibfield  {journal} {\bibinfo  {journal} {Journal of Fluid Mechanics}\ }\textbf {\bibinfo {volume} {818}},\ \bibinfo {pages} {716--743} (\bibinfo {year} {2017})}\BibitemShut {NoStop}%
\bibitem [{\citenamefont {Kidanemariam}, \citenamefont {Scherer},\ and\ \citenamefont {Uhlmann}(2022)}]{kidanemariam2022open}%
  \BibitemOpen
  \bibfield  {author} {\bibinfo {author} {\bibfnamefont {A.~G.}\ \bibnamefont {Kidanemariam}}, \bibinfo {author} {\bibfnamefont {M.}~\bibnamefont {Scherer}}, \ and\ \bibinfo {author} {\bibfnamefont {M.}~\bibnamefont {Uhlmann}},\ }\bibfield  {title} {\enquote {\bibinfo {title} {Open-channel flow over evolving subaqueous ripples},}\ }\href@noop {} {\bibfield  {journal} {\bibinfo  {journal} {Journal of Fluid Mechanics}\ }\textbf {\bibinfo {volume} {937}},\ \bibinfo {pages} {A26} (\bibinfo {year} {2022})}\BibitemShut {NoStop}%
\bibitem [{\citenamefont {Scherer}\ \emph {et~al.}(2022)\citenamefont {Scherer}, \citenamefont {Uhlmann}, \citenamefont {Kidanemariam},\ and\ \citenamefont {Krayer}}]{scherer2022role}%
  \BibitemOpen
  \bibfield  {author} {\bibinfo {author} {\bibfnamefont {M.}~\bibnamefont {Scherer}}, \bibinfo {author} {\bibfnamefont {M.}~\bibnamefont {Uhlmann}}, \bibinfo {author} {\bibfnamefont {A.~G.}\ \bibnamefont {Kidanemariam}}, \ and\ \bibinfo {author} {\bibfnamefont {M.}~\bibnamefont {Krayer}},\ }\bibfield  {title} {\enquote {\bibinfo {title} {On the role of turbulent large-scale streaks in generating sediment ridges},}\ }\href@noop {} {\bibfield  {journal} {\bibinfo  {journal} {Journal of Fluid Mechanics}\ }\textbf {\bibinfo {volume} {930}},\ \bibinfo {pages} {A11} (\bibinfo {year} {2022})}\BibitemShut {NoStop}%
\bibitem [{\citenamefont {Vowinckel}\ \emph {et~al.}(2016)\citenamefont {Vowinckel}, \citenamefont {Jain}, \citenamefont {Kempe},\ and\ \citenamefont {Fr{\"o}hlich}}]{vowinckel2016entrainment}%
  \BibitemOpen
  \bibfield  {author} {\bibinfo {author} {\bibfnamefont {B.}~\bibnamefont {Vowinckel}}, \bibinfo {author} {\bibfnamefont {R.}~\bibnamefont {Jain}}, \bibinfo {author} {\bibfnamefont {T.}~\bibnamefont {Kempe}}, \ and\ \bibinfo {author} {\bibfnamefont {J.}~\bibnamefont {Fr{\"o}hlich}},\ }\bibfield  {title} {\enquote {\bibinfo {title} {Entrainment of single particles in a turbulent open-channel flow: A numerical study},}\ }\href@noop {} {\bibfield  {journal} {\bibinfo  {journal} {Journal of Hydraulic Research}\ }\textbf {\bibinfo {volume} {54}},\ \bibinfo {pages} {158--171} (\bibinfo {year} {2016})}\BibitemShut {NoStop}%
\bibitem [{\citenamefont {Zhu}\ \emph {et~al.}(2022)\citenamefont {Zhu}, \citenamefont {Hu}, \citenamefont {Lei}, \citenamefont {Shen},\ and\ \citenamefont {Zheng}}]{zhu2022particle}%
  \BibitemOpen
  \bibfield  {author} {\bibinfo {author} {\bibfnamefont {Z.}~\bibnamefont {Zhu}}, \bibinfo {author} {\bibfnamefont {R.}~\bibnamefont {Hu}}, \bibinfo {author} {\bibfnamefont {Y.}~\bibnamefont {Lei}}, \bibinfo {author} {\bibfnamefont {L.}~\bibnamefont {Shen}}, \ and\ \bibinfo {author} {\bibfnamefont {X.}~\bibnamefont {Zheng}},\ }\bibfield  {title} {\enquote {\bibinfo {title} {{Particle resolved simulation of sediment transport by a hybrid parallel approach}},}\ }\href@noop {} {\bibfield  {journal} {\bibinfo  {journal} {International Journal of Multiphase Flow}\ }\textbf {\bibinfo {volume} {152}},\ \bibinfo {pages} {104072} (\bibinfo {year} {2022})}\BibitemShut {NoStop}%
\bibitem [{\citenamefont {Jain}, \citenamefont {Tschisgale},\ and\ \citenamefont {Froehlich}(2020)}]{jain2020effect}%
  \BibitemOpen
  \bibfield  {author} {\bibinfo {author} {\bibfnamefont {R.}~\bibnamefont {Jain}}, \bibinfo {author} {\bibfnamefont {S.}~\bibnamefont {Tschisgale}}, \ and\ \bibinfo {author} {\bibfnamefont {J.}~\bibnamefont {Froehlich}},\ }\bibfield  {title} {\enquote {\bibinfo {title} {Effect of particle shape on bedload sediment transport in case of small particle loading},}\ }\href@noop {} {\bibfield  {journal} {\bibinfo  {journal} {Meccanica}\ }\textbf {\bibinfo {volume} {55}},\ \bibinfo {pages} {299--315} (\bibinfo {year} {2020})}\BibitemShut {NoStop}%
\bibitem [{\citenamefont {Roma}, \citenamefont {Peskin},\ and\ \citenamefont {Berger}(1999)}]{roma1999adaptive}%
  \BibitemOpen
  \bibfield  {author} {\bibinfo {author} {\bibfnamefont {A.~M.}\ \bibnamefont {Roma}}, \bibinfo {author} {\bibfnamefont {C.~S.}\ \bibnamefont {Peskin}}, \ and\ \bibinfo {author} {\bibfnamefont {M.~J.}\ \bibnamefont {Berger}},\ }\bibfield  {title} {\enquote {\bibinfo {title} {An adaptive version of the immersed boundary method},}\ }\href@noop {} {\bibfield  {journal} {\bibinfo  {journal} {Journal of computational physics}\ }\textbf {\bibinfo {volume} {153}},\ \bibinfo {pages} {509--534} (\bibinfo {year} {1999})}\BibitemShut {NoStop}%
\bibitem [{\citenamefont {Liu}, \citenamefont {Jun},\ and\ \citenamefont {Zhang}(1995)}]{liu1995reproducing}%
  \BibitemOpen
  \bibfield  {author} {\bibinfo {author} {\bibfnamefont {W.~K.}\ \bibnamefont {Liu}}, \bibinfo {author} {\bibfnamefont {S.}~\bibnamefont {Jun}}, \ and\ \bibinfo {author} {\bibfnamefont {Y.~F.}\ \bibnamefont {Zhang}},\ }\bibfield  {title} {\enquote {\bibinfo {title} {Reproducing kernel particle methods},}\ }\href@noop {} {\bibfield  {journal} {\bibinfo  {journal} {International journal for numerical methods in fluids}\ }\textbf {\bibinfo {volume} {20}},\ \bibinfo {pages} {1081--1106} (\bibinfo {year} {1995})}\BibitemShut {NoStop}%
\bibitem [{\citenamefont {Pinelli}\ \emph {et~al.}(2010)\citenamefont {Pinelli}, \citenamefont {Naqavi}, \citenamefont {Piomelli},\ and\ \citenamefont {Favier}}]{pinelli2010immersed}%
  \BibitemOpen
  \bibfield  {author} {\bibinfo {author} {\bibfnamefont {A.}~\bibnamefont {Pinelli}}, \bibinfo {author} {\bibfnamefont {I.}~\bibnamefont {Naqavi}}, \bibinfo {author} {\bibfnamefont {U.}~\bibnamefont {Piomelli}}, \ and\ \bibinfo {author} {\bibfnamefont {J.}~\bibnamefont {Favier}},\ }\bibfield  {title} {\enquote {\bibinfo {title} {{Immersed-boundary methods for general finite-difference and finite-volume Navier--Stokes solvers}},}\ }\href@noop {} {\bibfield  {journal} {\bibinfo  {journal} {Journal of Computational Physics}\ }\textbf {\bibinfo {volume} {229}},\ \bibinfo {pages} {9073--9091} (\bibinfo {year} {2010})}\BibitemShut {NoStop}%
\bibitem [{\citenamefont {Akiki}\ and\ \citenamefont {Balachandar}(2016)}]{akiki2016immersed}%
  \BibitemOpen
  \bibfield  {author} {\bibinfo {author} {\bibfnamefont {G.}~\bibnamefont {Akiki}}\ and\ \bibinfo {author} {\bibfnamefont {S.}~\bibnamefont {Balachandar}},\ }\bibfield  {title} {\enquote {\bibinfo {title} {{Immersed boundary method with non-uniform distribution of Lagrangian markers for a non-uniform Eulerian mesh}},}\ }\href@noop {} {\bibfield  {journal} {\bibinfo  {journal} {Journal of Computational Physics}\ }\textbf {\bibinfo {volume} {307}},\ \bibinfo {pages} {34--59} (\bibinfo {year} {2016})}\BibitemShut {NoStop}%
\bibitem [{\citenamefont {Jang}\ and\ \citenamefont {Lee}(2017)}]{jang2017immersed}%
  \BibitemOpen
  \bibfield  {author} {\bibinfo {author} {\bibfnamefont {J.}~\bibnamefont {Jang}}\ and\ \bibinfo {author} {\bibfnamefont {C.}~\bibnamefont {Lee}},\ }\bibfield  {title} {\enquote {\bibinfo {title} {An immersed boundary method for nonuniform grids},}\ }\href@noop {} {\bibfield  {journal} {\bibinfo  {journal} {Journal of Computational Physics}\ }\textbf {\bibinfo {volume} {341}},\ \bibinfo {pages} {1--12} (\bibinfo {year} {2017})}\BibitemShut {NoStop}%
\bibitem [{\citenamefont {Berger}\ and\ \citenamefont {Oliger}(1984)}]{berger1984adaptive}%
  \BibitemOpen
  \bibfield  {author} {\bibinfo {author} {\bibfnamefont {M.~J.}\ \bibnamefont {Berger}}\ and\ \bibinfo {author} {\bibfnamefont {J.}~\bibnamefont {Oliger}},\ }\bibfield  {title} {\enquote {\bibinfo {title} {{Adaptive mesh refinement for hyperbolic partial differential equations}},}\ }\href@noop {} {\bibfield  {journal} {\bibinfo  {journal} {J. Comput. Phys.}\ }\textbf {\bibinfo {volume} {53}},\ \bibinfo {pages} {484--512} (\bibinfo {year} {1984})}\BibitemShut {NoStop}%
\bibitem [{\citenamefont {Berger}\ and\ \citenamefont {Colella}(1989)}]{berger1989local}%
  \BibitemOpen
  \bibfield  {author} {\bibinfo {author} {\bibfnamefont {M.~J.}\ \bibnamefont {Berger}}\ and\ \bibinfo {author} {\bibfnamefont {P.}~\bibnamefont {Colella}},\ }\bibfield  {title} {\enquote {\bibinfo {title} {{Local adaptive mesh refinement for shock hydrodynamics}},}\ }\href@noop {} {\bibfield  {journal} {\bibinfo  {journal} {J. Comput. Phys.}\ }\textbf {\bibinfo {volume} {82}},\ \bibinfo {pages} {64--84} (\bibinfo {year} {1989})}\BibitemShut {NoStop}%
\bibitem [{\citenamefont {Guittet}, \citenamefont {Theillard},\ and\ \citenamefont {Gibou}(2015)}]{guittet2015stable}%
  \BibitemOpen
  \bibfield  {author} {\bibinfo {author} {\bibfnamefont {A.}~\bibnamefont {Guittet}}, \bibinfo {author} {\bibfnamefont {M.}~\bibnamefont {Theillard}}, \ and\ \bibinfo {author} {\bibfnamefont {F.}~\bibnamefont {Gibou}},\ }\bibfield  {title} {\enquote {\bibinfo {title} {{A stable projection method for the incompressible Navier--Stokes equations on arbitrary geometries and adaptive Quad/Octrees}},}\ }\href@noop {} {\bibfield  {journal} {\bibinfo  {journal} {J. Comput. Phys.}\ }\textbf {\bibinfo {volume} {292}},\ \bibinfo {pages} {215--238} (\bibinfo {year} {2015})}\BibitemShut {NoStop}%
\bibitem [{\citenamefont {Mirzadeh}\ \emph {et~al.}(2016)\citenamefont {Mirzadeh}, \citenamefont {Guittet}, \citenamefont {Burstedde},\ and\ \citenamefont {Gibou}}]{mirzadeh2016parallel}%
  \BibitemOpen
  \bibfield  {author} {\bibinfo {author} {\bibfnamefont {M.}~\bibnamefont {Mirzadeh}}, \bibinfo {author} {\bibfnamefont {A.}~\bibnamefont {Guittet}}, \bibinfo {author} {\bibfnamefont {C.}~\bibnamefont {Burstedde}}, \ and\ \bibinfo {author} {\bibfnamefont {F.}~\bibnamefont {Gibou}},\ }\bibfield  {title} {\enquote {\bibinfo {title} {{Parallel level-set methods on adaptive tree-based grids}},}\ }\href@noop {} {\bibfield  {journal} {\bibinfo  {journal} {J. Comput. Phys.}\ }\textbf {\bibinfo {volume} {322}},\ \bibinfo {pages} {345--364} (\bibinfo {year} {2016})}\BibitemShut {NoStop}%
\bibitem [{\citenamefont {Popinet}(2003)}]{popinet2003gerris}%
  \BibitemOpen
  \bibfield  {author} {\bibinfo {author} {\bibfnamefont {S.}~\bibnamefont {Popinet}},\ }\bibfield  {title} {\enquote {\bibinfo {title} {{Gerris: a tree-based adaptive solver for the incompressible Euler equations in complex geometries}},}\ }\href@noop {} {\bibfield  {journal} {\bibinfo  {journal} {J. Comput. Phys.}\ }\textbf {\bibinfo {volume} {190}},\ \bibinfo {pages} {572--600} (\bibinfo {year} {2003})}\BibitemShut {NoStop}%
\bibitem [{\citenamefont {Burstedde}, \citenamefont {Wilcox},\ and\ \citenamefont {Ghattas}(2011)}]{BursteddeWilcoxGhattas11}%
  \BibitemOpen
  \bibfield  {author} {\bibinfo {author} {\bibfnamefont {C.}~\bibnamefont {Burstedde}}, \bibinfo {author} {\bibfnamefont {L.~C.}\ \bibnamefont {Wilcox}}, \ and\ \bibinfo {author} {\bibfnamefont {O.}~\bibnamefont {Ghattas}},\ }\bibfield  {title} {\enquote {\bibinfo {title} {{\texttt{p4est}}: scalable algorithms for parallel adaptive mesh refinement on forests of octrees},}\ }\href {\doibase 10.1137/100791634} {\bibfield  {journal} {\bibinfo  {journal} {SIAM J. Sci. Comput.}\ }\textbf {\bibinfo {volume} {33}},\ \bibinfo {pages} {1103--1133} (\bibinfo {year} {2011})}\BibitemShut {NoStop}%
\bibitem [{\citenamefont {Kirk}\ \emph {et~al.}(2006)\citenamefont {Kirk}, \citenamefont {Peterson}, \citenamefont {Stogner},\ and\ \citenamefont {Carey}}]{libMeshPaper}%
  \BibitemOpen
  \bibfield  {author} {\bibinfo {author} {\bibfnamefont {B.~S.}\ \bibnamefont {Kirk}}, \bibinfo {author} {\bibfnamefont {J.~W.}\ \bibnamefont {Peterson}}, \bibinfo {author} {\bibfnamefont {R.~H.}\ \bibnamefont {Stogner}}, \ and\ \bibinfo {author} {\bibfnamefont {G.~F.}\ \bibnamefont {Carey}},\ }\bibfield  {title} {\enquote {\bibinfo {title} {{\texttt{libMesh}: A C++ Library for Parallel Adaptive Mesh Refinement/Coarsening Simulations}},}\ }\href@noop {} {\bibfield  {journal} {\bibinfo  {journal} {Engineering with Computers}\ }\textbf {\bibinfo {volume} {22}},\ \bibinfo {pages} {237--254} (\bibinfo {year} {2006})},\ \bibinfo {note} {\url{https://doi.org/10.1007/s00366-006-0049-3}}\BibitemShut {NoStop}%
\bibitem [{\citenamefont {Fryxell}\ \emph {et~al.}(2000)\citenamefont {Fryxell}, \citenamefont {Olson}, \citenamefont {Ricker}, \citenamefont {Timmes}, \citenamefont {Zingale}, \citenamefont {Lamb}, \citenamefont {MacNeice}, \citenamefont {Rosner}, \citenamefont {Truran},\ and\ \citenamefont {Tufo}}]{fryxell2000flash}%
  \BibitemOpen
  \bibfield  {author} {\bibinfo {author} {\bibfnamefont {B.}~\bibnamefont {Fryxell}}, \bibinfo {author} {\bibfnamefont {K.}~\bibnamefont {Olson}}, \bibinfo {author} {\bibfnamefont {P.}~\bibnamefont {Ricker}}, \bibinfo {author} {\bibfnamefont {F.~X.}\ \bibnamefont {Timmes}}, \bibinfo {author} {\bibfnamefont {M.}~\bibnamefont {Zingale}}, \bibinfo {author} {\bibfnamefont {D.}~\bibnamefont {Lamb}}, \bibinfo {author} {\bibfnamefont {P.}~\bibnamefont {MacNeice}}, \bibinfo {author} {\bibfnamefont {R.}~\bibnamefont {Rosner}}, \bibinfo {author} {\bibfnamefont {J.}~\bibnamefont {Truran}}, \ and\ \bibinfo {author} {\bibfnamefont {H.}~\bibnamefont {Tufo}},\ }\bibfield  {title} {\enquote {\bibinfo {title} {{FLASH: An adaptive mesh hydrodynamics code for modeling astrophysical thermonuclear flashes}},}\ }\href@noop {} {\bibfield  {journal} {\bibinfo  {journal} {The Astrophysical Journal Supplement Series}\ }\textbf {\bibinfo {volume} {131}},\ \bibinfo {pages} {273} (\bibinfo {year} {2000})}\BibitemShut {NoStop}%
\bibitem [{\citenamefont {Gunney}\ and\ \citenamefont {Anderson}(2016)}]{gunney2016advances}%
  \BibitemOpen
  \bibfield  {author} {\bibinfo {author} {\bibfnamefont {B.~T.}\ \bibnamefont {Gunney}}\ and\ \bibinfo {author} {\bibfnamefont {R.~W.}\ \bibnamefont {Anderson}},\ }\bibfield  {title} {\enquote {\bibinfo {title} {{Advances in patch-based adaptive mesh refinement scalability}},}\ }\href@noop {} {\bibfield  {journal} {\bibinfo  {journal} {J. Parallel Distrib. Comput.}\ }\textbf {\bibinfo {volume} {89}},\ \bibinfo {pages} {65--84} (\bibinfo {year} {2016})}\BibitemShut {NoStop}%
\bibitem [{\citenamefont {Zhang}\ \emph {et~al.}(2019)\citenamefont {Zhang}, \citenamefont {Almgren}, \citenamefont {Beckner}, \citenamefont {Bell}, \citenamefont {Blaschke}, \citenamefont {Chan}, \citenamefont {Day}, \citenamefont {Friesen}, \citenamefont {Gott}, \citenamefont {Graves} \emph {et~al.}}]{zhang2019amrex}%
  \BibitemOpen
  \bibfield  {author} {\bibinfo {author} {\bibfnamefont {W.}~\bibnamefont {Zhang}}, \bibinfo {author} {\bibfnamefont {A.}~\bibnamefont {Almgren}}, \bibinfo {author} {\bibfnamefont {V.}~\bibnamefont {Beckner}}, \bibinfo {author} {\bibfnamefont {J.}~\bibnamefont {Bell}}, \bibinfo {author} {\bibfnamefont {J.}~\bibnamefont {Blaschke}}, \bibinfo {author} {\bibfnamefont {C.}~\bibnamefont {Chan}}, \bibinfo {author} {\bibfnamefont {M.}~\bibnamefont {Day}}, \bibinfo {author} {\bibfnamefont {B.}~\bibnamefont {Friesen}}, \bibinfo {author} {\bibfnamefont {K.}~\bibnamefont {Gott}}, \bibinfo {author} {\bibfnamefont {D.}~\bibnamefont {Graves}},  \emph {et~al.},\ }\bibfield  {title} {\enquote {\bibinfo {title} {{AMReX: a framework for block-structured adaptive mesh refinement}},}\ }\href@noop {} {\bibfield  {journal} {\bibinfo  {journal} {J. Open Source Softw.}\ }\textbf {\bibinfo {volume} {4}} (\bibinfo {year} {2019})}\BibitemShut {NoStop}%
\bibitem [{\citenamefont {Zhang}\ \emph {et~al.}(2020)\citenamefont {Zhang}, \citenamefont {Myers}, \citenamefont {Gott}, \citenamefont {Almgren},\ and\ \citenamefont {Bell}}]{zhang2020amrex}%
  \BibitemOpen
  \bibfield  {author} {\bibinfo {author} {\bibfnamefont {W.}~\bibnamefont {Zhang}}, \bibinfo {author} {\bibfnamefont {A.}~\bibnamefont {Myers}}, \bibinfo {author} {\bibfnamefont {K.}~\bibnamefont {Gott}}, \bibinfo {author} {\bibfnamefont {A.}~\bibnamefont {Almgren}}, \ and\ \bibinfo {author} {\bibfnamefont {J.}~\bibnamefont {Bell}},\ }\bibfield  {title} {\enquote {\bibinfo {title} {{AMReX: Block-Structured Adaptive Mesh Refinement for Multiphysics Applications}},}\ }\href@noop {} {\bibfield  {journal} {\bibinfo  {journal} {arXiv preprint arXiv:2009.12009}\ } (\bibinfo {year} {2020})}\BibitemShut {NoStop}%
\bibitem [{\citenamefont {Burstedde}\ \emph {et~al.}(2014)\citenamefont {Burstedde}, \citenamefont {Calhoun}, \citenamefont {Mandli},\ and\ \citenamefont {Terrel}}]{burstedde2014forestclaw}%
  \BibitemOpen
  \bibfield  {author} {\bibinfo {author} {\bibfnamefont {C.}~\bibnamefont {Burstedde}}, \bibinfo {author} {\bibfnamefont {D.}~\bibnamefont {Calhoun}}, \bibinfo {author} {\bibfnamefont {K.}~\bibnamefont {Mandli}}, \ and\ \bibinfo {author} {\bibfnamefont {A.~R.}\ \bibnamefont {Terrel}},\ }\bibfield  {title} {\enquote {\bibinfo {title} {Forestclaw: Hybrid forest-of-octrees amr for hyperbolic conservation laws},}\ }in\ \href@noop {} {\emph {\bibinfo {booktitle} {Parallel Computing: Accelerating Computational Science and Engineering (CSE)}}}\ (\bibinfo  {publisher} {IOS Press},\ \bibinfo {year} {2014})\ pp.\ \bibinfo {pages} {253--262}\BibitemShut {NoStop}%
\bibitem [{\citenamefont {Colella}\ \emph {et~al.}(2009)\citenamefont {Colella}, \citenamefont {Graves}, \citenamefont {Ligocki}, \citenamefont {Martin}, \citenamefont {Modiano}, \citenamefont {Serafini},\ and\ \citenamefont {Van~Straalen}}]{colella2009chombo}%
  \BibitemOpen
  \bibfield  {author} {\bibinfo {author} {\bibfnamefont {P.}~\bibnamefont {Colella}}, \bibinfo {author} {\bibfnamefont {D.~T.}\ \bibnamefont {Graves}}, \bibinfo {author} {\bibfnamefont {T.}~\bibnamefont {Ligocki}}, \bibinfo {author} {\bibfnamefont {D.}~\bibnamefont {Martin}}, \bibinfo {author} {\bibfnamefont {D.}~\bibnamefont {Modiano}}, \bibinfo {author} {\bibfnamefont {D.}~\bibnamefont {Serafini}}, \ and\ \bibinfo {author} {\bibfnamefont {B.}~\bibnamefont {Van~Straalen}},\ }\bibfield  {title} {\enquote {\bibinfo {title} {{Chombo software package for AMR applications design document}},}\ }\href@noop {} {\bibfield  {journal} {\bibinfo  {journal} {Available at the Chombo website: http://seesar. lbl. gov/ANAG/chombo/(September 2008)}\ }\textbf {\bibinfo {volume} {2}} (\bibinfo {year} {2009})}\BibitemShut {NoStop}%
\bibitem [{\citenamefont {Bhalla}\ \emph {et~al.}(2013)\citenamefont {Bhalla}, \citenamefont {Bale}, \citenamefont {Griffith},\ and\ \citenamefont {Patankar}}]{bhalla2013unified}%
  \BibitemOpen
  \bibfield  {author} {\bibinfo {author} {\bibfnamefont {A.~P.~S.}\ \bibnamefont {Bhalla}}, \bibinfo {author} {\bibfnamefont {R.}~\bibnamefont {Bale}}, \bibinfo {author} {\bibfnamefont {B.~E.}\ \bibnamefont {Griffith}}, \ and\ \bibinfo {author} {\bibfnamefont {N.~A.}\ \bibnamefont {Patankar}},\ }\bibfield  {title} {\enquote {\bibinfo {title} {{A unified mathematical framework and an adaptive numerical method for fluid--structure interaction with rigid, deforming, and elastic bodies}},}\ }\href@noop {} {\bibfield  {journal} {\bibinfo  {journal} {J. Comput. Phys.}\ }\textbf {\bibinfo {volume} {250}},\ \bibinfo {pages} {446--476} (\bibinfo {year} {2013})}\BibitemShut {NoStop}%
\bibitem [{\citenamefont {Zeng}, \citenamefont {Bhalla},\ and\ \citenamefont {Shen}(2022)}]{zeng2022subcycling}%
  \BibitemOpen
  \bibfield  {author} {\bibinfo {author} {\bibfnamefont {Y.}~\bibnamefont {Zeng}}, \bibinfo {author} {\bibfnamefont {A.~P.~S.}\ \bibnamefont {Bhalla}}, \ and\ \bibinfo {author} {\bibfnamefont {L.}~\bibnamefont {Shen}},\ }\bibfield  {title} {\enquote {\bibinfo {title} {{A subcycling/non-subcycling time advancement scheme-based DLM immersed boundary method framework for solving single and multiphase fluid--structure interaction problems on dynamically adaptive grids}},}\ }\href@noop {} {\bibfield  {journal} {\bibinfo  {journal} {Comput. Fluids}\ ,\ \bibinfo {pages} {105358}} (\bibinfo {year} {2022})}\BibitemShut {NoStop}%
\bibitem [{\citenamefont {Bhalla}\ \emph {et~al.}(2014)\citenamefont {Bhalla}, \citenamefont {Bale}, \citenamefont {Griffith},\ and\ \citenamefont {Patankar}}]{bhalla2014fully}%
  \BibitemOpen
  \bibfield  {author} {\bibinfo {author} {\bibfnamefont {A.~P.~S.}\ \bibnamefont {Bhalla}}, \bibinfo {author} {\bibfnamefont {R.}~\bibnamefont {Bale}}, \bibinfo {author} {\bibfnamefont {B.~E.}\ \bibnamefont {Griffith}}, \ and\ \bibinfo {author} {\bibfnamefont {N.~A.}\ \bibnamefont {Patankar}},\ }\bibfield  {title} {\enquote {\bibinfo {title} {{Fully resolved immersed electrohydrodynamics for particle motion, electrolocation, and self-propulsion}},}\ }\href@noop {} {\bibfield  {journal} {\bibinfo  {journal} {Journal of Computational Physics}\ }\textbf {\bibinfo {volume} {256}},\ \bibinfo {pages} {88--108} (\bibinfo {year} {2014})}\BibitemShut {NoStop}%
\bibitem [{\citenamefont {Li}\ and\ \citenamefont {Kong}(2009)}]{li2009mesh}%
  \BibitemOpen
  \bibfield  {author} {\bibinfo {author} {\bibfnamefont {Y.}~\bibnamefont {Li}}\ and\ \bibinfo {author} {\bibfnamefont {S.-C.}\ \bibnamefont {Kong}},\ }\bibfield  {title} {\enquote {\bibinfo {title} {{Mesh refinement algorithms in an unstructured solver for multiphase flow simulation using discrete particles}},}\ }\href@noop {} {\bibfield  {journal} {\bibinfo  {journal} {Journal of Computational Physics}\ }\textbf {\bibinfo {volume} {228}},\ \bibinfo {pages} {6349--6360} (\bibinfo {year} {2009})}\BibitemShut {NoStop}%
\bibitem [{\citenamefont {Nangia}, \citenamefont {Patankar},\ and\ \citenamefont {Bhalla}(2019)}]{nangia2019dlm}%
  \BibitemOpen
  \bibfield  {author} {\bibinfo {author} {\bibfnamefont {N.}~\bibnamefont {Nangia}}, \bibinfo {author} {\bibfnamefont {N.~A.}\ \bibnamefont {Patankar}}, \ and\ \bibinfo {author} {\bibfnamefont {A.~P.~S.}\ \bibnamefont {Bhalla}},\ }\bibfield  {title} {\enquote {\bibinfo {title} {{A DLM immersed boundary method based wave-structure interaction solver for high density ratio multiphase flows}},}\ }\href@noop {} {\bibfield  {journal} {\bibinfo  {journal} {J. Comput. Phys.}\ }\textbf {\bibinfo {volume} {398}},\ \bibinfo {pages} {108804} (\bibinfo {year} {2019})}\BibitemShut {NoStop}%
\bibitem [{\citenamefont {Peskin}(2002)}]{peskin2002immersed}%
  \BibitemOpen
  \bibfield  {author} {\bibinfo {author} {\bibfnamefont {C.~S.}\ \bibnamefont {Peskin}},\ }\bibfield  {title} {\enquote {\bibinfo {title} {{The immersed boundary method}},}\ }\href@noop {} {\bibfield  {journal} {\bibinfo  {journal} {Acta Numer.}\ }\textbf {\bibinfo {volume} {11}},\ \bibinfo {pages} {479--517} (\bibinfo {year} {2002})}\BibitemShut {NoStop}%
\bibitem [{\citenamefont {Chorin}(1967)}]{chorin1967numerical}%
  \BibitemOpen
  \bibfield  {author} {\bibinfo {author} {\bibfnamefont {A.~J.}\ \bibnamefont {Chorin}},\ }\bibfield  {title} {\enquote {\bibinfo {title} {{The numerical solution of the Navier-Stokes equations for an incompressible fluid}},}\ }\href@noop {} {\bibfield  {journal} {\bibinfo  {journal} {Bulletin of the American Mathematical Society}\ }\textbf {\bibinfo {volume} {73}},\ \bibinfo {pages} {928--931} (\bibinfo {year} {1967})}\BibitemShut {NoStop}%
\bibitem [{\citenamefont {Sussman}\ \emph {et~al.}(1999)\citenamefont {Sussman}, \citenamefont {Almgren}, \citenamefont {Bell}, \citenamefont {Colella}, \citenamefont {Howell},\ and\ \citenamefont {Welcome}}]{sussman1999adaptive}%
  \BibitemOpen
  \bibfield  {author} {\bibinfo {author} {\bibfnamefont {M.}~\bibnamefont {Sussman}}, \bibinfo {author} {\bibfnamefont {A.~S.}\ \bibnamefont {Almgren}}, \bibinfo {author} {\bibfnamefont {J.~B.}\ \bibnamefont {Bell}}, \bibinfo {author} {\bibfnamefont {P.}~\bibnamefont {Colella}}, \bibinfo {author} {\bibfnamefont {L.~H.}\ \bibnamefont {Howell}}, \ and\ \bibinfo {author} {\bibfnamefont {M.~L.}\ \bibnamefont {Welcome}},\ }\bibfield  {title} {\enquote {\bibinfo {title} {{An adaptive level set approach for incompressible two-phase flows}},}\ }\href@noop {} {\bibfield  {journal} {\bibinfo  {journal} {J. Comput. Phys.}\ }\textbf {\bibinfo {volume} {148}},\ \bibinfo {pages} {81--124} (\bibinfo {year} {1999})}\BibitemShut {NoStop}%
\bibitem [{\citenamefont {Sverdrup}, \citenamefont {Nikiforakis},\ and\ \citenamefont {Almgren}(2018)}]{sverdrup2018highly}%
  \BibitemOpen
  \bibfield  {author} {\bibinfo {author} {\bibfnamefont {K.}~\bibnamefont {Sverdrup}}, \bibinfo {author} {\bibfnamefont {N.}~\bibnamefont {Nikiforakis}}, \ and\ \bibinfo {author} {\bibfnamefont {A.}~\bibnamefont {Almgren}},\ }\bibfield  {title} {\enquote {\bibinfo {title} {Highly parallelisable simulations of time-dependent viscoplastic fluid flow with structured adaptive mesh refinement},}\ }\href@noop {} {\bibfield  {journal} {\bibinfo  {journal} {Phys. Fluids}\ }\textbf {\bibinfo {volume} {30}},\ \bibinfo {pages} {093102} (\bibinfo {year} {2018})}\BibitemShut {NoStop}%
\bibitem [{\citenamefont {Zeng}\ \emph {et~al.}(2023)\citenamefont {Zeng}, \citenamefont {Liu}, \citenamefont {Gao}, \citenamefont {Almgren}, \citenamefont {Bhalla},\ and\ \citenamefont {Shen}}]{zeng2023consistent}%
  \BibitemOpen
  \bibfield  {author} {\bibinfo {author} {\bibfnamefont {Y.}~\bibnamefont {Zeng}}, \bibinfo {author} {\bibfnamefont {H.}~\bibnamefont {Liu}}, \bibinfo {author} {\bibfnamefont {Q.}~\bibnamefont {Gao}}, \bibinfo {author} {\bibfnamefont {A.}~\bibnamefont {Almgren}}, \bibinfo {author} {\bibfnamefont {A.~P.~S.}\ \bibnamefont {Bhalla}}, \ and\ \bibinfo {author} {\bibfnamefont {L.}~\bibnamefont {Shen}},\ }\bibfield  {title} {\enquote {\bibinfo {title} {{A consistent adaptive level set framework for incompressible two-phase flows with high density ratios and high Reynolds numbers}},}\ }\href@noop {} {\bibfield  {journal} {\bibinfo  {journal} {J. Comput. Phys.}\ }\textbf {\bibinfo {volume} {478}},\ \bibinfo {pages} {111971} (\bibinfo {year} {2023})}\BibitemShut {NoStop}%
\bibitem [{\citenamefont {Zeng}\ \emph {et~al.}(2022)\citenamefont {Zeng}, \citenamefont {Xuan}, \citenamefont {Blaschke},\ and\ \citenamefont {Shen}}]{zeng2022aparallel}%
  \BibitemOpen
  \bibfield  {author} {\bibinfo {author} {\bibfnamefont {Y.}~\bibnamefont {Zeng}}, \bibinfo {author} {\bibfnamefont {A.}~\bibnamefont {Xuan}}, \bibinfo {author} {\bibfnamefont {J.}~\bibnamefont {Blaschke}}, \ and\ \bibinfo {author} {\bibfnamefont {L.}~\bibnamefont {Shen}},\ }\bibfield  {title} {\enquote {\bibinfo {title} {A parallel cell-centered adaptive level set framework for efficient simulation of two-phase flows with subcycling and non-subcycling},}\ }\href@noop {} {\bibfield  {journal} {\bibinfo  {journal} {J. Comput. Phys.}\ }\textbf {\bibinfo {volume} {448}},\ \bibinfo {pages} {110740} (\bibinfo {year} {2022})}\BibitemShut {NoStop}%
\bibitem [{\citenamefont {Almgren}, \citenamefont {Bell},\ and\ \citenamefont {Szymczak}(1996)}]{almgren1996numerical}%
  \BibitemOpen
  \bibfield  {author} {\bibinfo {author} {\bibfnamefont {A.~S.}\ \bibnamefont {Almgren}}, \bibinfo {author} {\bibfnamefont {J.~B.}\ \bibnamefont {Bell}}, \ and\ \bibinfo {author} {\bibfnamefont {W.~G.}\ \bibnamefont {Szymczak}},\ }\bibfield  {title} {\enquote {\bibinfo {title} {{A numerical method for the incompressible Navier--Stokes equations based on an approximate projection}},}\ }\href@noop {} {\bibfield  {journal} {\bibinfo  {journal} {SIAM J. Sci. Comput.}\ }\textbf {\bibinfo {volume} {17}},\ \bibinfo {pages} {358--369} (\bibinfo {year} {1996})}\BibitemShut {NoStop}%
\bibitem [{\citenamefont {Rider}(1995)}]{rider1995approximate}%
  \BibitemOpen
  \bibfield  {author} {\bibinfo {author} {\bibfnamefont {W.~J.}\ \bibnamefont {Rider}},\ }\href@noop {} {\enquote {\bibinfo {title} {Approximate projection methods for incompressible flow: Implementation, variants and robustness},}\ }\bibinfo {type} {LANL Unclassified Report}\ \bibinfo {number} {LA-UR-94-2000}\ (\bibinfo  {institution} {Los Alamos National Laboratory},\ \bibinfo {year} {1995})\BibitemShut {NoStop}%
\bibitem [{\citenamefont {Balaras}\ and\ \citenamefont {Vanella}(2009)}]{balaras2009adaptive}%
  \BibitemOpen
  \bibfield  {author} {\bibinfo {author} {\bibfnamefont {E.}~\bibnamefont {Balaras}}\ and\ \bibinfo {author} {\bibfnamefont {M.}~\bibnamefont {Vanella}},\ }\bibfield  {title} {\enquote {\bibinfo {title} {Adaptive mesh refinement strategies for immersed boundary methods},}\ }in\ \href@noop {} {\emph {\bibinfo {booktitle} {47th AIAA aerospace sciences meeting including the new horizons forum and aerospace exposition}}}\ (\bibinfo {year} {2009})\ p.\ \bibinfo {pages} {162}\BibitemShut {NoStop}%
\bibitem [{\citenamefont {Cui}\ \emph {et~al.}(2018)\citenamefont {Cui}, \citenamefont {Yang}, \citenamefont {Jiang}, \citenamefont {Huang},\ and\ \citenamefont {Shen}}]{cui2018sharp}%
  \BibitemOpen
  \bibfield  {author} {\bibinfo {author} {\bibfnamefont {Z.}~\bibnamefont {Cui}}, \bibinfo {author} {\bibfnamefont {Z.}~\bibnamefont {Yang}}, \bibinfo {author} {\bibfnamefont {H.-Z.}\ \bibnamefont {Jiang}}, \bibinfo {author} {\bibfnamefont {W.-X.}\ \bibnamefont {Huang}}, \ and\ \bibinfo {author} {\bibfnamefont {L.}~\bibnamefont {Shen}},\ }\bibfield  {title} {\enquote {\bibinfo {title} {A sharp-interface immersed boundary method for simulating incompressible flows with arbitrarily deforming smooth boundaries},}\ }\href@noop {} {\bibfield  {journal} {\bibinfo  {journal} {Int. J. Comput. Methods}\ }\textbf {\bibinfo {volume} {15}},\ \bibinfo {pages} {1750080} (\bibinfo {year} {2018})}\BibitemShut {NoStop}%
\bibitem [{\citenamefont {He}\ \emph {et~al.}(2022)\citenamefont {He}, \citenamefont {Yang}, \citenamefont {Sotiropoulos},\ and\ \citenamefont {Shen}}]{he2022numerical}%
  \BibitemOpen
  \bibfield  {author} {\bibinfo {author} {\bibfnamefont {S.}~\bibnamefont {He}}, \bibinfo {author} {\bibfnamefont {Z.}~\bibnamefont {Yang}}, \bibinfo {author} {\bibfnamefont {F.}~\bibnamefont {Sotiropoulos}}, \ and\ \bibinfo {author} {\bibfnamefont {L.}~\bibnamefont {Shen}},\ }\bibfield  {title} {\enquote {\bibinfo {title} {{Numerical simulation of interaction between multiphase flows and thin flexible structures}},}\ }\href@noop {} {\bibfield  {journal} {\bibinfo  {journal} {J. Comput. Phys.}\ }\textbf {\bibinfo {volume} {448}},\ \bibinfo {pages} {110691} (\bibinfo {year} {2022})}\BibitemShut {NoStop}%
\bibitem [{\citenamefont {Griffith}\ \emph {et~al.}(2007)\citenamefont {Griffith}, \citenamefont {Hornung}, \citenamefont {McQueen},\ and\ \citenamefont {Peskin}}]{griffith2007adaptive}%
  \BibitemOpen
  \bibfield  {author} {\bibinfo {author} {\bibfnamefont {B.~E.}\ \bibnamefont {Griffith}}, \bibinfo {author} {\bibfnamefont {R.~D.}\ \bibnamefont {Hornung}}, \bibinfo {author} {\bibfnamefont {D.~M.}\ \bibnamefont {McQueen}}, \ and\ \bibinfo {author} {\bibfnamefont {C.~S.}\ \bibnamefont {Peskin}},\ }\bibfield  {title} {\enquote {\bibinfo {title} {{An adaptive, formally second order accurate version of the immersed boundary method}},}\ }\href@noop {} {\bibfield  {journal} {\bibinfo  {journal} {J. Comput. Phys.}\ }\textbf {\bibinfo {volume} {223}},\ \bibinfo {pages} {10--49} (\bibinfo {year} {2007})}\BibitemShut {NoStop}%
\bibitem [{\citenamefont {Min}\ \emph {et~al.}(2022)\citenamefont {Min}, \citenamefont {Brazell}, \citenamefont {Tomboulides}, \citenamefont {Churchfield}, \citenamefont {Fischer},\ and\ \citenamefont {Sprague}}]{min2022towards}%
  \BibitemOpen
  \bibfield  {author} {\bibinfo {author} {\bibfnamefont {M.}~\bibnamefont {Min}}, \bibinfo {author} {\bibfnamefont {M.}~\bibnamefont {Brazell}}, \bibinfo {author} {\bibfnamefont {A.}~\bibnamefont {Tomboulides}}, \bibinfo {author} {\bibfnamefont {M.}~\bibnamefont {Churchfield}}, \bibinfo {author} {\bibfnamefont {P.}~\bibnamefont {Fischer}}, \ and\ \bibinfo {author} {\bibfnamefont {M.}~\bibnamefont {Sprague}},\ }\bibfield  {title} {\enquote {\bibinfo {title} {{Towards exascale for wind energy simulations}},}\ }\href@noop {} {\bibfield  {journal} {\bibinfo  {journal} {The International Journal of High Performance Computing Applications}\ ,\ \bibinfo {pages} {10943420241252511}} (\bibinfo {year} {2022})}\BibitemShut {NoStop}%
\bibitem [{\citenamefont {Yao}\ \emph {et~al.}(2022)\citenamefont {Yao}, \citenamefont {Jambunathan}, \citenamefont {Zeng},\ and\ \citenamefont {Nonaka}}]{yao2022massively}%
  \BibitemOpen
  \bibfield  {author} {\bibinfo {author} {\bibfnamefont {Z.}~\bibnamefont {Yao}}, \bibinfo {author} {\bibfnamefont {R.}~\bibnamefont {Jambunathan}}, \bibinfo {author} {\bibfnamefont {Y.}~\bibnamefont {Zeng}}, \ and\ \bibinfo {author} {\bibfnamefont {A.}~\bibnamefont {Nonaka}},\ }\bibfield  {title} {\enquote {\bibinfo {title} {{A massively parallel time-domain coupled electrodynamics--micromagnetics solver}},}\ }\href@noop {} {\bibfield  {journal} {\bibinfo  {journal} {The International Journal of High Performance Computing Applications}\ }\textbf {\bibinfo {volume} {36}},\ \bibinfo {pages} {167--181} (\bibinfo {year} {2022})}\BibitemShut {NoStop}%
\bibitem [{\citenamefont {Young}\ \emph {et~al.}(2009)\citenamefont {Young}, \citenamefont {Lin}, \citenamefont {Fan},\ and\ \citenamefont {Chiu}}]{Young2009TheMO}%
  \BibitemOpen
  \bibfield  {author} {\bibinfo {author} {\bibfnamefont {D.~L.}\ \bibnamefont {Young}}, \bibinfo {author} {\bibfnamefont {Y.~C.}\ \bibnamefont {Lin}}, \bibinfo {author} {\bibfnamefont {C.~M.}\ \bibnamefont {Fan}}, \ and\ \bibinfo {author} {\bibfnamefont {C.~L.}\ \bibnamefont {Chiu}},\ }\bibfield  {title} {\enquote {\bibinfo {title} {The method of fundamental solutions for solving incompressible navier–stokes problems},}\ }\href {https://api.semanticscholar.org/CorpusID:122231170} {\bibfield  {journal} {\bibinfo  {journal} {Engineering Analysis With Boundary Elements}\ }\textbf {\bibinfo {volume} {33}},\ \bibinfo {pages} {1031--1044} (\bibinfo {year} {2009})}\BibitemShut {NoStop}%
\bibitem [{\citenamefont {Schiller}(1933)}]{schiller1933uber}%
  \BibitemOpen
  \bibfield  {author} {\bibinfo {author} {\bibfnamefont {V.~L.}\ \bibnamefont {Schiller}},\ }\bibfield  {title} {\enquote {\bibinfo {title} {{On the basic calculations in gravity processing}},}\ }\href@noop {} {\bibfield  {journal} {\bibinfo  {journal} {Z. Association of German Engineers}\ }\textbf {\bibinfo {volume} {77}},\ \bibinfo {pages} {318--321} (\bibinfo {year} {1933})}\BibitemShut {NoStop}%
\bibitem [{\citenamefont {Gong}\ \emph {et~al.}(2023)\citenamefont {Gong}, \citenamefont {Wu}, \citenamefont {An}, \citenamefont {Zhang},\ and\ \citenamefont {Fu}}]{gong2023cp3d}%
  \BibitemOpen
  \bibfield  {author} {\bibinfo {author} {\bibfnamefont {Z.}~\bibnamefont {Gong}}, \bibinfo {author} {\bibfnamefont {Z.}~\bibnamefont {Wu}}, \bibinfo {author} {\bibfnamefont {C.}~\bibnamefont {An}}, \bibinfo {author} {\bibfnamefont {B.}~\bibnamefont {Zhang}}, \ and\ \bibinfo {author} {\bibfnamefont {X.}~\bibnamefont {Fu}},\ }\bibfield  {title} {\enquote {\bibinfo {title} {{CP3d: A comprehensive Euler-Lagrange solver for direct numerical simulation of particle-laden flows}},}\ }\href@noop {} {\bibfield  {journal} {\bibinfo  {journal} {Computer Physics Communications}\ }\textbf {\bibinfo {volume} {286}},\ \bibinfo {pages} {108666} (\bibinfo {year} {2023})}\BibitemShut {NoStop}%
\bibitem [{\citenamefont {Tschisgale}, \citenamefont {Kempe},\ and\ \citenamefont {Fr{\"o}hlich}(2017)}]{tschisgale2017non}%
  \BibitemOpen
  \bibfield  {author} {\bibinfo {author} {\bibfnamefont {S.}~\bibnamefont {Tschisgale}}, \bibinfo {author} {\bibfnamefont {T.}~\bibnamefont {Kempe}}, \ and\ \bibinfo {author} {\bibfnamefont {J.}~\bibnamefont {Fr{\"o}hlich}},\ }\bibfield  {title} {\enquote {\bibinfo {title} {{A non-iterative immersed boundary method for spherical particles of arbitrary density ratio}},}\ }\href@noop {} {\bibfield  {journal} {\bibinfo  {journal} {Journal of Computational Physics}\ }\textbf {\bibinfo {volume} {339}},\ \bibinfo {pages} {432--452} (\bibinfo {year} {2017})}\BibitemShut {NoStop}%
\bibitem [{\citenamefont {ten Cate}\ \emph {et~al.}(2002)\citenamefont {ten Cate}, \citenamefont {Nieuwstad}, \citenamefont {Derksen},\ and\ \citenamefont {Van~den Akker}}]{tenCate2002}%
  \BibitemOpen
  \bibfield  {author} {\bibinfo {author} {\bibfnamefont {A.}~\bibnamefont {ten Cate}}, \bibinfo {author} {\bibfnamefont {C.~H.}\ \bibnamefont {Nieuwstad}}, \bibinfo {author} {\bibfnamefont {J.~J.}\ \bibnamefont {Derksen}}, \ and\ \bibinfo {author} {\bibfnamefont {H.~E.~A.}\ \bibnamefont {Van~den Akker}},\ }\bibfield  {title} {\enquote {\bibinfo {title} {{Particle imaging velocimetry experiments and lattice-Boltzmann simulations on a single sphere settling under gravity}},}\ }\href@noop {} {\bibfield  {journal} {\bibinfo  {journal} {Physics of Fluids}\ }\textbf {\bibinfo {volume} {14}},\ \bibinfo {pages} {4012--4025} (\bibinfo {year} {2002})}\BibitemShut {NoStop}%
\bibitem [{\citenamefont {Apte}, \citenamefont {Martin},\ and\ \citenamefont {Patankar}(2009)}]{Apte2009ANM}%
  \BibitemOpen
  \bibfield  {author} {\bibinfo {author} {\bibfnamefont {S.~V.}\ \bibnamefont {Apte}}, \bibinfo {author} {\bibfnamefont {M.}~\bibnamefont {Martin}}, \ and\ \bibinfo {author} {\bibfnamefont {N.~A.}\ \bibnamefont {Patankar}},\ }\bibfield  {title} {\enquote {\bibinfo {title} {A numerical method for fully resolved simulation (frs) of rigid particle-flow interactions in complex flows},}\ }\href {https://api.semanticscholar.org/CorpusID:17875968} {\bibfield  {journal} {\bibinfo  {journal} {J. Comput. Phys.}\ }\textbf {\bibinfo {volume} {228}},\ \bibinfo {pages} {2712--2738} (\bibinfo {year} {2009})}\BibitemShut {NoStop}%
\bibitem [{\citenamefont {Liao}\ \emph {et~al.}(2015)\citenamefont {Liao}, \citenamefont {Hsiao}, \citenamefont {Lin},\ and\ \citenamefont {Lin}}]{Liao2015SimulationsOT}%
  \BibitemOpen
  \bibfield  {author} {\bibinfo {author} {\bibfnamefont {C.-C.}\ \bibnamefont {Liao}}, \bibinfo {author} {\bibfnamefont {W.-W.}\ \bibnamefont {Hsiao}}, \bibinfo {author} {\bibfnamefont {T.}~\bibnamefont {Lin}}, \ and\ \bibinfo {author} {\bibfnamefont {C.-A.}\ \bibnamefont {Lin}},\ }\bibfield  {title} {\enquote {\bibinfo {title} {Simulations of two sedimenting-interacting spheres with different sizes and initial configurations using immersed boundary method},}\ }\href {https://api.semanticscholar.org/CorpusID:121741655} {\bibfield  {journal} {\bibinfo  {journal} {Computational Mechanics}\ }\textbf {\bibinfo {volume} {55}},\ \bibinfo {pages} {1191--1200} (\bibinfo {year} {2015})}\BibitemShut {NoStop}%
\bibitem [{\citenamefont {Zhu}, \citenamefont {Hu},\ and\ \citenamefont {Zheng}(2023)}]{zhu2023multiple}%
  \BibitemOpen
  \bibfield  {author} {\bibinfo {author} {\bibfnamefont {Z.}~\bibnamefont {Zhu}}, \bibinfo {author} {\bibfnamefont {R.}~\bibnamefont {Hu}}, \ and\ \bibinfo {author} {\bibfnamefont {X.}~\bibnamefont {Zheng}},\ }\bibfield  {title} {\enquote {\bibinfo {title} {{A multiple-time-step integration algorithm for particle-resolved simulation with physical collision time}},}\ }\href@noop {} {\bibfield  {journal} {\bibinfo  {journal} {International Journal of Multiphase Flow}\ }\textbf {\bibinfo {volume} {163}},\ \bibinfo {pages} {104411} (\bibinfo {year} {2023})}\BibitemShut {NoStop}%
\bibitem [{\citenamefont {Liu}\ \emph {et~al.}(2024)\citenamefont {Liu}, \citenamefont {He}, \citenamefont {Cheng},\ and\ \citenamefont {Zeng}}]{liu2024investigate}%
  \BibitemOpen
  \bibfield  {author} {\bibinfo {author} {\bibfnamefont {D.}~\bibnamefont {Liu}}, \bibinfo {author} {\bibfnamefont {S.}~\bibnamefont {He}}, \bibinfo {author} {\bibfnamefont {H.}~\bibnamefont {Cheng}}, \ and\ \bibinfo {author} {\bibfnamefont {Y.}~\bibnamefont {Zeng}},\ }\bibfield  {title} {\enquote {\bibinfo {title} {{Investigate the efficiency of incompressible flow simulations on CPUs and GPUs with BSAMR}},}\ }\href@noop {} {\bibfield  {journal} {\bibinfo  {journal} {arXiv preprint arXiv:2405.07148}\ } (\bibinfo {year} {2024})}\BibitemShut {NoStop}%
\bibitem [{\citenamefont {Wu}, \citenamefont {Shu},\ and\ \citenamefont {Wan}(2024)}]{wu2024implicit}%
  \BibitemOpen
  \bibfield  {author} {\bibinfo {author} {\bibfnamefont {B.}~\bibnamefont {Wu}}, \bibinfo {author} {\bibfnamefont {C.}~\bibnamefont {Shu}}, \ and\ \bibinfo {author} {\bibfnamefont {M.}~\bibnamefont {Wan}},\ }\bibfield  {title} {\enquote {\bibinfo {title} {{An implicit immersed boundary method for Robin boundary condition}},}\ }\href@noop {} {\bibfield  {journal} {\bibinfo  {journal} {International Journal of Mechanical Sciences}\ }\textbf {\bibinfo {volume} {261}},\ \bibinfo {pages} {108694} (\bibinfo {year} {2024})}\BibitemShut {NoStop}%
\bibitem [{\citenamefont {Sverdrup}, \citenamefont {Almgren},\ and\ \citenamefont {Nikiforakis}(2019)}]{sverdrup2019embedded}%
  \BibitemOpen
  \bibfield  {author} {\bibinfo {author} {\bibfnamefont {K.}~\bibnamefont {Sverdrup}}, \bibinfo {author} {\bibfnamefont {A.}~\bibnamefont {Almgren}}, \ and\ \bibinfo {author} {\bibfnamefont {N.}~\bibnamefont {Nikiforakis}},\ }\bibfield  {title} {\enquote {\bibinfo {title} {{An embedded boundary approach for efficient simulations of viscoplastic fluids in three dimensions}},}\ }\href@noop {} {\bibfield  {journal} {\bibinfo  {journal} {Physics of Fluids}\ }\textbf {\bibinfo {volume} {31}} (\bibinfo {year} {2019})}\BibitemShut {NoStop}%
\bibitem [{\citenamefont {Costa}\ \emph {et~al.}(2015)\citenamefont {Costa}, \citenamefont {Boersma}, \citenamefont {Westerweel},\ and\ \citenamefont {Breugem}}]{costa2015collision}%
  \BibitemOpen
  \bibfield  {author} {\bibinfo {author} {\bibfnamefont {P.}~\bibnamefont {Costa}}, \bibinfo {author} {\bibfnamefont {B.~J.}\ \bibnamefont {Boersma}}, \bibinfo {author} {\bibfnamefont {J.}~\bibnamefont {Westerweel}}, \ and\ \bibinfo {author} {\bibfnamefont {W.-P.}\ \bibnamefont {Breugem}},\ }\bibfield  {title} {\enquote {\bibinfo {title} {{Collision model for fully resolved simulations of flows laden with finite-size particles}},}\ }\href@noop {} {\bibfield  {journal} {\bibinfo  {journal} {Physical Review E}\ }\textbf {\bibinfo {volume} {92}},\ \bibinfo {pages} {053012} (\bibinfo {year} {2015})}\BibitemShut {NoStop}%
\bibitem [{\citenamefont {Gan}, \citenamefont {Zhou},\ and\ \citenamefont {Yu}(2016)}]{gan2016cfd}%
  \BibitemOpen
  \bibfield  {author} {\bibinfo {author} {\bibfnamefont {J.}~\bibnamefont {Gan}}, \bibinfo {author} {\bibfnamefont {Z.}~\bibnamefont {Zhou}}, \ and\ \bibinfo {author} {\bibfnamefont {A.}~\bibnamefont {Yu}},\ }\bibfield  {title} {\enquote {\bibinfo {title} {{CFD--DEM modeling of gas fluidization of fine ellipsoidal particles}},}\ }\href@noop {} {\bibfield  {journal} {\bibinfo  {journal} {AIChE Journal}\ }\textbf {\bibinfo {volume} {62}},\ \bibinfo {pages} {62--77} (\bibinfo {year} {2016})}\BibitemShut {NoStop}%
\end{thebibliography}%

\end{document}